\documentclass[11pt,a4paper]{article}
\pdfoutput=1
\usepackage{amssymb,amsmath,amsfonts, mathtools, mathrsfs}
\usepackage[utf8]{inputenc} 
\usepackage[dvipsnames]{xcolor}
\newif\ifnatbibsort\natbibsorttrue
\relax
\ifnatbibsort\RequirePackage[numbers,sort&compress]{natbib}\else\RequirePackage[numbers,compress]{natbib}\fi
\RequirePackage[colorlinks=true
,urlcolor=blue
,anchorcolor=blue
,citecolor=blue
,filecolor=blue
,linkcolor=blue
,menucolor=blue
,linktocpage=true
,pdfproducer=medialab
,pdfa=true
]{hyperref}
\usepackage{tikz}
\usepackage{mathrsfs}
\usepackage{xcolor}
\usepackage{graphicx,subcaption}
\usepackage{caption}
\usepackage{tensor}
\usepackage{enumerate}
\usepackage{dsfont}
\usepackage{nicefrac}
\usepackage{euscript,stmaryrd,braket}
\usepackage{verbatim}
\usetikzlibrary{arrows,decorations.markings,patterns,arrows.meta,calc,positioning}
\usepackage{slashed}

\newcommand{\bea}{\begin{eqnarray}}
\newcommand{\eea}{\end{eqnarray}}

\newcommand{\be}{\begin{equation}}
\newcommand{\ee}{\end{equation}}
\newcommand{\Tr}{{\text{Tr}}}
\newcommand{\maj}{{\Psi}}
\newcommand{\parti}{{\mathcal{Z}}}
\newcommand{\id}{{0}}
\newcommand{\e}{{\frac{1}{2}}}
\newcommand{\s}{{\frac{1}{16}}}
\newcommand{\x}{{\xi}}
\newcommand{\iup}{{\boldsymbol{1}^{\!\textrm{$\triangleleft$}}}}
\newcommand{\ilow}{{\boldsymbol{1}_{\!\textrm{$\triangleright$}}}}
\definecolor{myBeige}{RGB}{230, 220, 210}

\newcommand{\ThetaN}[2]{\Theta \bigg[\begin{array}{c}\vspace{-.08cm} #1 \\  \vspace{-.03cm}  #2 \end{array} \bigg] }
\newcommand{\ThetaF}[2]{\Theta \Big[\begin{array}{c}\vspace{-.15cm} #1 \\  \vspace{.05cm}  #2 \end{array} \Big] }

\makeatletter
\let\old@startsection=\@startsection
\let\oldl@section=\l@section
\renewcommand{\@startsection}[6]{\old@startsection{#1}{#2}{#3}{#4}{#5}{#6\mathversion{bold}}}
\renewcommand{\l@section}[2]{\oldl@section{\mathversion{bold}#1}{#2}}
\makeatother

\numberwithin{equation}{section}

\newcommand{\genusnlong}[9]{%
\begin{tikzpicture}[baseline={(current bounding box.center)}, scale=1.3]
  \draw[thick] (0,0.5) circle (0.25);
  \draw[thick] (0.25,0.5) -- (0.75,0.5);
  \draw[thick] (1.0,0.5) circle (0.25);
  \draw[thick] (1.25,0.5) -- (1.75,0.5); 
  \node at (2.0,0.5) {$\cdots$};
  \draw[thick] (2.2,0.5) -- (2.7,0.5);
  \draw[thick] (2.95,0.5) circle (0.25);
  \node at (0,0.9) {#1};
  \node at (0,0.1) {#2};
  \node at (0.48,0.7) {#3};
  \node at (1.0,0.9) {#4};
  \node at (1.0,0.1) {#5};
  \node at (1.42,0.7) {#6};
  \node at (2.5,0.7) {#7};
  \node at (2.95,0.9) {#8};
  \node at (2.95,0.1) {#9};
\end{tikzpicture}%
}
\newcommand{\genusii}[3]{%
\begin{tikzpicture}[baseline={(current bounding box.center)}, scale=1.3]
  \draw[thick] (0,0) circle (0.25);
  \draw[thick] (0.25,0) -- (0.5,0);
  \draw[thick] (0.75,0) circle (0.25);
  \node at (0,0.47) {\scriptsize #1};
  \node at (0.37,0.23) {\scriptsize #2};
  \node at (0.75,0.47) {\scriptsize #3};
\node at (0.4,-0.5) {};
\end{tikzpicture}%
}
\newcommand{\loopi}[4]{%
\begin{tikzpicture}[baseline={(current bounding box.center)}, scale=1.3]
  \draw[thick] (0.25,0.5) -- (0.5,0.5);
  \draw[thick] (0.75,0.5) circle (0.25);
  \draw[thick] (1,0.5) -- (1.25,0.5); 
  \node at (0.37,0.7) {\scriptsize #1};
  \node at (0.75,0.95) {\scriptsize #2};
  \node at (0.75,0.05) {\scriptsize #3};
  \node at (1.12,0.7) {\scriptsize #4};
\end{tikzpicture}%
}

\usepackage{tocloft} 
\definecolor{csheet}{RGB}{243, 221, 219}
\begin{document}
	\renewcommand{\thefootnote}{\arabic{footnote}}
	
	\overfullrule=0pt
	\parskip=2pt
	\parindent=12pt
	\headheight=0in \headsep=0in \topmargin=0in \oddsidemargin=0in

	\vspace{ -3cm} \thispagestyle{empty} \vspace{-1cm}
	\begin{flushright} 
		\footnotesize
		\textcolor{red}{\phantom{print-report}}
	\end{flushright}

\begin{center}
	\vspace{.0cm}

	{\Large\bf \mathversion{bold}
	Jones index from R\'enyi entropies
	}
	\\
	\vspace{.25cm}
	\noindent
	{\Large\bf \mathversion{bold}
	in the Ising conformal field theory
    }

	\vspace{0.8cm}
		Valentin Benedetti$^{\,a,c}$,
        Isai Davila-Cuba$^{\,b,c}$
		and 
        Erik Tonni$^{\,b,c}$ 	
	\vskip  0.7cm
	
	\small
	{\em

    \vskip 0.05in

		$^{a}\,$The Abdus Salam International Centre for Theoretical Physics, \\
		Strada Costiera 11, Trieste 34151, Italy
		\vskip 0.09cm
        $^{b}\,$SISSA, Via Bonomea 265, 34136, Trieste, Italy 
        \vskip 0.09cm
		$^{c}\,$INFN, Sezione di Trieste, Via Valerio 2, I-34127 Trieste, Italy
      
	}  
	\normalsize

\end{center}

\vspace{0.3cm}
\begin{abstract} 

We study the relation between the Jones index 
and the R\'enyi entropies of two disjoint intervals on the line
and of the ground state
for a generic value of the R\'enyi index
in the two conformal field theory models 
given by the Ising model and a free Majorana fermion,
where Haag duality is satisfied. 
The analytic expressions of the crossing asymmetry 
for all the submodels displaying a violation of the Haag duality
that are closed under the fusion rules are obtained. 
In the limiting regime where the two intervals become adjacent,
the leading term of the expansion of the crossing asymmetry 
provides the Jones global index, 
for any finite value of the R\'enyi index.

\end{abstract}

\newpage

\vspace{-3cm}
\tableofcontents

\section{Introduction}
\label{sec_intro}

The Hilbert space of a Quantum Field theory (QFT) 
can be partitioned into superselection sectors
whenever the set made by all the local operators 
does not generate all the states 
with finite energy obtained from the vacuum \cite{haag2012local,araki1999mathematical}. 

The Doplicher-Haag-Roberts (DHR) superselection sectors \cite{DHR1,DHR2,DHR3}
are the prototypical example of non trivial superselection sectors 
and occur in a theory where the algebra associated to every spacetime region is the neutral one under some zero-form symmetry 
existing in a more complete theory that contains it. In this setup,  the charged sectors in the complete model 
are represented as endomorphisms of the local algebras 
of the smaller, non complete model.
In algebraic QFT, the existence of these DHR superselection sectors 
is known to be related to the violation of the Haag duality 
of the von Neumann algebra associated to 
the union of disjoint regions \cite{Casini:2019kex,Casini:2020rgj,Casini:2025lfn}. 
This violation occurs whenever this algebra is strictly included 
into the algebra made by all operators that commute with the ones in the causally complementary region. 
This proper inclusion is quantified by the Jones index, 
which is a standard tool to study the inclusions of von Neumann algebras  \cite{Jones83,KOSAKI1986123,L11,L12,LongoXu}. 
For this inclusion, the Jones index 
is known as the global index \cite{LongoNets}, and 
probes the DHR endomorphisms \cite{BenedettiMinimal}. 
For a non-complete model,
these endomorphisms are isomorphic 
to the representations of a group  
in $d> 2$ spacetime dimensions  \cite{DHR1,DHR2,DHR3}
and to a braided category in $d=2$ \cite{FRS1,FRS2,RehrenChiral,LongoKawaBook}. 
The value of the index always coincides 
with the total quantum dimension 
of the category of DHR sectors \cite{LongoNets}.

In the particular case of a 
conformal field theory in two spacetime dimensions (CFT$_2$), 
the modular invariance of the model defined on a generic Riemann surface 
is an important consistency condition \cite{CardyOperator,Segal:1988}
that provides various important results.  
In algebraic QFT, the modular invariance of a CFT$_2$
has a local interpretation in terms of local algebras associated to spacetime regions. 
In particular, it has been shown that the modular invariance implies 
the absence of superselection sectors and 
the absence of the Haag duality violations.
Moreover, 
the global Jones index is equal to one whenever modular invariance occurs  \cite{Rehren:2000ti,LongoKawaMuger,Muger:2009sq,LongoBlack,Benedetti:2024dku}.
 It is insightful to explore 
 consistent submodels of a modular invariant CFT$_2$
 whose local algebras in a Minkowski spacetime 
 obey all the axioms of algebraic QFT \cite{haag2012local,araki1999mathematical}. 
 These models,
 whose spectrum does not arise 
 from a modular invariant CFT$_2$ \cite{Rehren:2000ti},
 display non trivial  DHR superselection sectors;
 hence the above mentioned global Jones index associated to their inclusion 
 in a modular invariant CFT$_2$ is larger than one. 
For a rational CFT$_2$ with central charge $c<1$,  
all the possible values of the Jones index 
and all their DHR categories have been determined 
\cite{LongoKawa,BenedettiMinimal}.
Recently, this analysis has been carried out also 
for $c>1$, in specific coset CFT$_2$ theories \cite{Benedetti:2026drn}.

The global Jones index can be obtained also through 
some quantities introduced in quantum information theory
\cite{Xu:2018uxc,Casini:2019kex,Casini:2020rgj,MaganPontello,Hollands:2020owv}. 
We are interested in a CFT$_2$ model $\mathcal{T}$ 
on a line and in its ground state, 
when the line is bipartite by the union of two disjoint 
spatial intervals $R_1\equiv[u_1,v_1]$ and $R_2\equiv[u_2,v_2]$,
with $v_1 < u_2$.
The R\'enyi entropies $S_n(R_1\cup R_2)$ for this setup, 
that have been widely explored in the literature 
\cite{Calabrese:2004eu,Caraglio:2008pk,Furukawa:2008uk,Casini:2008wt,Casini:2009vk,Calabrese:2009ez,Calabrese:2010he,Headrick:2010zt,Headrick:2012fk,
Coser:2013qda,DeNobili:2015dla,Herzog:2016ohd,Grava:2021yjp},
read
\be
S_n(R_1\cup R_2)  
= 
\frac{c}{6} \left(1 + \frac{1}{n}\right)
\log\!\left(\frac{(v_1-u_1)(v_2-u_2)(1-\x)}{\epsilon^2}\right) 
+ 
\frac{\log \! \big[ \mathcal{F}_{\mathcal{T} \!, \, n} (\x) \big]}{1-n}
+
\frac{2\, \log C_n}{1-n} 
\label{entropydisjoint}
\ee 
where $c$ is the central charge, 
$\epsilon$ is the UV cutoff,
the integer $n \geqslant 2$ is the R\'enyi index 
and $\mathcal{F}_{\mathcal{T} \!, \, n}(\x)$ is a model dependent function 
of the cross ratio $\x \in (0,1)$ 
associated to the four endpoints of $R_1\cup R_2$.
The normalization constant $C_n$,
that depends only on $n$, 
allows us to set $\mathcal{F}_{\mathcal{T} \!, \, n}(0)=1$,
while $\mathcal{F}_{\mathcal{T} \!, \, n}(\x)$ is a highly non trivial function that encodes all the conformal data of the CFT$_2$ model. 
The interplay among the modular invariance, Haag duality and the transformation properties of $\mathcal{F}_{\mathcal{T} \!, \, n}(\x)$ 
has been first highlighted in \cite{Calabrese:2009ez}.
In \cite{Benedetti:2024dku}, it has been found that 
the R\'enyi entropies $S_n(R_1\cup R_2)$ 
allow to explore also the DHR sectors 
through the following quantity 
dubbed crossing asymmetry
\be
A_{\mathcal{T} \!, \, n}(\x)
\equiv 
\frac{1}{n-1}
\log\! 
\left(\frac{\mathcal{F}_{\mathcal{T} \!, \, n}(\x)}{\mathcal{F}_{\mathcal{T} \!, \, n}(1-\x)}\right) .
\label{asymetrydefintro}
\ee
In a CFT$_2$ displaying modular invariance, 
$\mathcal{F}_{\mathcal{T} \!, \, n}(\x)$ is symmetric under $\x\mapsto 1-\x$;
hence $A_n(\x)$ vanishes identically in $\x \in (0,1)$.
Instead, whenever the DHR sectors are not trivial, 
the crossing asymmetry (\ref{asymetrydefintro})
provides the global Jones index $\mu_\mathcal{T} $ as follows 
\cite{Benedetti:2024dku}
\be 
\label{jones-index-limit-asymmetry-intro}
\lim_{\x\to 0} A_{\mathcal{T} \!, \, n}(\x)
\,=\,
-\lim_{\x\to 1} A_{\mathcal{T} \!, \, n}(\x)
\,=\,
\frac{1}{2}\log{\mu}_\mathcal{T} 
\ee 
independently of the value of the R\'enyi index $n$,
where the limits $\x\to 0$ and $\x\to 1$ correspond to 
the asymptotic regimes given by 
large separation distance and adjacent intervals 
respectively.
The validity of (\ref{jones-index-limit-asymmetry-intro}) for $n>2$ 
has been argued in \cite{Benedetti:2024dku}.
However, for a generic value of the R\'enyi index,
explicit examples have not been explored.

In this manuscript, we explore further the relation 
(\ref{jones-index-limit-asymmetry-intro}) 
for a generic finite value of the R\'enyi index $n$,
by considering the special cases given by 
the Ising CFT$_2$,
a well-known rational CFT$_2$ with $c=1/2$ 
and three primary fields
\cite{Belavin:1984vu,Ginsparg:1988ui,DiFrancesco:1997nk, Mussardo:2020rxh}
and the free Majorana fermion whose entanglement entropies 
have been studied in \cite{Casini:2005rm,Casini:2009vk},
which is not invariant under all the modular transformations \cite{Headrick:2012fk}. 
Our analysis exploits the fact that the R\'enyi entropies 
(\ref{entropydisjoint}) for these models 
are known explicitly \cite{Calabrese:2010he,Casini:2009vk}
and can be written through the partition function of the model on a specific Riemann surface of genus $n-1$.

The outline of this paper is as follows. 
In Sec.\,\ref{subfactors}, the basic notions of algebraic QFT 
and DHR superselection sectors theory are reviewed, 
focussing on the Ising CFT$_2$. 
In Sec.\,\ref{JonesIsing},  we describe the determination of 
the global index through the Rényi mutual information, which can be written as a partition function on a specific Riemann surface of genus $g=n-1$.
In Sec.\,\ref{JonesIsing2} and Sec.\,\ref{JonesIsingn},
these notions are applied to the $n=2$ case and 
to the case of generic integer $n \geqslant 2$ respectively,
obtaining an expression for (\ref{asymetrydefintro}) 
in the subtheories of the Ising CFT$_2$. 
These results are also employed to study the properties of the crossing asymmetry in this example, 
like the monotonicity and the $n\to + \infty$ limit. 
In Sec.\,\ref{Fermions}, 
the crossing asymmetry (\ref{asymetrydefintro}) for the submodels of the 
free Majorana fermion is explored. 
Some conclusions are drawn in Sec.\,\ref{sec-conclusions}.
Further technical details supporting some parts of the main text 
are reported in the 
appendices\;\ref{app-details0higher-genus} and \ref{modular}.

\section{Jones index in the Ising CFT$_2$
\label{subfactors}}

In this section,  we describe the tools we employ to study  von Neumann algebras in the Ising CFT$_2$ 
and their corresponding submodels. 
In Sec.\,\ref{completenessqft}, we discuss the assignations of algebras 
to regions in QFT and in  CFT$_2$. 
In Sec.\,\ref{jones} a brief summary of the relation 
between the theory of superselection sectors and the Jones index
is provided. 
Finally, in Sec.\,\ref{Ising} we discuss the example 
of the Ising CFT$_2$ in terms of von Neumann algebras.

\subsection{Completeness and modular invariance 
\label{completenessqft} }

In the Haag-Araki approach to algebraic QFT 
\cite{HAraki1, HAraki2},
a von Neumann algebra of the observable bounded operators $\mathcal{A}(R)$ 
is assigned to each codimension one spatial region $R$ 
contained in a constant time Cauchy slice $\Sigma$. 
In our analysis, we always consider this formulation in the 
$(d+1)$-dimensional Minkowski spacetime $\mathbb{R}^{d,1}$.
The assignment of a von Neumann algebra 
to a set of complementary regions is known as a choice of net,
and every net must satisfy certain axioms, in order to provide a consistent QFT model \cite{haag2012local,araki1999mathematical}.  
One of these axioms imposes the causality,
which is implemented by requiring 
that the algebras associated to regions that are 
spacelike separated commute. 
This means that, given two spatial regions $ R_1 $ and $R_2$ 
that are spacelike separated, 
$ \big[ a_1, a_2 \big] = 0$ for any choice of 
$a_1\in \mathcal{A}(R_1)$ and $a_2 \in  \mathcal{A}(R_2)$.
The causality constraint can be equivalently expressed 
in terms of a single region $R$ as follows
\be
\label{causality}
\mathcal{A}( R) \subseteq \mathcal{A}'(R')
\ee
where the $'$ acting over an algebra $\mathcal{A}(R)$ 
represents the commutant $\mathcal{A}'(R)$, 
namely the set of all the operators 
that commute with every operator in $\mathcal{A}(R)$;
while $'$ acting over the spatial region $R$ 
represents the causal complement of $R$, 
which is the set of points that are spacelike separated from $R$. 
Considering $R$ in a constant time Cauchy slice $\Sigma$, 
the region $R'$ is the set of points not included in $R$, 
since $R\cup R' = \Sigma$.\footnote{We assume that the algebra associated with a spatial region $R$ coincides with the one of its causal completion ${R}''$. This property is known as the local time slice axiom $\mathcal{A}(R'')=\mathcal{A}( R)$ and is satisfied in theories with a well-defined Hamiltonian density (excluding the generalized free fields \cite{Gff1,Gff2,Gff3}). In addition, the fact that the algebras that we are considering are von Neumann algebras implies that $\mathcal{A}({R} )=\mathcal{A}''({R} )$. 
By the von Neumann's double commutant theorem, 
the latter property is equivalent to the closure under weak topology 
in an algebra that includes the identity.}

The causality constraint (\ref{causality}) is a necessary property for every bosonic net, and it cannot be violated. 
However, the saturation of causality, namely
\be
\mathcal{A}(R) = \mathcal{A}'(R')
\label{haagduality}
\ee
which is usually called Haag duality,
is not a standard requirement for all regions. 
Furthermore, Haag duality violations can be considered a diagnosis of a theory 
containing non local operators \cite{Casini:2020rgj,Casini:2021zgr,Benedetti:2022zbb,Shao:2025mfj,Harlow:2025cqc}.

In a given theory  $\mathcal{T}$, 
the assignment of an algebra to a given region $R$
can always be done by considering 
the algebra $\mathcal{A}_\mathcal{T}(R)$ 
made by the operators that are locally generated inside the region, 
which can be formally defined by gluing the algebras of balls $B_i$ 
covering the region $R$ as follows
\be 
\mathcal{A}_\mathcal{T}(R) 
\,\equiv \!
\bigvee_{B_i \in R } \mathcal{A}_\mathcal{T} (B_i)
= 
\Bigg(\,
\bigcup_{B_i \in R } \mathcal{A}_\mathcal{T} (B_i)
\Bigg)'' .
\label{additivealgebra}
\ee
The choice of this algebra for every region $R$ provides the additive net \cite{Casini:2020rgj} because 
(\ref{additivealgebra}) always satisfies the additivity property
\be 
\mathcal{A}_\mathcal{T} (R_1\cup R_2)=\mathcal{A}_\mathcal{T} (R_1)\vee \mathcal{A}_\mathcal{T} (R_2)= 
\big(\mathcal{A}_\mathcal{T} (R_1)\cup \mathcal{A}_\mathcal{T} (R_2) 
\big)' \,.
\label{additivity}
\ee
Since the additive algebra $\mathcal{A}_\mathcal{T}(R)$
is the smallest algebra that can be assigned to the region $R$,
the causality condition (\ref{causality}) implies that 
the largest one is its commutant, namely
\be
\hat{\mathcal{A}}_\mathcal{T}(R)\equiv\mathcal{A}_\mathcal{T}(R')' \,.
\label{maximalalgebra}
\ee
The algebra $\hat{\mathcal{A}}_\mathcal{T}(R)$ is the maximal algebra of $R$ and generically includes all possible non local operators in the region.

In some theories, 
the additive algebra and the maximal one coincide for every region, 
namely
\be 
\hat{\mathcal{A}}_\mathcal{T}(R)=\mathcal{A}_\mathcal{T}(R)
\;\;\;\qquad\;\;\;
\forall \,R \in \Sigma
\label{haag-duality-any-R}
\ee
meaning that the additive algebra (\ref{additivealgebra}) satisfies 
the Haag duality (\ref{haagduality}) for all possible regions. 
The requirement (\ref{haag-duality-any-R}) 
is the completeness condition for the QFT model $\mathcal{T}$,
which can be understood as a spectrum completeness, 
meaning that  $\mathcal{T}$ 
contains enough charged operators  to break all non local operators 
into local ones \cite{Casini:2021zgr,Benedetti:2022ofj}. 
Hereafter, $\mathcal{C}$ corresponds to a complete model.  
We remark that the breaking of the completeness condition (\ref{haag-duality-any-R}) does not lead to pathological models.  Indeed, for instance, the free Maxwell theory is not a complete model  because 
the Wilson loops and the 't Hooft loops violate the Haag duality 
in regions with the topology of the annulus.

In a rational CFT$_2$, the  Haag duality 
has been widely explored \cite{L11,L12,LongoXu,LongoNets,BenedettiMinimal,RehrenChiral,LongoKawaBook,Rehren:2000ti,LongoKawaMuger,Muger:2009sq,LongoBlack,Benedetti:2024dku,LongoKawa,FRS1,FRS2}.
Interestingly, it has been shown that the modular invariance 
of a CFT$_2$ can be related to its completeness \cite{Rehren:2000ti,LongoKawaMuger,Muger:2009sq,
LongoBlack,Benedetti:2024dku}. 
A rational CFT$_2$ model $\mathcal{T}$ with central charge $c$ 
can be constructed from the finite number of the chiral representations 
of the Virasoro algebra, 
which are labeled by their conformal weights $(h,\bar{h})$. 
This implies that the full Hilbert space $\mathcal{H}_\mathcal{T}$ of  $\mathcal{T}$  decomposes into the corresponding Verma modules as follows \cite{Belavin:1984vu}
\be 
\label{H-T-dec}
\mathcal{H}_\mathcal{T}
= \bigoplus_{h,\bar{h}} 
Z^\mathcal{T}_{h,\bar{h}} 
\,V_h\otimes\overline{V}_{\bar{h}}
\ee
where the Verma module $V_h$ is generated  
by the primary state corresponding to the primary field with conformal dimension $h$ and its descendants,
while $Z_{h,\bar{h}}^\mathcal{T}$ are non-negative integers.
In this work, we extract some information 
about the completeness of a model 
by employing the partition functions of a CFT$_2$ 
on higher genus Riemann surfaces.
It is very instructive to consider first the case of the torus.

The partition function of a CFT$_2$ on a torus 
characterised by the modular parameter $\tau$, 
whose imaginary part is strictly positive, 
can be written as follows  \cite{CardyOperator,DiFrancesco:1997nk} 
\be 
\parti_{\mathcal{T}\!,1} 
= 
\Tr_{\mathcal{H}_\mathcal{T}}\!
\left( q^{L_0-\tfrac{c}{24}}\;\bar{q}^{\bar{L}_0-\tfrac{c}{24}}\right)
= 
\sum_{h,\bar{h}} Z^{\mathcal{T}}_{h,\bar{h}} \;\chi_h(\tau)\,\bar{\chi}_{\bar{h}}(\bar{\tau})
\;\;\qquad\;\; 
q\equiv \textrm{e}^{2\pi \textrm{i} \tau}
\label{toruspart}
\ee
where $\parti_{\mathcal{T}\!,g}$ denotes the partition function of the model $ \mathcal{T}$ on the genus $g$ Riemann surface. 
In the case of the torus, where $g=1$,
the expression is given in terms of 
the Virasoro character $\chi_h(\tau)$ associated to $V_h$,
defined by 
\be 
\chi_h(\tau) \equiv 
\Tr_{V_h}\!\left( q^{L_0-\tfrac{c}{24}}\right) .
\ee

Modular invariance is a consistency condition 
for the partition function on the torus. 
It corresponds to the invariance under the 
$\textrm{Sp}(2,\mathbb{Z})$ transformations of the modular parameter, 
which are generated by 
$T : \tau \mapsto \tau + 1$ 
and  $S : \tau \mapsto -1/\tau$ \cite{CardyOperator,Siegelsym}.
Thus, the modular invariance of (\ref{toruspart}) 
is guaranteed by the validity of the following two conditions 
\be
\parti_{\mathcal{T}\!,1} (\tau +1)=\parti_{\mathcal{T}\!,1} (\tau)
\;\;\;\;\qquad\;\;\;\;
\parti_{\mathcal{T}\!,1} (-1/\tau) =\parti_{\mathcal{T}\!,1} (\tau) \,.
\label{tsinv}
\ee
In rational CFT$_2$, the transformations are implemented over the characters in (\ref{toruspart}) through 
the matrices $T$ and $S$, defined respectively by 
\be 
\chi_h\big(\tau+1\big)=\sum_{h'} T_{h,h'} \,\chi_h (\tau)
\;\;\;\qquad\;\;\;
\chi_h(-1/\tau)
=
\sum_{h'} S_{h,h'} \, \chi_h(\tau )  \,.
\label{ST}
\ee

Both properties in (\ref{tsinv}) 
have a local interpretation in terms of algebras in Minkowski spacetime. 
The invariance under $\tau \mapsto \tau + 1$ enforces 
the causality constraint (\ref{causality}) 
of the fields in bosonic models \cite{Rehren:2000ti},
which is a crucial property of every bosonic model 
that cannot be violated. 
Instead, the invariance under $\tau \mapsto -1/\tau$
is equivalent to the completeness condition (\ref{haag-duality-any-R}) 
of the model \cite{LongoKawaMuger,Benedetti:2024dku}. 
Thus, the models that are invariant under $\tau \mapsto -1/\tau$
are complete, providing a local interpretation of this invariance
\cite{Benedetti:2024dku}.

\subsection{The Jones index as a measure of completeness}
\label{jones}

In order to understand the subalgebras 
of the Ising CFT$_2$ that are not complete, 
we introduce the Jones index as a way to quantify completeness
and also some useful notions involving 
subalgebra inclusions and modular categories
(see \cite{haag2012local,LongoKawaBook} for reviews).

It is instructive to describe 
the setup of Sec.\,\ref{completenessqft} 
in terms of DHR endomorphisms \cite{DHR1,DHR2,DHR3}. 
Consider a complete model $\mathcal{C}$ and a submodel $\mathcal{T}$ defining a strict inclusion $\mathcal{T}\subsetneq \mathcal{C}$.
Hence, $\mathcal{C}$ contains some irreducible charged sectors 
$\psi_{r}$ (labelled by the index $r$) that do not occur in $\mathcal{T}$.
Extra indexes might occur in the sector corresponding to $r$,
providing $\psi_r^i$,
whose occurrence and range depend on $r$,  
but these indexes are suppressed in our analysis 
to enlighten the notation. 
Since the definition of $\mathcal{T}$ implies its neutrality under 
all the elements of any charged sector $\psi_{r}$, 
we can employ the operators $\psi_r$ to construct the 
endomorphisms $\rho_r$ of the local algebras of $\mathcal{T}$ given by $\psi_{r}\, t = \rho_r(t) \,\psi_{r}$,
for all $t\in\mathcal{T}$. 
These endomorphisms obey some fusion rules that can be schematically written as \cite{DHR1,DHR2,DHR3,LongoKawaBook,RehrenChiral,FRS1,FRS2} 
\be 
\big(\rho_r \circ \rho_{r'} \big) (t)
\equiv 
\rho_r \big(\rho_{r'} (t)\big)= \bigoplus_{r''} \tilde{N}_{r,r'}^{r''}  \rho_{r''} (t) 
\;\;\;\qquad\;\;\;
t\in\mathcal{T} \,.
\label{fusiondhr}
\ee
The embedding  of $\mathcal{T}$ into $\mathcal{C}$ 
can be described through a canonical endomorphism $\rho_{\mathcal{C}:\mathcal{T}}$  
that can be written as a combination of the irreducible endomorphisms $\rho_r(t)$ 
as follows \cite{LongoNets}
\be 
\rho_{\mathcal{C}:\mathcal{T}}
= \bigoplus_{r} n_r \, \rho_r 
\label{ecanonical}
\ee
where $n_r$ are positive integers 
taking into account the multiplicities of the occurrence of $\rho_r$.

 The Jones index $\big[\, \mathcal{A}_2 \! : \! \mathcal{A}_1 \big]$ is defined for any inclusion of von Neumann algebras 
 $\mathcal{A}_1 \subseteq \mathcal{A}_2 $ 
 and quantifies the relative size of 
 $\mathcal{A}_2$ with respect to $\mathcal{A}_1$ \cite{Jones83}. 
 Since in our analysis we are interested in the algebras of operators that are locally generated inside a region $R$ (see (\ref{additivealgebra})),
 we focus on the inclusion relation  
 $ \mathcal{A}_\mathcal{T} (R)  \subseteq \mathcal{A}_\mathcal{C} (R)$, where $\mathcal{T}$ is a non complete submodel 
 included in the complete model $\mathcal{C}$.
 The corresponding Jones index is 
\be 
\lambda_{\mathcal{C}:\mathcal{T}}
\equiv 
\big[\,
\mathcal{A}_\mathcal{C} (R) 
: 
\mathcal{A}_\mathcal{T} (R)
\,\big] \,.
\label{jonesinddef}
\ee
When $R$ is an interval, (\ref{jonesinddef}) provides 
the dimension of the canonical endomorphism (\ref{ecanonical})
\cite{L11,L12,LongoXu}. 
The index (\ref{jonesinddef}) 
can be expressed as follows 
\be 
\lambda_{\mathcal{C}:\mathcal{T}}= \sum_r n_r \,\tilde{d}_r
\label{icanonical}
\ee
where $\tilde{d}_r$ is the quantum dimension of the irreducible endomorphism defined as the largest eigenvalue of $\tilde{N}_{r,r'}^{r''}$ in (\ref{fusiondhr}) at fixed $r$.

Any Jones index of the form (\ref{jonesinddef}) 
can take only specific values, 
constrained as follows \cite{Jones83}:
\\
{\bf (a)}
The bound $\lambda_{\mathcal{C}:\mathcal{T}}\geqslant 1$ holds.
The value $\lambda_{\mathcal{C}:\mathcal{T}}=1$ 
corresponds to $\mathcal{A}_\mathcal{T} (R)= \mathcal{A}_\mathcal{C} (R)$, while the values $\lambda_{\mathcal{C}:\mathcal{T}}>1$ correspond 
to strict inclusions $\mathcal{A}_\mathcal{T} (R) \subset \mathcal{A}_\mathcal{C} (R) $.
\\
{\bf (b)}
The values $\lambda_{\mathcal{C}:\mathcal{T}}<4$ 
    belong to the discrete sequence   $\lambda_{\mathcal{C}:\mathcal{T}} = 
    4 [ \cos(\pi/n)]^2$ for an integer $n \geqslant 3$. 
    Any real number $\lambda_{\mathcal{C}:\mathcal{T}}\geqslant 4$  
    is allowed.
    \\
{\bf (c)}
$\lambda_{\mathcal{C}:\mathcal{T}}$ can  become infinite.

Let us discuss two relevant examples. 
First, consider
 $\mathcal{T}=\mathcal{C}/G$ given by  the fixed-point algebra under the action of a finite group $G$ acting over $\mathcal{C}$.
 In this case, the canonical endomorphism reads \cite{DHR1,DHR2,DHR3} 
\be 
\rho_{\mathcal{C}:\mathcal{C}/G}= \bigoplus_{r\in G} \tilde{d}_r \rho_r \label{egroup}
\ee
where $\tilde{d}_r$ are defined as in (\ref{icanonical}), 
by diagonalizing the group fusion rules. 
The total quantum dimension of the endomorphism, 
that gives the index, 
is the order of the group $|G|$. 
More precisely, comparing (\ref{ecanonical}) with (\ref{egroup}) and applying (\ref{icanonical}), we have
\be
\lambda_{\mathcal{C}:\mathcal{C}/G}= 
\sum_{r\in G}  \tilde{d}_r^{\, 2} = |G|  \,.
\label{igroup}
\ee 
The second example is the sector $\mathcal{T}=\mathcal{T}_0$, 
made only by the identity sector of a CFT$_2$.
In this case, the irreducible endomorphisms $\rho_{(h,\bar{h})}$ are labeled by the conformal dimensions $(h,\bar{h})$ and can be written as the product of chiral endomorphisms, 
i.e.  $\rho_{(h,\bar{h})}=\rho_{h}\otimes \rho_{\bar{h}}$, 
where $n_{(h,\bar{h})}=1$. 
In this case, the canonical endomorphism 
takes the following form \cite{LongoKawaMuger,LongoKawaBook} 
\be 
\rho_{\mathcal{C}:\mathcal{T}_0}
= 
\bigoplus_{h,\bar{h}} Z_{h,\bar{h}}^{\mathcal{C}} \,\rho_{(h,\bar{h} )}
\label{evirasoro}
\ee
where $Z_{h,\bar{h}}^{\mathcal{C}}$  
is the matrix defining the torus partition function 
of a complete model $\mathcal{C}$ containing $\mathcal{T}_0$, 
like in (\ref{toruspart}) for $\mathcal{T}$.
In this example, following (\ref{icanonical}) again, 
the index reads
\be 
\lambda_{\mathcal{C}:\mathcal{T}_0}=\sum_{(h,\bar{h})} Z^\mathcal{C}_{h,\bar{h}} \,d_{h} \, d_{\bar{h}}  \,.
\label{singlecft}
\ee 
In this expression, we have employed the fact that, in this case,
the quantum dimensions of the endomorphisms $\tilde{d}_h$ in (\ref{icanonical}) coincide with the ones of the primary fields ${d}_h$ defined through the matrix occurring in the second transformation rule 
in (\ref{ST}) \cite{LongoKawaMuger} as follows
\be 
{d}_h=\frac{S_{h,0}}{S_{0,0}} \label{dim-general}
\ee
where the index $0$ corresponds to the identity sector. 
Note that (\ref{dim-general}) is 
a consequence of obtaining the chiral fusion rules through the Verlinde formula 
\cite{Verlinde-lines,CardyVerlinde}.

The index (\ref{jonesinddef}) that we discussed so far
as a measure of completeness is based on the single interval 
and requires the specification of a complete theory. 
Instead, we can consider the union of two 
disjoint intervals $R= R_1 \cup R_2$ , 
and focus on the inclusion of the additive algebra 
$\mathcal{A}_\mathcal{T}(R_1 \cup R_2)$ 
into the maximal one $\hat{\mathcal{A}}_\mathcal{T}(R_1 \cup R_2)$
(see (\ref{additivealgebra}) and (\ref{maximalalgebra}) respectively).
This leads us to introduce the global index defined by 
\cite{L11,L12,LongoNets}
\be
\label{mu-global-index-def}
\mu^{\phantom{x}}_{\mathcal{T}}
\equiv 
\big[\,
\hat{\mathcal{A}}_\mathcal{T} (R_1 \cup R_2)
:
\mathcal{A}_\mathcal{T} (R_1 \cup R_2) 
\,\big] \,.
\ee
This index measures the size of the Haag duality violation,
and it is independent of the completion $\mathcal{C}$.
It should be understood as the index associated to 
the braided category of DHR localizable superselection sectors \cite{DHR1,DHR2,DHR3}. 
These are states in a given model that differ from the vacuum only within a bounded region and cannot be generated from the vacuum 
through local operators. 
They can be characterized as endomorphisms, like in (\ref{fusiondhr}).
Given an inclusion $\mathcal{T}_1 \subseteq \mathcal{T}_2$,
the corresponding global indexes (\ref{mu-global-index-def}) 
and single interval (local) index $\lambda_{\mathcal{T}_2:\mathcal{T}_1}$ 
(see (\ref{jonesinddef}))
are not independent quantities. 
Indeed, they are related as follows \cite{LongoKawaMuger} 
\be 
\mu^{\phantom{x}}_{\mathcal{T}_1}
=
\lambda^2_{\mathcal{T}_1:\mathcal{T}_2} \, 
\mu^{\phantom{x}}_{\mathcal{T}_2}  \,.
\label{global1}
\ee
When $\mathcal{T}_2=\mathcal{C}$ is a complete model
and $\mathcal{T}_1=\mathcal{T}$,
the relation (\ref{global1}) becomes 
\be 
\mu^{\phantom{x}}_{\mathcal{T}}
=
\lambda^2_{\mathcal{C}:\mathcal{T}} 
=\Big(\sum_{r} \tilde{d}_r^{\, 2}\Big)^2 \label{lsq}
\ee
where the sum runs over the 
irreducible endomorphisms included in the canonical one (\ref{ecanonical}). 

In a CFT$_2$, 
we have that $n_r=1$ holds and the dimensions are squared because of the inclusion of both chiral parts (see also (\ref{singlecft})).
Moreover, since in this case 
(\ref{singlecft}) holds for any model $\mathcal{T}$
(complete or not) 
that trivially obeys $\mathcal{T}_0\subseteq\mathcal{T}$,
the index $\mu^{\phantom{x}}_{\mathcal{T}}$ can be expressed 
in terms of the quantum dimensions of the Virasoro representations. 
In particular, by employing (\ref{global1}) and (\ref{lsq}), 
one finds 
\be 
\mu^{\phantom{x}}_{\mathcal{T}}
=
\lambda^{-2}_{\mathcal{T}:\mathcal{T}_0} \,\mu^{\phantom{x}}_{\mathcal{T}_0} 
= 
\left(\frac{\sum_{h',\bar{h}'} Z^\mathcal{C}_{h',\bar{h}'}{d}_{h'}{d}_{\bar{h'}}}{\sum_{h,\bar{h}}  Z^\mathcal{T}_{h,\bar{h}}{d}_{h}{d}_{\bar{h}}}\right)^2  .
\label{global3}
\ee
The same quantity can be found through  
the high temperature limit of the torus partition function (\ref{toruspart}) as follows 
\cite{LongoBlack,Benedetti:2024dku}
\be 
\mu^{-1/2}_{\mathcal{T}}
\,=\,
\lim_{ \beta \to 0}
\left( 
\,\frac{\parti_{\mathcal{T}\!,1}(\tau)}{\parti_{\mathcal{T}\!,1}(-1/\tau)} 
\,\bigg|_{\tau = \textrm{i} \beta}
\, \right) .
\label{global2}
\ee
This relation implies that $\mu_{\mathcal{T}} = 1$
for a modular invariant model. 
From (\ref{lsq}),
this corresponds to the fact that the invariance 
under $\tau \mapsto -1/\tau$
implies the absence of DHR  superselection sectors,
as first observed in \cite{Rehren:2000ti}.

\subsection{The subalgebras of the Ising CFT$_2$}
\label{Ising}

The categories of the DHR superselection sectors for a CFT$_2$ 
with $c<1$ have been classified \cite{LongoKawa,BenedettiMinimal}. 
We are interested in the special case given by the Ising CFT$_2$ model,
where $c=1/2$ and the chiral representations of the Virasoro algebra have scaling dimensions  $h \in \big\{ 0,1/16,1/2 \big\}$ and spin zero, 
i.e. $h = \bar{h}$.
This model has three primaries, which are 
the identity operator $1$, the spin operator $\sigma$
and the energy operator $\varepsilon$, 
whose conformal dimensions are defined by 
$h=0$, $h=1/16$ and $h=1/2$ respectively 
\cite{Friedan:1983xq,Belavin:1984vu}.  
The free Majorana fermion with antisymmetric Neveu–Schwarz (NS) 
boundary conditions, 
which is another important CFT$_2$ model with $c=1/2$, 
is considered in Sec.\,\ref{Fermions}.

In the Ising CFT$_2$, the non trivial fusion rules 
(i.e. the ones where the identity does not occur in the l.h.s.'s) are
\cite{DiFrancesco:1997nk}
\be 
\sigma\times \sigma = 1+\varepsilon
\;\;\;\qquad \;\;\;
\varepsilon\times\varepsilon=1
\;\;\;\qquad \;\;\;
\varepsilon\times\sigma=\sigma\times \varepsilon = \sigma \,.
\label{fusionrules}
\ee
These can be obtained by applying the Verlinde formula 
\cite{Verlinde-lines} to the Ising CFT$_2$, 
whose matrix occurring in the second relation of (\ref{ST})
reads \cite{CardyVerlinde,DiFrancesco:1997nk}
\be 
S_{\text{\tiny Ising}}= \frac{1}{2} \left(\,
\begin{matrix} 
1 & \sqrt{2}&  1\\ \sqrt{2} & 0& -\sqrt{2}\\
    1 & -\sqrt{2}& 1 \end{matrix}
    \right) .
    \label{Sising}
\ee
From the elements of this matrix and (\ref{dim-general}),
one obtains the following quantum dimensions
\be 
d_{\id} = 1
\;\;\;\qquad\;\;\; 
d_{\s} = \sqrt{2}
\;\;\;\qquad\;\;\; 
d_{\e} = 1 \,.
\label{isingdims}
\ee

The additive algebras of the Ising model are generated 
by the products of these fields and their descendants 
in the corresponding region $R$. 
Since the Ising CFT$_2$ model is modular invariant, 
it is also complete. 
Thus, for every choice of $R$,
the additive algebras satisfy the Haag duality 
and therefore coincide with the maximal algebra 
$\mathcal{A}_{\text{\tiny Ising}}(R)=\mathcal{A}'_{\text{\tiny Ising}}(R')=\hat{\mathcal{A}}_{\text{\tiny Ising}}(R)$, 
as follows from the condition (a) in Sec.\,\ref{jones} 
and the relation (\ref{global2}).

As for the additive subalgebras of 
$\mathcal{A}_{\text{\tiny Ising}}(R)$, 
since any algebra must be closed under product, 
we can construct submodels that are not modular invariant 
by employing the fusion rules (\ref{fusionrules}).
It is straightforward to realise 
that only two submodels occur that are closed under the fusion rules (\ref{fusionrules}):
the submodel $\mathcal{T}_{\mathbb{Z}_2}$, 
generated by the sectors corresponding to the identity and $\varepsilon$, 
and the submodel $\mathcal{T}_{\textrm{su}(2)_2}$, 
containing only the sector corresponding to the identity,
namely
\be 
\mathcal{T}_{\mathbb{Z}_2}\equiv\{1,\varepsilon\}
\;\;\;\;\qquad \;\;\;\;
\mathcal{T}_{\textrm{su}(2)_2}\equiv\{1\} 
\label{theories}
\ee
where the curly brackets enclose 
only the primary fields generating 
the corresponding additive algebras (\ref{additivealgebra}).

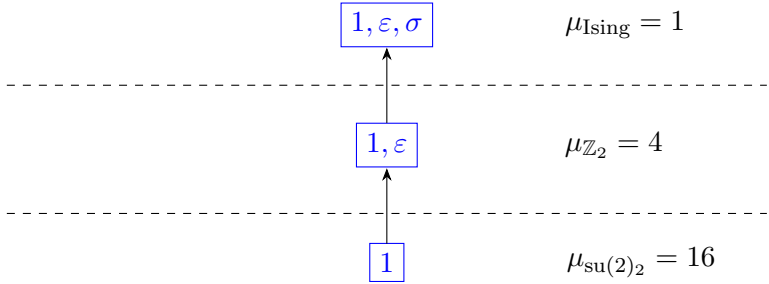
\begin{figure}[t!]
\centering
\begin{tikzpicture}[>=Stealth, node distance=1.0cm, every node/.style={font=\normalsize}]
\node[draw, rectangle,blue] (top) {$1,\varepsilon,\sigma$};
\node[draw, rectangle, below=of top,blue] (mid) {$1,\varepsilon$};
\node[draw, rectangle, below=of mid,blue] (bot) {$1$};
\draw[->] (0,-1.3)-- (0,-0.3);
\draw[->] (0,-2.9)-- (0,-1.9);
\draw[dashed] (-5,-0.8)-- (5,-0.8);
\draw[dashed] (-5,-2.5)-- (5,-2.5);
\node[right=1.6cm of top] {$\mu_{\text{Ising}} = 1$};
\node[right=1.8cm of mid] {$\mu_{\mathbb{Z}_2} = 4$};
\node[right=2cm of bot] {$\mu_{\textrm{su}(2)_2} = 16$};
\end{tikzpicture}
\caption{
Subalgebras of the Ising CFT$_2$, enclosed by the blue boxes,
and the corresponding values of the global Jones index 
(\ref{mu-global-index-def}).}
\label{sub1}
\end{figure}

The global indexes (\ref{mu-global-index-def}) 
of the submodels (\ref{theories})
can be computed through the techniques discussed in Sec.\,\ref{jones}. 
As for $\mathcal{T}_{\mathbb{Z}_2}$, notice that the fusion rules (\ref{fusionrules}) have a $\mathbb{Z}_2$ symmetry \cite{DiFrancesco:1997nk}
acting in a non trivial way over the spin field, such that  $\sigma \mapsto -\sigma$. 
The local algebras $\mathcal{A}_{\mathbb{Z}_2}(R)$ corresponding to $\mathcal{T}_{\mathbb{Z}_2}$ can be understood as the neutral elements of $\mathcal{A}_{\text{\tiny Ising}}(R)$ under a finite $\mathbb{Z}_2$ group. Thus, the endomorphisms obey the fusion rules (\ref{fusiondhr}) 
corresponding to the ones of this group \cite{DHR1,DHR2,DHR3}.  
This leads us to compute the global Jones index as follows
\be 
\mu^{\phantom{x}}_{\mathbb{Z}_2}
= \Big(1^2 + 1^2\Big)^2 
= 
\left(\frac{1^2 + \sqrt{2}^2+ 1^2}{1^2 + 1^2} \right)^2=\,4 
\label{mu4}
\ee
where the first equality is obtained by applying (\ref{igroup}) and (\ref{lsq}), considering that the quantum dimensions obtained from the ${\mathbb{Z}_2}$ fusion rules are equal to one. 
The second equality in (\ref{mu4}) is found by using  
(\ref{isingdims}) in  (\ref{global3}) specialised to  $Z^\mathcal{C}=\text{diag}(1,1,1)$ and $Z^\mathcal{T}=\text{diag}(1,0,1)$, which take into account the fact that the field $\sigma$
does not occur in $\mathcal{T}_{\mathbb{Z}_2}$.
The submodel $\mathcal{T}_{\textrm{su}(2)_2}$ contains only the identity sector, and therefore the fusion rules (\ref{fusiondhr}) of the endomorphisms are the ones of an affine lie algebra $\textrm{su}(2)_2$ \cite{LongoKawa}. 
From the quantum dimensions (\ref{isingdims}), 
we find that the global Jones index is
\be 
\mu^{\phantom{x}}_{\textrm{su}(2)_2}
=
\Big(1^2 + 1^2+ \sqrt{2}^2\,\Big)^2 =\left(\frac{1^2 + \sqrt{2}^2 + 1^2 }{1^2} \right)^2=16 \label{mu16}
\ee
where (\ref{singlecft}) and (\ref{global3}) have been applied in the first and second equality, respectively. 
Notice that the first equality can also be understood as a sum over the three $\textrm{su}(2)_2$ quantum dimensions.
 In Fig.\,\ref{sub1} we show the hierarchy of the subalgebras
in the Ising CFT$_2$, with the corresponding value of the 
global Jones index 
(\ref{mu-global-index-def}).

Let us consider also a different perspective,
based on the topological defect lines \cite{Petkova:2000ip}.
The topological defect lines are defects that commute with all Virasoro generators separately.
Each algebra can be defined by a set of topological defect lines  
that commutes with all the primaries of the model \cite{Shao:2025mfj,Benedetti:2026drn}. 
In diagonal rational CFT$_2$, where all primaries obey $h=\bar{h}$, the topological defect lines can be spanned using the basis of the Verlinde lines, which are labeled by the scaling dimensions of the primaries and obey the same fusion rules  \cite{Verlinde-lines}. 
The line $\mathcal{L}_h$ commutes with the primary field having 
scaling dimension $h'$ if and only if \cite{Chang:2018iay}
\be
d_h = \frac{S_{h,h'}}{S_{0,h'}} \label{commuting-condition}
\ee
where $d_h$ is the quantum dimension defined in (\ref{dim-general}).

In the Ising CFT$_2$,
the topological defect lines are characterized by the three Verlinde lines $\mathcal{L}_{1},\mathcal{L}_{\varepsilon}$ and $\mathcal{L}_{\sigma}$. 
These lines indeed obey fusion rules that are equivalent 
to the ones in (\ref{fusionrules}), satisfied by the chiral primaries, namely 
\be 
\label{Verlinde-lines-Ising-composition}\mathcal{L}_{\sigma}\times \mathcal{L}_{\sigma}=\mathcal{L}_1+\mathcal{L}_{\varepsilon}
\;\;\qquad\;\;
\mathcal{L}_{\varepsilon}\times \mathcal{L}_{\varepsilon}=\mathcal{L}_1
\;\;\qquad\;\;
\mathcal{L}_{\varepsilon}\times \mathcal{L}_{\sigma} = \mathcal{L}_{\sigma} \times \mathcal{L}_{\varepsilon}= \mathcal{L}_{\sigma} \,.
\ee

The line $\mathcal{L}_{1}$ acts in a trivial way and leaves all the fields invariant. 
Instead, the line $\mathcal{L}_{\varepsilon}$ implements the $\mathbb{Z}_2$ symmetry; hence it transforms the local field $\sigma$ into $-\sigma$. 
From (\ref{commuting-condition}),
we can check  that $\mathcal{L}_{\varepsilon}$ commutes with the identity $1$ and $\varepsilon$.
In other words, the local algebra $\mathcal{A}_{\mathbb{Z}_2}(R)$, corresponding to  $\mathcal{T}_{\mathbb{Z}_2}$, is made up of the operators that are invariant under $\mathcal{L}_1$ and $\mathcal{L}_\varepsilon$.
The topological defect line $\mathcal{L}_\sigma$ 
is non-invertible and implements the  Kramers-Wannier duality 
\cite{KW} in the Ising CFT$_2$ at the self-dual point. 
At that point, the duality degenerates to a non-invertible symmetry \cite{Shao:2023gho,Seiberg:2023cdc}. 
The important point for our purposes is that the local algebras $\mathcal{A}_{\textrm{su}(2)_2}(R)$  containing only the identity primary (corresponding to the theory $\mathcal{T}_{\textrm{su}(2)_2}$) commute with the action of all lines. 
The fusion rules of the $\textrm{su}(2)_k$ category are \cite{Bockenhauer:1998in}
\be
    j_1 \times j_2 
    \,=\!\! 
    \sum_{j=|j_1 - j_2|}^{j_{\textrm{\tiny max}}} j
    \;\;\;\qquad\;\;\;
    j_{\textrm{\tiny max}} \equiv 
    \min\!\big(j_1 + j_2, \, k - (j_1 + j_2)\big)
\ee
where $j_1, j_2, j\in \big\{0,1/2,1,\dots,k/2 \big\}$,
and the fusion rules (\ref{Verlinde-lines-Ising-composition}) provide a realization of the fusion rules of the $\textrm{su}(2)_2$ tensor category.

\section{Jones index from the crossing asymmetry in CFT$_2$}
\label{JonesIsing}

In this section, we describe the relation between the global Jones index (\ref{mu-global-index-def}) and the R\'enyi entropies, found in \cite{Benedetti:2024dku}.

The definitions of the entanglement entropies for a spatial region $R$, 
namely of the R\'enyi entropies and of the entanglement entropy, 
that are given respectively by 
\be 
S_n(\mathcal{A}(R))
\equiv 
\frac{1}{1-n}\, \log\!\big(\, \Tr \,\rho_R^n\,\big)
\;\;\;\qquad \;\;\;
S_n(\mathcal{A}_{\mathcal{T}}(R)) = \lim_{n \to 1} S_n(\mathcal{A}_{\mathcal{T}}(R))
\ee
where $\rho_R$ is the reduced density matrix of the subregion $R$ 
and the integer $n \geqslant 2$ is the R\'enyi index, 
require the choice of the algebra $\mathcal{A}_{\mathcal{T}}(R)$ on $R$ 
and of the global state of the whole system 
\cite{Casini:2013rba,Xu:2018uxc,Hollands:2020owv,MaganPontello,Casini:2020rgj,Huerta:2022cqw,Abate:2025ywp}. 
The crucial role played by the choice of the algebra of the region $R$
suggests that the Jones index should also provide some useful information. 
When the global state of the whole system is pure, 
the entanglement entropies associated with commuting algebras 
are equivalent
\cite{petzuse}, i.e. 
\be 
S_n\big(\mathcal{A}_{\mathcal{T}}(R)\big)
=
S_n\big(\mathcal{A}_{\mathcal{T}}'(R)\big) \,.
\label{pure-state-S-AB}
\ee
This makes it evident that the validity of the Haag duality for the choice of $\mathcal{A}_{\mathcal{T}}(R)$ and $\mathcal{A}_{\mathcal{T}}(R')$ applied to (\ref{pure-state-S-AB}) leads to
\be
S_n(\mathcal{A}_{\mathcal{T}}\big(R)\big)
=
S_n\big(\mathcal{A}_{\mathcal{T}}(R')\big)  \,.
\hspace{.5cm} 
\label{entropieshaag}
\ee
This identity holds in a complete model. 
The fact that $S_n(R)\neq S_n(R')$ for some choice of algebra 
when the system is in a pure state occurs 
in QFTs with non trivial superselection sectors. 
This includes models with higher form symmetries \cite{Casini:2020rgj,Huerta:2022cqw}
and CFT$_2$'s that are not modular invariant, like
e.g. the free chiral scalar current \cite{Hollands:2019hje,Arias:2018tmw, Arias:2026bqh}.

In the following, the invariance under the full modular group is not assumed 
for the CFT$_2$ model; hence the algebras associated to a given $R$ are not unique,
as discussed in Sec.\,\ref{completenessqft}.
We denote by $S_{n}(R)\equiv S_n\big(\mathcal{A}_{\mathcal{T}}(R)\big)$ the R\'enyi entropies associated to the 
additive algebra $\mathcal{A}_{\mathcal{T}}(R)$ of the theory ${\mathcal{T}}$ when the whole system is in the vacuum. When  
$R$ is an interval of length $r$ 
and the underlying model is a CFT$_2$ on the line, 
it is well known that \cite{Holzhey:1994we,Calabrese:2004eu}
\be
S_{n}(R)= \frac{c}{6} \left(1 + \frac{1}{n}\right)
\log(r/\epsilon) + \frac{\log C_n}{1-n}
\label{SAone-int-CFT}
\ee
where $\epsilon \to 0^+$ is the UV cutoff and 
$C_n$ is a constant satisfying the condition $C_1 = 1$
that provides the normalization of the two-point function in the approach based on the branch-point twist fields
\cite{Calabrese:2004eu,Calabrese:2009qy}.

When $R = R_1 \cup R_2$ is the union of two disjoint intervals
$R_1 \equiv [u_1, v_1]$ and $R_2 \equiv [u_2, v_2]$, 
with $u_1 < v_1 < u_2 <v_2$,
an important quantity is 
\be 
\mathcal{I}_n(R_1,R_2) \equiv 
S_n(R_1) + S_n(R_2) - S_n(R_1\cup R_2) 
=
\frac{1}{n-1} \, \log \! \left(
\frac{ 
\Tr\, \rho_{R_1 \cup R_2}^n
}{ 
\big(\Tr\,\rho_{R_1}^n\big) \big(\Tr\,\rho_{R_2}^n\big) 
}
\right)
\label{mutualdef}
\ee
which is UV finite and 
whose analytic continuation $n \to 1$ provides the mutual information. 
For a CFT$_2$ on the line and in its vacuum, 
$S_n(R_1)$ and $S_n(R_2)$ in (\ref{mutualdef}) are given by 
(\ref{SAone-int-CFT}),
while the term $S_n(R_1\cup R_2) $ is obtained from the following moments \cite{Furukawa:2008uk,Calabrese:2009ez,Calabrese:2010he}
\be
 \Tr\big(\rho_{R_1 \cup R_2}^n\big)
 \,=\, 
 C_n^2 \,
 \big| P(u_1, v_1, u_2, v_2)\big|^{-\tfrac{c}{6} (n -1/n)}
 \,\mathcal{F}_{\mathcal{T} \!, \, n}(\x)
 \label{Fn}
\ee
where $\x$ is the cross ratio constructed from the four endpoints of $R$, namely
\be 
\xi \equiv \frac{(v_1-u_1)(v_2-u_2)}{(u_2-u_1)(v_2-v_1)}
\,\in\,(0,1)
\label{cross}
\ee
and the rational function within the absolute value is defined as follows
\be 
P(u_1, v_1, u_2, v_2)
\equiv 
\frac{(v_1-u_1)(v_2-u_2)}{\epsilon^2}\, (1-\x) \,.
\label{P-prefactor-def}
\ee
From (\ref{Fn}), it is straightforward to obtain 
(\ref{entropydisjoint}).
The factor $C_n^2$ in (\ref{Fn}) guarantees that 
$\mathcal{F}_{\mathcal{T}\!, \, n}(\x) \to 1$ in the limit of large separation distance between the two intervals, namely when $\x \to 0^+$.
From (\ref{SAone-int-CFT}) and (\ref{Fn}), 
it is straightforward to realise that  
$\mathcal{I}_n(R_1,R_2)$ in (\ref{mutualdef}) becomes 
a function of $\x$ only;
hence it is denoted by $\mathcal{I}_n(\x)$ hereafter. 

In this setup, the exchange $R \leftrightarrow R'$ 
is implemented as follows
\be
u_1 \mapsto v_1 
\;\;\qquad\;\;
v_1 \mapsto u_2
\;\;\qquad\;\;
u_2 \mapsto v_2
\;\;\qquad\;\;
v_2 \mapsto u_1
 \label{Haag-exchange-UV-def}
\ee
which implies that $\x \mapsto 1-\x$ for the cross ratio (\ref{cross}).

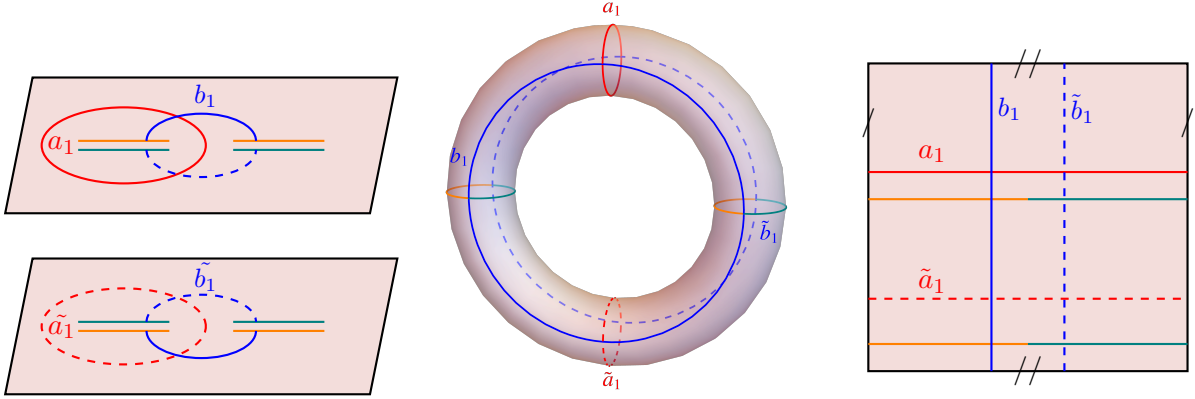
\begin{figure}[t!]
\centering
\makebox[\textwidth][c]{
\begin{subfigure}{.33\textwidth}
\centering
    \begin{tikzpicture}[scale=1.2,>=Stealth,thick]
\draw[fill=csheet, draw=black] (-3.3,0.7) -- (0.7,0.7) -- (1,2.2) -- (-3,2.2) -- cycle;
\draw[orange, thick] (-2.5,1.5) -- (-1.5,1.5);
\draw[teal, thick] (-2.5,1.4) -- (-1.5,1.4);
\draw[orange, thick] (-0.8,1.5) -- (0.2,1.5); 
\draw[teal, thick] (-0.8,1.4) -- (0.2,1.4); 
\node[red, thick] at (-2.68,1.45) {$a_1$};
\draw[red, thick] (-2,1.45) ellipse (0.9cm and 0.42cm);
\draw[blue, thick]
  (-1.75,1.5) arc[start angle=180, end angle=0, x radius=0.6cm, y radius=0.3cm];
\node[blue, thick] at (-1.1,2.0) {\small{$b_1$}};
\draw[blue, dashed]
  (-1.75,1.4) arc[start angle=180, end angle=360, x radius=0.6cm, y radius=0.3cm];
\draw[fill=csheet, draw=black] (-3.3,-1.3) -- (0.7,-1.3) -- (1,0.2) -- (-3,0.2) -- cycle;
\draw[teal, thick] (-2.5,-0.5) -- (-1.5,-0.5);
\draw[orange, thick] (-2.5,-0.6) -- (-1.5,-0.6);
\draw[teal, thick] (-0.8,-0.5) -- (0.2,-0.5); 
\draw[orange, thick] (-0.8,-0.6) -- (0.2,-0.6); 
\draw[blue, thick]
  (-1.75,-0.6) arc[start angle=180, end angle=360, x radius=0.6cm, y radius=0.3cm];
  \draw[blue, dashed]
  (-1.75,-0.5) arc[start angle=180, end angle=0, x radius=0.6cm, y radius=0.3cm];
  \node[red, thick] at (-2.68,-0.55) {$\tilde{a_1}$};
 \node[blue, thick] at (-1.1,0) {\small{$\tilde{b_1}$}};
\draw[red, dashed] (-2,-0.55) ellipse (0.9cm and 0.42cm);
\end{tikzpicture}
\end{subfigure}
\hspace{0.2cm}
\begin{subfigure}{.31\textwidth}
\centering
    \includegraphics[width=1.0\textwidth]{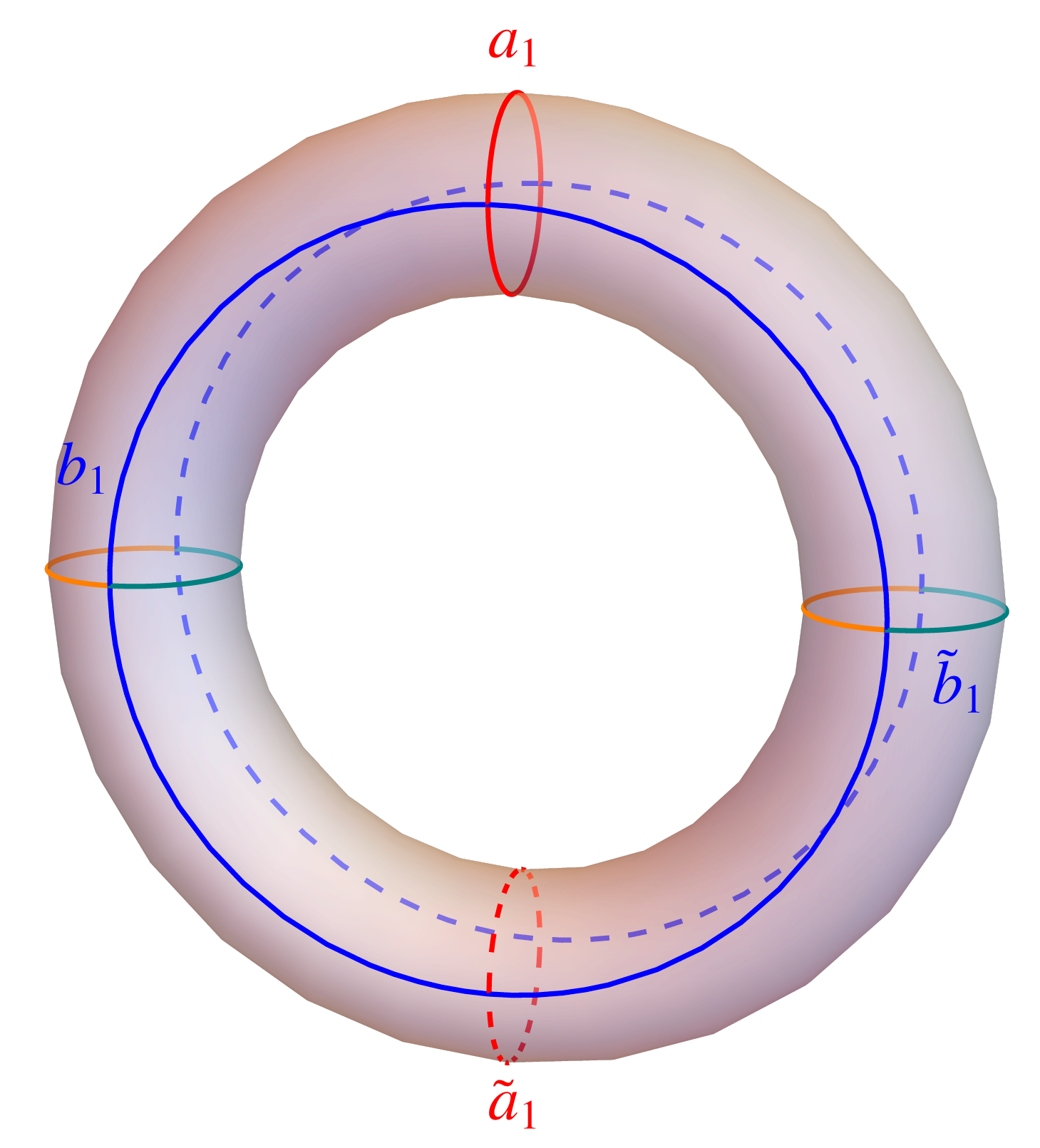}
\end{subfigure}
\hspace{0.2cm}
\begin{subfigure}{.33\textwidth}
\centering
    \begin{tikzpicture}[scale=1.2,>=Stealth,thick]
\draw[fill=csheet, draw=black] (2.1,-1.2) -- (2.1,2.2) --  (5.6,2.2) -- (5.6,-1.2) -- cycle ;
\draw[orange] (2.1,-0.9) -- (3.85,-0.9);
\draw[teal] (3.85,-0.9) -- (5.6,-0.9);
\draw[orange] (2.1,0.7) -- (3.85,0.7) ;
\draw[teal] (3.85,0.7) -- (5.6,0.7) ;
\draw[red,thick] (2.1,1) -- (5.6,1);
\draw[blue,thick] (3.45,-1.2) -- (3.45,2.2);
\draw[blue,dashed] (4.25,-1.2) -- (4.25,2.2);
\draw[red,dashed] (2.1,-0.4) -- (5.6,-0.4);
\node at (3.85,-1.2) {$//$};
\node at (3.85,2.2) {$//$};
\node at (2.1,1.55) {$/$};
\node at (5.6,1.55) {$/$};
 \node[red, thick] at (2.8,1.2) {$a_1$};
\node[red, thick] at (2.8,-0.2) {$\tilde{a}_1$};
  \node[blue, thick] at (3.65,1.7) {\small{$b_1$}};  
  \node[blue, thick] at (4.45,1.7) {\small{$\tilde{b}_1$}};
\end{tikzpicture}
\end{subfigure}
}
\vspace{.4cm}
\caption{
The torus corresponding to the Riemann surface $\mathscr{R}_2$, 
with two equivalent choices for the canonical homology basis,
denoted by $\big\{ (a_1, b_1) \big\}$ 
and $\big\{ (\tilde{a}_1, \tilde{b}_1) \big\}$.
}
\label{replica}
\end{figure}

\begin{figure}[t!]
\centering
\makebox[\textwidth][c]{
 \begin{subfigure}{.5\textwidth}
 \centering
        \includegraphics[width=1.15\textwidth]{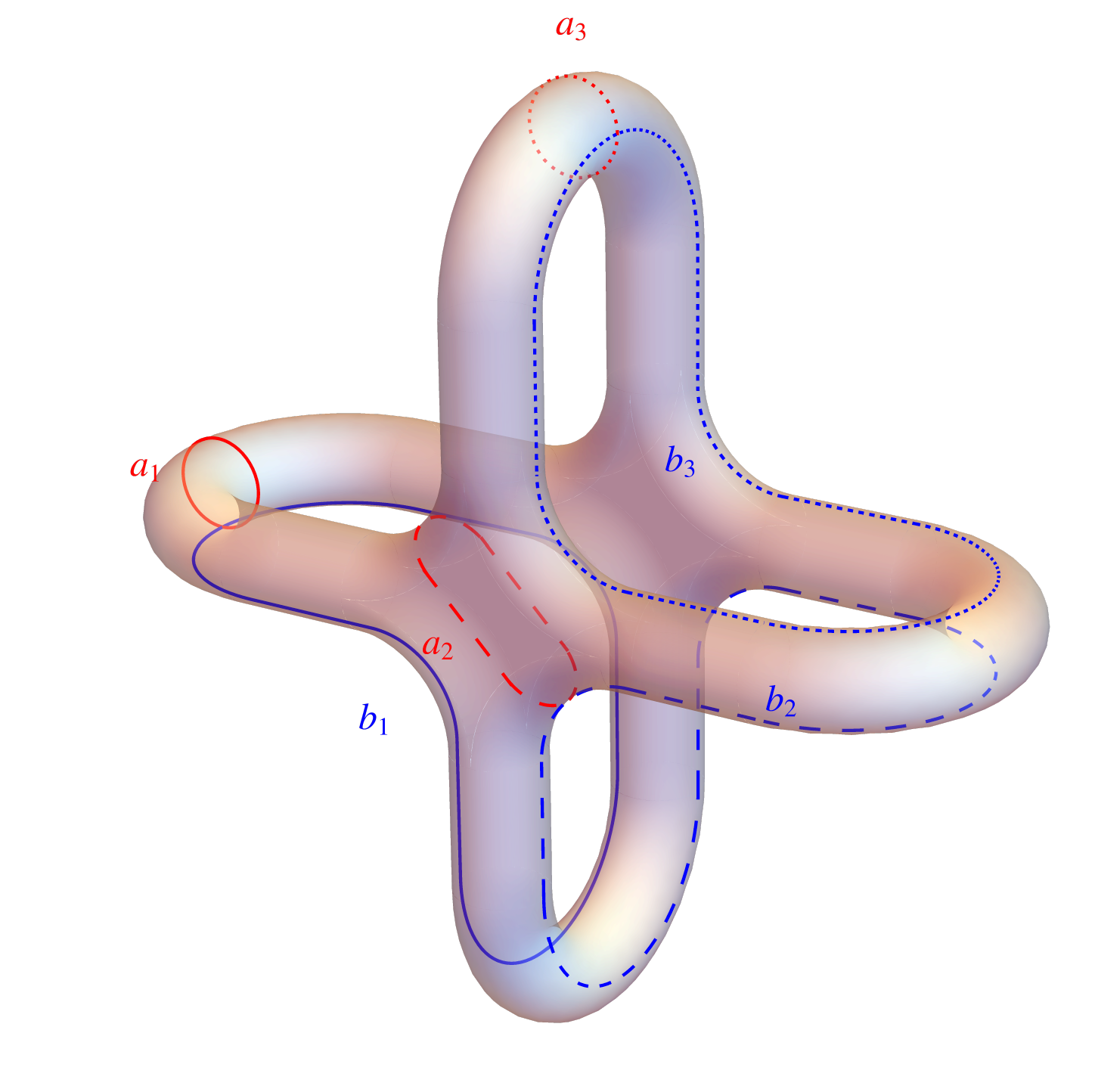}
  \end{subfigure}
  \hspace{1cm}
  \begin{subfigure}{.5\textwidth}
  \centering
\begin{tikzpicture}[scale=1.2,>=Stealth,thick]
\draw[fill=csheet, draw=black] (-3.3,0.7) -- (0.7,0.7) -- (1,2.2) -- (-3,2.2) -- cycle;
\draw[orange, thick] (-2.5,1.5) -- (-1.5,1.5);
\draw[teal, thick] (-2.5,1.4) -- (-1.5,1.4);
\draw[orange, thick] (-0.8,1.5) -- (0.2,1.5); 
\draw[teal, thick] (-0.8,1.4) -- (0.2,1.4); 
\node[red, thick] at (-2.68,1.44) {${a}_1$};
\draw[red, thick] (-2,1.45) ellipse (0.9cm and 0.42cm);
\draw[blue, thick]
  (-1.75,1.5) arc[start angle=180, end angle=0, x radius=0.6cm, y radius=0.3cm];
\node[blue, thick] at (-1.1,2) {\small{${b}_1$}};

\begin{scope}[shift={(0,0.3)}]
\draw[fill=csheet, draw=black] (-3.3,-1.3) -- (0.7,-1.3) -- (1,0.2) -- (-3,0.2) -- cycle;
\draw[cyan, thick] (-2.5,-0.5) -- (-1.5,-0.5);
\draw[orange, thick] (-2.5,-0.6) -- (-1.5,-0.6);
\draw[cyan, thick] (-0.8,-0.5) -- (0.2,-0.5); 
\draw[orange, thick] (-0.8,-0.6) -- (0.2,-0.6); 
\draw[blue, dashed]
  (-1.75,-0.5) arc[start angle=180, end angle=0, x radius=0.6cm, y radius=0.3cm];
\node[blue, thick] at (-1.1,0) {\small{${b}_2$}};
\draw[blue, thick]
  (-1.75,-0.6) arc[start angle=180, end angle=360, x radius=0.6cm, y radius=0.3cm];
\end{scope}

\begin{scope}[shift={(-5,-3.4)}]
\draw[fill=csheet, draw=black] (1.7,0.7) -- (5.7,0.7) -- (6,2.2) -- (2,2.2) -- cycle;
\draw[magenta, thick] (2.5,1.5) -- (3.5,1.5);
\draw[cyan, thick] (2.5,1.4) -- (3.5,1.4);
\draw[magenta, thick] (4.2,1.5) -- (5.2,1.5); 
\draw[cyan, thick] (4.2,1.4) -- (5.2,1.4); 
\node[red, thick] at (2.3,1.2){${a}_2$};
\draw[red, dashed] (2.4,1.5) arc[start angle=180, end angle=0, x radius=0.1cm, y radius=0.1cm];
\draw[red, dashed] (3.4,1.5) arc[start angle=180, end angle=0, x radius=0.1cm, y radius=0.1cm];
\draw[red, dashed] (2.4,1.5) arc[start angle=180, end angle=360, x radius=0.6cm, y radius=0.5cm];
\draw[blue, dotted]
  (3.25,1.5) arc[start angle=180, end angle=0, x radius=0.6cm, y radius=0.3cm];
\node[blue, thick] at (3.9,2) {\small{${b}_3$}};
\draw[blue, dashed]
  (3.25,1.4) arc[start angle=180, end angle=360, x radius=0.6cm, y radius=0.3cm];
\end{scope}

\begin{scope}[shift={(-5,-3.1)}]
\draw[fill=csheet, draw=black] (1.7,-1.3) -- (5.7,-1.3) -- (6,0.2) -- (2,0.2) -- cycle;
\draw[teal, thick] (2.5,-0.5) -- (3.5,-0.5);
\draw[magenta, thick] (2.5,-0.6) -- (3.5,-0.6);
\draw[teal, thick] (4.2,-0.5) -- (5.2,-0.5); 
\draw[magenta, thick] (4.2,-0.6) -- (5.2,-0.6); 
\draw[red, dashed] (2.4,-0.6) arc[start angle=180, end angle=360, x radius=0.1cm, y radius=0.1cm];
\draw[red, dashed] (3.4,-0.6) arc[start angle=180, end angle=360, x radius=0.1cm, y radius=0.1cm];
\draw[red, dashed] (2.4,-0.6) arc[start angle=180, end angle=0, x radius=0.6cm, y radius=0.5cm];
  \draw[blue, dotted]
  (3.25,-0.6) arc[start angle=180, end angle=360, x radius=0.6cm, y radius=0.3cm];
\node[red, thick] at (2.25,-0) {${a}_3$};
\draw[red, dotted] (3,-0.55) ellipse (0.9cm and 0.55cm);
\end{scope}
\end{tikzpicture}
  \end{subfigure}}
  \caption{
   The Riemann surface $\mathscr{R}_4$ (see (\ref{eq:replicaSurface})) 
   with a specific choice of the canonical homology basis that provides the period matrix (\ref{periodm}).
  }
\label{replica4tilde}
\end{figure}
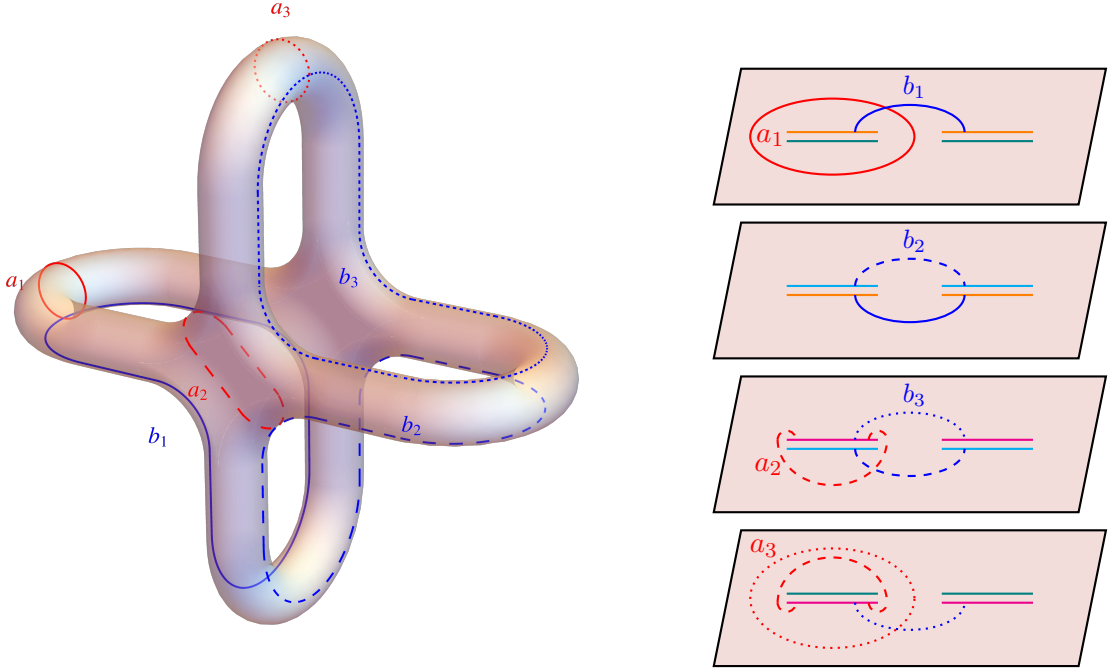

The expression (\ref{Fn}) can be interpreted also as 
the partition function $\parti_{\mathcal{T} \!, \, n-1}$
of the underlying CFT$_2$ on a specific 
Riemann surface $\mathscr{R}_n$ with genus $g=n-1$ as follows
\cite{Calabrese:2009ez,Calabrese:2010he,Coser:2013qda}
\be 
\Tr\big(\rho_{R_1 \cup R_2}^n\big)
 =
 \frac{\parti_{\mathcal{T} \!, \, n-1}
 }{
 \big( \parti_{\mathcal{T} \!, \, 0} \big)^n
 }
 \label{replicaeq}
\ee
We remark that $\mathscr{R}_n$ is a very specific Riemann surface 
obtained from the replica.
Indeed, $\mathscr{R}_n$ displays the $\mathbb{Z}_n$ invariance provided by the replica construction and 
can be written as the following complex curve \cite{GravaAbel}
\be 
\Big\{(y,z)\in \mathbb{C}^2 \,\big| \, y^n=z(z-1)(z-\x)^{n-1}\Big\} \,.
\label{eq:replicaSurface}
\ee
The Riemann surfaces $\mathscr{R}_2$ and $\mathscr{R}_4$ 
for equal intervals 
are shown in the middle panel of Fig.\,\ref{replica}
and in the left panel of Fig.\,\ref{replica4tilde}
respectively.

In this setup, 
the validity of (\ref{entropieshaag}) corresponds to 
the modular invariance of $\mathscr{R}_n$, 
as first observed in \cite{Calabrese:2009ez}.
By using (\ref{Haag-exchange-UV-def}) in (\ref{P-prefactor-def}), 
we have that $P(v_1, u_2, v_2, u_1) = - P(u_1, v_1, u_2, v_2)$;
hence (\ref{Fn}) and (\ref{replicaeq}) lead to 
\be
\frac{ \parti_{\mathcal{T} \!, \, n-1} }{
\parti_{\mathcal{T} \!, \, n-1} \big|_{R \leftrightarrow R'}}
=
\frac{\mathcal{F}_{\mathcal{T} \!, \, n}(\x)}{\mathcal{F}_{\mathcal{T} \!, \, n}(1-\x)} \,.
\label{partitionF}
\ee
The relation between the Haag duality and the modular invariance 
can be made more precise by introducing \cite{Benedetti:2024dku}
\be
A_{\mathcal{T} \!, \, n}(\x)
\,\equiv\,
\frac{1}{n-1}
\log\!\left(
\frac{ \parti_{\mathcal{T} \!, \, n-1} }{
\parti_{\mathcal{T} \!, \, n-1} \big|_{R \leftrightarrow R'}}
\right) 
\label{asymmetrypart}
\ee
which has been dubbed crossing asymmetry.
In the CFT$_2$ setup that we are considering, 
where (\ref{SAone-int-CFT}) and (\ref{Fn}) hold, 
by employing also (\ref{partitionF}), 
one finds that (\ref{asymmetrypart}) 
becomes (\ref{asymetrydefintro}), 
which can also be written as
\be
A_{\mathcal{T} \!, \, n}(\x)
\,=\,
\frac{c}{6} \left(1 + \frac{1}{n}\right)
\log\!\left(\frac{1-\x}{\x}\right) 
+
\mathcal{I}_{\mathcal{T} \!, \, n}(\x)
-
\mathcal{I}_{\mathcal{T} \!, \, n}(1-\x) \,.
\label{asymetrydef-mutual} 
\ee
In  \cite{Benedetti:2024dku} it has been found that
the crossing asymmetry (\ref{asymetrydefintro}) in the limits of small and large separation distance between the intervals, when their lengths are kept fixed, 
provides the global Jones index (\ref{mu-global-index-def})
through the limit (\ref{jones-index-limit-asymmetry-intro}).

An important quantity associated to the Riemann surface $\mathscr{R}_n$
that typically occurs in $\parti_{\mathcal{T} \!, \, n-1}$
is the complex $g \times g$ period matrix $\tau_n(\x)$, where $g=n-1$,
which depends only on the cross ratio (\ref{cross}) 
and has a positive definite imaginary part. 
Its construction, discussed in 
\cite{Dijkgraaf:1987vp,Calabrese:2009ez,GravaAbel,Coser:2013qda},
requires to introduce (through (\ref{eq:replicaSurface}))
a normalized basis of holomorphic differentials 
$\big\{ \omega_j \, ; 1 \leqslant j \leqslant g \big\}$
and a canonical homology basis, 
which is a set $\big\{ (a_j, b_j) \, ; 1 \leqslant j \leqslant g \big\}$
made by $2g$ closed oriented curves (a.k.a. cycles) 
on $\mathscr{R}_n$
that cannot be contracted to a point 
and whose intersections satisfy  
\be
a_i \cap a_j = b_i \cap b_j =  \emptyset
\;\;\;\qquad\;\;\;
\sharp (a_i \cap b_j ) =  \delta_{i,j}
\label{canonicalhom}
\ee
where $\sharp(M)$ denotes the cardinality of the set $M$. 
The period matrix $\tau_n$ is defined by 
\be
\oint_{a_i}\omega_j=\delta_{i,j}
\;\;\;\;\qquad\;\;\;
\oint_{b_i}\omega_j= \big[\tau_n\big]_{i,j}
\label{period-matrix-def}
\ee
where the first equation imposes the normalization condition,
while the second one provides the generic element 
of the period matrix $\tau_n$. 
Some convenient choices for the canonical homology basis 
have been discussed in \cite{Coser:2013qda}\footnote{We adopt a different notation from \cite{Coser:2013qda} for the cycles of the canonical homology bases and period matrices. 
In particular, with respect to \cite{Coser:2013qda},
we interchange $a_i\leftrightarrow\tilde{a}_i$, $b_i\leftrightarrow\tilde{b}_i$ and, consequently, also ${\tau}_n\leftrightarrow\tilde{\tau}_n$.} 
and one of them 
(which is shown in Fig.\,\ref{replica4tilde} for the $n=4$ case)
leads to the
period matrix found in \cite{Calabrese:2009ez}, namely
\be
\big[ {\tau}_n \big]_{i,j}
= \,\textrm{i}\,
\frac{2}{n}\,\sum_{k=1}^{n-1}
\sin(\pi k/n) \, 
\frac{ F_{k/n}(1-\x) }{ F_{k/n}(\x) } \,
\cos\!\big(2\pi k(i-j)/n\big)
\label{periodm}
\ee
where $ F_{r}(\x) \equiv \,{}_2F_1(r,1-r; 1;\x)$, in terms of the 
hypergeometric function ${}_2F_1(\alpha ,\beta ; \gamma;x)$.

\section{Ising CFT$_2$: Jones index from the 
$n=2$ R\'enyi entropy
}
\label{JonesIsing2}

In this section, we discuss the relation (\ref{jones-index-limit-asymmetry-intro}) 
between the Jones index and the R\'enyi entropies 
in the special case of the Ising CFT$_2$ and for the simplest value of the R\'enyi index, i.e. $n=2$.

When $n=2$, the crossing asymmetry (\ref{asymmetrypart}) 
becomes \cite{Benedetti:2024dku}
\be
A_{\mathcal{T} \!, \, 2}(\x)=\log\left(\frac{ \parti_{\mathcal{T} \!, \, 1} }{
\parti_{\mathcal{T} \!, \, 1} \big|_{R \leftrightarrow R'}}\right)
\label{asymetry2}
\ee
where the torus partition function occurs,
which depends on the modular parameter of the torus 
obtained 
by specialising the period matrix (\ref{periodm}) to $n=2$, namely 
\be
\tau_2
\,=\, 
\textrm{i} \,\frac{F_{1/2}(1-\x)}{F_{1/2}(\x)} 
\, = \,
\textrm{i} \,\frac{K(1-\x)}{K(\x)}
\label{periodm2}
\ee
in terms of the elliptic integral of the first kind.
The Riemann surface $\mathscr{R}_2$ is the torus 
illustrated in Fig.\,\ref{replica},
where the left panel highlights the role of the two sheets
(see also the appendix\;A in \cite{Calabrese:2010he}). 
From the modular parameter (\ref{periodm2}),
it is important to observe \cite{Calabrese:2009ez} 
that the map $\xi \mapsto 1-\xi$, corresponding to the exchange $R \leftrightarrow R'$ (see (\ref{Haag-exchange-UV-def})),
induces the modular transformation of type $S$ given by 
$\tau_2 \mapsto \tau_2\big|_{\x\mapsto 1-\x} = -1/\tau_2$.

The partition function of the Ising CFT$_2$ on the torus 
occurring in the crossing asymmetry (\ref{asymetry2}) reads
\cite{DiFrancesco:1997nk}
\be 
\parti_{\text{\tiny Ising},  1}=\frac{1}{2}\bigg(
 \left|\frac{\theta_2(\tau_2)}{\eta(\tau_2)}\right|+ \left|\frac{\theta_3(\tau_2)}{\eta(\tau_2)}\right|+ \left|\frac{\theta_4(\tau_2)}{\eta(\tau_2)}\right|\bigg)
\label{partising}
\ee
in terms of the modular parameter (\ref{periodm2}). 
This partition function provides 
the function $\mathcal{F}_{\mathcal{T} \!, \, 2}(\x)$ 
for the Ising CFT$_2$  as follows
\cite{Calabrese:2010he}
\be 
\mathcal{F}_{\text{\tiny Ising},  2}(\x)
= \frac{1}{2}\bigg(1+ \left|\frac{\theta_2(\tau_2)}{\theta_3(\tau_2)}\right|+ \left|\frac{\theta_4(\tau_2)}{\theta_3(\tau_2)}\right|\bigg)
\label{fising}
\ee
which satisfies the condition 
$\mathcal{F}_{\text{\tiny Ising},  2}(\x=0)=1$
(see the text below (\ref{P-prefactor-def})).

It is well known \cite{DiFrancesco:1997nk}
that the partition function (\ref{partising}) is modular invariant for any allowed value of the modular parameter $\tau$ and that (\ref{fising}) is crossing symmetric. These properties can be checked through  
the transformations of the Jacobi theta functions \cite{Faygenus},
given by 
\be
\label{transf-jacobi-T}
\theta_2(\tau+1)
= \textrm{e}^{\textrm{i} \pi/4} \theta_2(\tau)
\;\;\qquad\;\; 
\theta_3(\tau+1) = \theta_4(\tau) 
\;\;\qquad\;\; 
\theta_4(\tau+1) = \theta_3(\tau)  
\ee
and 
\be
\label{transf-jacobi-S}
\theta_2(-1/\tau)= \sqrt{-\textrm{i} \tau}\;\theta_4(\tau)
\;\qquad\;
\theta_3(-1/\tau)=\sqrt{-\textrm{i} \tau} \;\theta_3(\tau)
\;\qquad\;\
\theta_4(-1/\tau)=\sqrt{-\textrm{i} \tau}\; \theta_2(\tau)
\ee
combined with the corresponding transformations   
for the Dedekind eta function
\be
\label{transf-eta-dedekind}
\eta(\tau+1)= \textrm{e}^{\textrm{i} \pi/12} \, \eta(\tau) 
\;\;\;\qquad\;\;\;
\eta(-1/\tau)=\sqrt{-\textrm{i} \tau}\, \eta(\tau)  \,.
\ee

Combining (\ref{jones-index-limit-asymmetry-intro}) with 
the invariance of (\ref{fising}) under $\tau_2 \mapsto -1/\tau_2$,
it is straightforward to realise that the Jones index for the complete Ising CFT$_2$ is
\be
\mu^{\phantom{x}}_{\textrm{\tiny Ising}}=1 \,.
\ee

In order to identify some terms in the partition function (\ref{partising}) of the complete model that correspond to the subalgebras 
$\mathcal{T}_{\mathbb{Z}_2}$ and $\mathcal{T}_{\textrm{su}(2)_2}$
(see Fig.\,\ref{sub1}),
let us consider the expressions of the characters 
for the irreducible representations of the Virasoro algebra
corresponding to $c=1/2$ and $h \in \big\{ 0,1/2,1/16 \big\}$.
These characters can be written as follows 
\cite{Itzykson:1986pj,DiFrancesco:1997nk}
\bea
\chi_0(\tau)\,
& = &
\frac{1}{2}\left(\sqrt{\frac{\theta_3(\tau)}{\eta(\tau)}}+\sqrt{\frac{\theta_4(\tau)}{\eta(\tau)}} \;\right)
\label{idchar}
\\ 
\rule{0pt}{.9cm}
\chi_{\frac{1}{16}}(\tau)
& = &
\sqrt{\frac{\theta_2(\tau)}{2\eta(\tau)}}
\label{sigmachar} 
\\
\rule{0pt}{.9cm}
\chi_{\frac{1}{2}}(\tau)\,
& = & 
\frac{1}{2}\left(\sqrt{\frac{\theta_3(\tau)}{\eta(\tau)}}
-\sqrt{\frac{\theta_4(\tau)}{\eta(\tau)}}\;\right)  .
\label{epchar} 
\eea
The partition function  of the complete Ising CFT$_2$ in (\ref{partising})
can be equivalently written in terms of the characters 
(\ref{idchar})-(\ref{epchar}) as follows
\cite{Belavin:1984vu,Cappelli:1986hf}
\be 
\parti_{\text{\tiny Ising},  1}
= 
\chi_0(\tau_2) \, \bar{\chi}_0(\bar{\tau}_2)
+ \chi_{\frac{1}{16}}(\tau_2) \,\bar{\chi}_{\frac{1}{16}}(\bar{\tau}_2)+
\chi_{\frac{1}{2}}(\tau_2) \,\bar{\chi}_{\frac{1}{2}}(\bar{\tau}_2)
\label{partising-character}
\ee
which is (\ref{toruspart}) specialised to 
the Ising CFT$_2$ and $\tau = \tau_2$.

The partition functions of the subalgebras 
$\mathcal{T}_{\mathbb{Z}_2}$ and $\mathcal{T}_{\textrm{su}(2)_2}$
can be constructed by considering only the characters 
corresponding to the primaries contained 
in each subalgebra (see Fig.\,\ref{sub1}).
This gives respectively
\be
\parti_{\mathbb{Z}_2,1}
\equiv 
\chi_0(\tau_2) \, \bar{\chi}_0(\bar{\tau}_2)
+
\chi_{\frac{1}{2}}(\tau_2) \,\bar{\chi}_{\frac{1}{2}}(\bar{\tau}_2)
\,=\,  
\frac{1}{2} \bigg(\left|\frac{\theta_3(\tau_2)}{\eta(\tau_2)}\right|+ \left|\frac{\theta_4(\tau_2)}{\eta(\tau_2)}\right|\bigg)
\label{z21}
\ee
and
\be
\parti_{\textrm{su}(2)_2,1}
\equiv 
\chi_0(\tau_2)\,\bar{\chi}_0(\bar{\tau}_2)
\,=\,  
\frac{\big| \big( \sqrt{\theta_3(\tau_2)}+\sqrt{\theta_4(\tau_2)}\,\big)\big|^2
}{4\, |{\eta(\tau_2)}|}
\label{z22}
\ee
where $\tau_2$ is (\ref{periodm2})
and $\bar{\tau}_2$ its complex conjugate. 
From (\ref{z21}) and (\ref{z22}), 
we obtain respectively
\be
\mathcal{F}_{\mathbb{Z}_2,2}(\x)
\,=\,  
\frac{1}{2}\bigg(1+ \left|\frac{\theta_4(\tau_2)}{\theta_3(\tau_2)}\right|\bigg)
\;\;\;\qquad\;\;\;
\mathcal{F}_{\textrm{su}(2)_2,2}(\x)
\,=\,  
\frac{\big| \big( \sqrt{\theta_3(\tau_2)}+\sqrt{\theta_4(\tau_2)}\,\big)\big|^2
}{4\, |{\theta_3(\tau_2)}|} \;.
\label{f22}
\ee

By employing (\ref{transf-jacobi-S})-(\ref{transf-eta-dedekind}), 
one realises that, 
while both (\ref{z21}) and (\ref{z22}) are invariant under $\tau_2 \mapsto \tau_2 + 1$, both of them are not invariant under $\tau_2 \mapsto -1/\tau_2$.
Hence, the crossing asymmetry (\ref{asymetry2}) 
specialised to (\ref{z21}) and (\ref{z22}) provides 
non trivial functions of the cross ratio (\ref{cross}),
that are given respectively by 
\be
A_{\mathbb{Z}_2,2}(\x)=
\log\!\left(
\frac{\mathcal{F}_{\mathbb{Z}_2,2}(\x)}{\mathcal{F}_{\mathbb{Z}_2,2}(1-\x)}
\right) =
\log\!\left(
\frac{\big|\theta_3(\tau_2)\big|+\big|\theta_4(\tau_2)\big|}{\big|\theta_3(\tau_2)\big|+\big|\theta_2(\tau_2)\big|}
\right) 
\label{asyz21}
\ee
and
\be
A_{\textrm{su}(2)_2,2}(\x)
= 
\log\!\left(
\frac{\mathcal{F}_{\textrm{su}(2)_2,2}(\x)}{\mathcal{F}_{\textrm{su}(2)_2,2}(1-\x)}
\right) 
=\,
2\,
\log\!
\left(\frac{
  \sqrt{\theta_3(\tau_2)}+\sqrt{\theta_4(\tau_2)}\,
}{
  \sqrt{\theta_3(\tau_2)}+\sqrt{\theta_2(\tau_2)}\, 
}\right) .
\label{asyz22}
\ee
By applying (\ref{jones-index-limit-asymmetry-intro}) 
to (\ref{asyz21}) and (\ref{asyz22}),
we find that their limit $\x\to 1^{-}$ gives respectively 
\be 
\mu^{\phantom{x}}_{\mathbb{Z}_2}=4 
\;\;\;\;\;\;\qquad\;\;\;\;\;
\mu^{\phantom{x}}_{\textrm{su}(2)_2}=16
\ee
which are the Jones indexes for 
$\mathcal{T}_{\mathbb{Z}_2}$ and $\mathcal{T}_{\textrm{su}(2)_2}$
obtained in (\ref{mu4}) and (\ref{mu16}) respectively.

We find it instructive to discuss also an expression that is not associated 
to a well defined local subalgebra.
In particular, 
by considering the fields $1$ and $\sigma$,
the procedure employed above provides 
the following partition function 
\be
\parti_{\{1\vee\sigma\},1}
\equiv 
\chi_0(\tau_2) \, \bar{\chi}_0(\bar{\tau}_2)
+
\chi_{\frac{1}{16}}(\tau_2) \,\bar{\chi}_{\frac{1}{16}}(\bar{\tau}_2)
\,=\,  
\frac{ 
\big|
\big(  \sqrt{\theta_3(\tau_2)}+\sqrt{\theta_4(\tau_2)}\,\big)\big|^2
+2\,\big|\theta_2(\tau_2)\big|
}{2 \, \eta(\tau_2)} \;.
\label{z2-fake}
\ee
The corresponding crossing asymmetry (\ref{asymetry2})  reads
\be 
A_{\{1\vee\sigma\},2}(\x)
\,=\,
\log\!
\left(\,
\frac{\big|\big(  \sqrt{\theta_3(\tau_2)}+\sqrt{\theta_4(\tau_2)}\, \big)\big|^2+2\,\big|\theta_2(\tau_2)\big|
}{
\big| \big(\sqrt{\theta_3(\tau_2)}+\sqrt{\theta_2(\tau_2)}\, \big)\big|^2
+2\,\big|\theta_4(\tau_2)\big|}
\,\right)
\ee
whose limit $\x\to1^-$ gives 
\be 
\mu^{\phantom{x}}_{\{1\vee\sigma\}}=\frac{16}{9}
\ee
which is not included in the set of the allowed values for the Jones index \cite{Jones83}.
This is consistent with the fact that the local subalgebra generated 
by  the fields $1$ and $\sigma$ is not closed under fusion.
Indeed,  from (\ref{fusionrules}), we have that 
also the field $\varepsilon$ occurs in $\sigma\times\sigma=1+ \varepsilon$.

In order to clarify the relation with the case of generic R\'enyi index $n \geqslant 2$, 
discussed in Sec.\,\ref{JonesIsingn}, 
we find it convenient to write the expressions 
(\ref{f22}) in terms of the Riemann theta function with characteristic. 
Given a $g \times g$ complex matrix $\tau$ 
whose imaginary part is positive definite,  
the Riemann theta function with characteristic 
and vanishing  argument $z=0$ is defined as follows
\be
\ThetaF{\boldsymbol{p}}{\boldsymbol{q}}(\tau)
\equiv 
\sum_{\pmb{m}\in \mathbb{Z}^g}
\textrm{e}^{\textrm{i} \pi (\pmb{m}+\boldsymbol{p})\cdot\tau\cdot(\pmb{m}+\boldsymbol{p})+ 2\textrm{i} \pi (\pmb{m}+\boldsymbol{p})\cdot \boldsymbol{q}} 
\;\;\;\qquad\;\;\;
\boldsymbol{p} \, , \boldsymbol{q} \in \mathbb{Z}^g/(2\mathbb{Z}^g)
\label{Thetafunc}
\ee 
where $\boldsymbol{p}$ and $\boldsymbol{q}$ 
are two $g$-dimensional vectors whose elements 
belong to $\big\{0\, , 1/2 \big\}$
and that define the characteristic of (\ref{Thetafunc}).
In the special case given by $g=1$,
the Riemann theta function (\ref{Thetafunc}) gives 
the Jacobi theta functions as follows
\be
\label{Jacobi234-spin}
\theta_2(\tau)=\ThetaN{\frac{1}{2}}{0}(\tau)
\;\;\qquad\;\;
\theta_3(\tau)=\ThetaN{0}{0}(\tau) 
\;\;\qquad\;\;
\theta_4(\tau)=\ThetaN{0}{\frac{1}{2}}(\tau) \,.
\ee
Hence, the functions in (\ref{f22}) can be written as
\bea
\mathcal{F}_{\mathbb{Z}_2,2}(\x)
&=&
\frac{1}{2\,
\bigg|\ThetaN{0}{0}\big(\tau_2\big) \bigg|}\;
\Bigg( \,
\bigg|\ThetaN{0}{0}\big(\tau_2\big) \bigg| + \bigg|\ThetaN{0}{\frac{1}{2}}\big(\tau_2\big) \bigg|
\,
\Bigg)
\label{F2z2} 
\\
\rule{0pt}{.9cm}
\mathcal{F}_{\textrm{su}(2)_2,2}(\x)
&=&
\frac{1}{4\,
\bigg|\ThetaN{0}{0}\big(\tau_2\big) \bigg|}\;
\left|\,
\sqrt{\ThetaN{0}{0}\big(\tau_2\big)}
+
\sqrt{\ThetaN{0}{\frac{1}{2}} \big(\tau_2\big)} \;
\right|^2 .
\label{F2su22}
\eea

We also find it instructive to obtain 
the partition functions (\ref{z21}) and (\ref{z22}) 
for the incomplete models corresponding to 
$\mathcal{T} \in \big\{ \mathbb{Z}_2\, , \textrm{su}(2)_2 \big\}$
from the partition function of the complete model 
(see (\ref{partising}) and (\ref{partising-character}))
by inserting specific projectors $\mathcal{P}_{\mathcal{T}}$.
 By using (\ref{toruspart}) specialized to the Ising CFT$_2$, 
let us consider 
\be 
\parti_{\mathcal{T}\!,1}=\Tr_{\mathcal{H}_{\text{\tiny Ising}}}\Big(
\mathcal{P}_{\mathcal{T}} \,
q^{L_0-\frac{1}{48}}\,
\bar{q}^{\bar{L}_0-\frac{1}{48}} \Big)
\label{torusprojector}
\ee
where each projector $\mathcal{P}_{\mathcal{T}}$ can be written in terms of 
the Verlinde lines. 
Given a cycle of $a$-type on the torus, 
the action of the Verlinde line along this cycle
(i.e. across the whole spatial direction, 
at a fixed value of the Euclidean time)
on the characters is \cite{Verlinde-lines}
\be
\mathcal{L}_h \, \chi_{h'} (\tau) 
=
\frac{S_{h,h'}}{S_{0,h'}} \, \chi_{h} (\tau) \,.
\label{L-action-on-chi}
\ee
Hence, from the partition function (\ref{toruspart}) of a CFT$_2$ 
(sub)model $\mathcal{T}$ on the torus, we get
\be 
 \Tr_{\mathcal{H}_{\mathcal{T}}}
 \Big( 
 \mathcal{L}_h \,q^{L_0-\frac{1}{48}} \,\bar{q}^{\bar{L}_0-\frac{1}{48}} 
 \Big)
 \,=\;\, 
 \begin{tikzpicture}[baseline={(current bounding box.center)}]
  \draw[thick] (0,0) rectangle (1.5,1.5);
  \draw[brown, thick] (0,0.75) -- (1.5,0.75);
  \node at (0.75,0.95) {\textcolor{brown}{$\mathcal{L}_h$}};
\end{tikzpicture}
\;\, =\,
\sum_{h'} \frac{S_{h,h'}}{S_{0,h'}} \;
\chi_{h'}(\tau)\,\bar{\chi}_{h'}(\bar{\tau})
\label{Tr-H-tau-L-a}
 \ee
in terms of the elements of the matrix introduced in (\ref{ST}).
In (\ref{Tr-H-tau-L-a}), 
notice that the Verlinde lines are in one-to-one correspondence with the primaries of the CFT$_2$ \cite{Verlinde-lines}.

In the special case of the Ising CFT$_2$, 
we can insert three Verlinde lines and,
by employing (\ref{Sising}), we get 
\bea
\label{1-verlinde-line-theta}
    \Tr_{\mathcal{H}_{\text{\tiny Ising}}}
    \Big( 
    \mathcal{L}_1 \,q^{L_0-\frac{1}{48}}\,\bar{q}^{\bar{L}_0-\frac{1}{48}} \Big)
   &=&
\begin{tikzpicture}[baseline={(current bounding box.center)}]
  \draw[thick] (0,0) rectangle (1.5,1.5);
\end{tikzpicture}
\;= \,
\frac{1}{2}\bigg(\left|\frac{\theta_3(\tau_2)}{\eta(\tau_2)}\right|+ \left|\frac{\theta_4(\tau_2)}{\eta(\tau_2)}\right|+ \left|\frac{\theta_2(\tau_2)}{\eta(\tau_2)}\right|\bigg)
\\
\rule{0pt}{1cm}
\label{eps-verlinde-line-theta}
\Tr_{\mathcal{H}_{\text{\tiny Ising}}}
\Big(
\mathcal{L}_\varepsilon \,q^{L_0-\frac{1}{48}}\,\bar{q}^{\bar{L}_0-\frac{1}{48}} \Big)
&=&
\begin{tikzpicture}[baseline={(current bounding box.center)}]
  \draw[thick] (0,0) rectangle (1.5,1.5);
  \draw[red, thick] (0,0.75) -- (1.5,0.75);
  \node at (0.75,0.95) {\textcolor{red}{$\mathcal{L}_\varepsilon$}};
\end{tikzpicture}
\;=\,
\frac{1}{2}\bigg(\left|\frac{\theta_3(\tau_2)}{\eta(\tau_2)}\right|+ \left|\frac{\theta_4(\tau_2)}{\eta(\tau_2)}\right|- \left|\frac{\theta_2(\tau_2)}{\eta(\tau_2)}\right|\bigg)
\\
\rule{0pt}{1cm}
\label{sigma-verlinde-line-theta}
\Tr_{\mathcal{H}_{\text{\tiny Ising}}}
\Big(
\mathcal{L}_\sigma \,q^{L_0-\frac{1}{48}}\,\bar{q}^{\bar{L}_0-\frac{1}{48}} \Big)
&=&
\begin{tikzpicture}[baseline={(current bounding box.center)}]
  \draw[thick] (0,0) rectangle (1.5,1.5);
  \draw[blue, thick] (0,0.75) -- (1.5,0.75);
  \node at (0.75,0.95) {\textcolor{blue}{$\mathcal{L}_\sigma$}};
\end{tikzpicture}
\;=\,
\frac{\sqrt{\,\theta_3(\tau_2)\,\theta_4(\bar{\tau}_2)} 
+ \sqrt{\,\theta_4(\tau_2)\,\theta_3(\bar{\tau}_2)}}{\sqrt{2}|{\eta(\tau_2)}|}\;.
\eea

As for the two subalgebras corresponding to 
$\mathcal{T} \in \big\{ \mathbb{Z}_2\, , \textrm{su}(2)_2 \big\}$,
the $\mathbb{Z}_2$ symmetry is implemented by $\mathcal{L}_\varepsilon$,
while for the $\textrm{su}(2)_2$ case 
we need a projector onto the identity sector, 
which is the only invariant Verma module under the combined action of $\mathcal{L}_\varepsilon$ and $\mathcal{L}_\sigma$.  
The projectors corresponding to 
$\mathcal{T} = \mathbb{Z}_2$ and $\mathcal{T} = \textrm{su}(2)_2 $
read respectively
\be
{\mathcal{P}}_{\mathbb{Z}_2}=
\frac{1}{2}
\big(1+\mathcal{L}_\varepsilon\big)
\;\;\;\;\;\qquad\;\;\;\;\;
{\mathcal{P}}_{\textrm{su}(2)_2}
=
\frac{1}{4}
\big(1+\mathcal{L}_\varepsilon+\sqrt{2}\,\mathcal{L}_\sigma\big)\,.
\label{projectors-ising}
\ee
By specialising (\ref{torusprojector}) to these projectors
and employing the expressions 
(\ref{1-verlinde-line-theta}), (\ref{eps-verlinde-line-theta}) and (\ref{sigma-verlinde-line-theta}), we find respectively 
\be
{\parti}_{\mathbb{Z}_2,1} = \frac{1}{2}\;
\begin{tikzpicture}[baseline={(current bounding box.center)}]
  \draw[thick] (0,0) rectangle (1.5,1.5);
\end{tikzpicture}
\;+\; \frac{1}{2}\,
\begin{tikzpicture}[baseline={(current bounding box.center)}]
  \draw[thick] (0,0) rectangle (1.5,1.5);
  \draw[red, thick] (0,0.75) -- (1.5,0.75);
  \node at (0.75,0.95) {\textcolor{red}{$\mathcal{L}_\varepsilon$}};
\end{tikzpicture}
\;=\,
\frac{1}{2}\left(\left|\frac{\theta_3(\tau_2)}{\eta(\tau_2)}\right|+ \left|\frac{\theta_4(\tau_2)}{\eta(\tau_2)}\right|\right)
\ee
and 
\be
{\parti}_{\textrm{su}(2)_2,1} 
\,=\,  \frac{1}{4}\;
\begin{tikzpicture}[baseline={(current bounding box.center)}]
  \draw[thick] (0,0) rectangle (1.5,1.5);
\end{tikzpicture}
\;+\, \frac{1}{4}\;
\begin{tikzpicture}[baseline={(current bounding box.center)}]
  \draw[thick] (0,0) rectangle (1.5,1.5);
  \draw[red, thick] (0,0.75) -- (1.5,0.75);
  \node at (0.75,0.95) {\textcolor{red}{$\mathcal{L}_\varepsilon$}};
\end{tikzpicture}
\;+\, \frac{\sqrt{2}}{4}\;
\begin{tikzpicture}[baseline={(current bounding box.center)}]
  \draw[thick] (0,0) rectangle (1.5,1.5);
  \draw[blue, thick] (0,0.75) -- (1.5,0.75);
  \node at (0.75,0.95) {\textcolor{blue}{$\mathcal{L}_\sigma$}};
\end{tikzpicture}
\;=\,  
\frac{\big| \big( \sqrt{\theta_3(\tau_2)}+\sqrt{\theta_4(\tau_2)}\,\big)\big|^2
}{4\, |{\eta(\tau_2)}|}
\label{z-su22-diagrams}
\ee
which coincide with (\ref{z21}) and (\ref{z22}).

\section{Ising CFT$_2$: 
Jones index from the $n\geqslant 2$ R\'enyi entropies
}
\label{JonesIsingn}

In this section, we discuss the expressions of the crossing asymmetry for a generic value of the R\'enyi index and for the submodels 
$\mathcal{T}_{\mathbb{Z}_2}$ and $\mathcal{T}_{\textrm{su}(2)_2}$, 
showing that the corresponding values of the Jones index are obtained 
from the limit (\ref{jones-index-limit-asymmetry-intro}).
The partition function of the Ising CFT$_2$ on $\mathscr{R}_n$
in terms of the characters for higher genus Riemann surfaces 
\cite{MooreSeiberg,ChoiKoh,Behera:1989gg,ChoiKim} 
and the expressions for the crossing asymmetry of 
$\mathcal{T}_{\mathbb{Z}_2}$ and $\mathcal{T}_{\textrm{su}(2)_2}$
are discussed in Sec.\,\ref{IsingGenus},
while their properties are explored in Sec.\,\ref{properties},  
including the relation with the corresponding values of the Jones index 
through the limit (\ref{jones-index-limit-asymmetry-intro}).

\subsection{Crossing asymmetry}
\label{IsingGenus}

The partition function of the Ising CFT$_2$ on a genus $g$ Riemann surface,
which can be obtained as a special case of the partition function of a rational CFT$_2$ on a generic Riemann surface
\cite{Alvarez-Gaume:1986rcs,Verlinde:1986kw,FriedanShenker,VerlindeVerlindeDijkgraaf,MooreSeiberg,ChoiKoh,Behera:1989gg,ChoiKim,Choi:su2,Mathur:1988xc,Gaberdiel:2010jf},
can be constructed from the characters 
for a genus $g$ Riemann surface
explored in 
\cite{MooreSeiberg,ChoiKoh,Behera:1989gg,ChoiKim,Choi:su2}. 
A convenient way to span the space of these characters exploits the
multiperipheral basis \cite{MooreSeiberg}, where 
a character is represented through the following diagram 
\be 
  \chi_{\pmb{i},\pmb{j}}^{\pmb{k}}(\tau)
  \,=\! 
  \begin{tikzpicture}[baseline={(current bounding box.center)}, scale=1.3]
  \draw[thick] (0,0.5) circle (0.25);
  \draw[thick] (0.25,0.5) -- (1.0,0.5);
  \draw[thick] (1.25,0.5) circle (0.25);
  \draw[thick] (1.5,0.5) -- (2.25,0.5); 
  \node at (2.5,0.5) {$\cdots$};
  \draw[thick] (2.65,0.5) -- (3.4,0.5);
  \draw[thick] (3.65,0.5) circle (0.25);
 \draw[thick] (3.9,0.5) -- (4.6,0.5);
 \draw[thick] (4.85,0.5) circle (0.25);
  \node at (0,0.9) {$i_1$};
  \node at (0,0.1) {$j_1=i_1$};
  \node at (0.6,0.7) {$k_1$};
  \node at (1.25,0.9) {$i_2$};
  \node at (1.25,0.1) {$j_2$};
  \node at (1.85,0.7) {$k_2$};
  \node at (3.0,0.7) {$k_{g-2}$};
  \node at (3.65,0.9) {$i_{g-1}$};
  \node at (3.65,0.1) {$j_{g-1}$};
  \node at (4.25,0.7) {$k_{g-1}$};
  \node at (4.85,0.9) {$i_g$};
  \node at (4.85,0.1) {$j_g=i_g$};
\end{tikzpicture}
\label{multi}
\ee
(see Fig.\,42 of \cite{MooreSeiberg}),
where $\tau$ is the period matrix of the Riemann surface,
while the components of the $g$-dimensional vectors 
$\pmb{i}=(i_1,i_2\dots,i_{g})$, 
$\pmb{j}=(j_1=i_1,j_2\dots,j_{g}=i_{g})$ 
and of the $g-1$-dimensional vector 
$\pmb{k}=(k_1,k_2\dots,k_{g-1})$ 
label the primary fields
and the vertices represent the fusion rules of the chiral CFT$_2$
(see (\ref{fusionrules}) for the case of the Ising model).
Since we are considering a rational CFT$_2$, 
the range of the components of these vectors is finite.

Explicit expressions for a generic character (\ref{multi}) are difficult to obtain 
and they are known only for a very few models
\cite{ChoiKoh,Behera:1989gg,ChoiKim,Choi:su2,Mathur:1988xc}.
In the case of the Ising CFT$_2$, 
they have been written in terms of the Riemann theta function (\ref{Thetafunc}) and read 
found \cite{ChoiKoh,Behera:1989gg,ChoiKim}
\be
\chi_{\pmb{i}}^{\pmb{k}} (\tau)
\,=\,
\frac{2^{\sum^g_{m=1} p_m(i_m,k_m)}}{2^{g} \, \sqrt{\parti_0}} \,
\sum_{\boldsymbol{l}} 
(-1)^{2\sum_{m=1}^g q_m(i_m,k_m) \delta_{i_m,\e}} \Big(
\ThetaF{\boldsymbol{p}(\pmb{i} ,\pmb{k})}{\boldsymbol{q}(\pmb{i} ,\pmb{k})}(\tau)
\Big)^{\frac{1}{2}} 
\label{charactersIsing}
\ee
where
the vectors $\boldsymbol{p}(\pmb{i} ,\pmb{k})$ 
and $\boldsymbol{q}(\pmb{i} ,\pmb{k})$ 
in the characteristic of the Riemann theta function
are determined by the vectors $\pmb{i}$ and $\pmb{k}$ 
introduced in (\ref{multi}) as follows
\be 
p_m(i_m,k_m)=\frac{1}{2}\,\delta_{i_m,\s}
\;\;\;\qquad\;\;\; 
q_m(i_m,k_m)=\frac{1}{2}\,
\delta_{i_m,\s} 
\big(1-\delta_{k_{m-1},k_m} \big)
+ 
\big(1-\delta_{i_m,\s} \big)\, l_m 
\label{characters-Ising-details}
\ee
where $l_m$ is the $m$-th component of the $g$-dimensional vector $\boldsymbol{l}$ (hence $l_m \in \{0,1/2\}$),
with
\be
\label{im-kr-values}
i_m \, , k_r \,\in \, 
\left\{ 0\,, \frac{1}{16}\, , \frac{1}{2} \, \right\}
\;\;\;\qquad\;\;\;
1 \leqslant m \leqslant g
\qquad
1 \leqslant r \leqslant g
\ee
and it is assumed that $k_0=k_g=\id$. 
In contrast with (\ref{multi}), 
we have not included $\pmb{j}$ because its elements 
are determined by the ones of $\pmb{i}$ and $\pmb{k}$
through the fusion rules (\ref{fusionrules}).
The characters (\ref{charactersIsing}) 
in the special case of $g=2$ are further discussed 
in the appendix\;\ref{g2characters}.
The normalization factor $\parti_0$ in (\ref{charactersIsing})
has been discussed in \cite{VerlindeVerlindeDijkgraaf}.

In term of the characters (\ref{charactersIsing}), 
the partition function of the Ising CFT$_2$
on the genus $n-1$ Riemann surface $\mathscr{R}_n$ 
(see (\ref{eq:replicaSurface})) reads 
\be
\parti_{\text{\tiny Ising},\,n-1}
=
\sum_{\pmb{i},\pmb{k}}
\big|
\chi_{\pmb{i}}^{\pmb{k}}({\tau}_n)
\big|^2
= \sum_{\pmb{i},\pmb{k}}
\Bigg|\;
\genusnlong{\small{$i_1$}}{\small{$i_1$}}{\small{$k_1$}}{\small{$i_2$}}{\small{$j_2(\pmb{i},\pmb{k})$}}{\small{$k_2$}}{\small{$k_{g-1}\,\,\,\,\,$}}{\small{$i_{g}$}}{\small{$i_{g}$}}
\Bigg|^2= \frac{1}{2^{n-1} \, \parti_0}\sum_{\boldsymbol{p},\boldsymbol{q}}
\left|
\ThetaF{\boldsymbol{p}}{\boldsymbol{q}}(\tau_n) 
\right|
\label{partg}
\ee
where $\tau_n$ is the period matrix (\ref{periodm})
and the sums over  $\pmb{i}$ and $\pmb{k}$ run over 
all possible values of $i_m$ and $k_r$ 
(see (\ref{im-kr-values}) with $g=n-1$)
that are allowed by the fusion rules  (\ref{fusionrules}). 
A complete list of the internal loops occurring in (\ref{partg})
is provided in the appendix\;\ref{Internal-loops}. 
Thus, (\ref{partg}) 
is the special case corresponding to $\mathscr{R}_n$
of the partition function of the Ising CFT$_2$ 
on a generic genus $g$ Riemann surface
\cite{VerlindeVerlindeDijkgraaf}.
The partition function (\ref{partg}) allows us to introduce 
\cite{Calabrese:2010he}
\be
\mathcal{F}_{\text{\tiny Ising},\,n}(\x)
= 
\frac{1}{2^{n-1}\,
\Big|\ThetaF{\pmb{0}}{\pmb{0}} \big(\tau_n\big) \Big|
}\;
\sum_{\boldsymbol{p},\boldsymbol{q}}
\left|\ThetaF{\boldsymbol{p}}{\boldsymbol{q}} \big(\tau_n \big) \right|
\label{FnIsing}
\ee
which satisfies the normalization condition 
$\mathcal{F}_{\text{\tiny Ising},\,n}(\x=0)=1$
and becomes (\ref{fising}) when $n=2$.
Combining (\ref{entropydisjoint}) specialised to $c=1/2$
and (\ref{FnIsing}), one obtains the R\'enyi entropies 
of two disjoint intervals for the Ising CFT$_2$ on the line and in the ground state \cite{Calabrese:2010he}.
We stress that, in the sum occurring in (\ref{partg}) and (\ref{FnIsing}),
only the Riemann theta functions with even values of $4 \boldsymbol{p} \cdot \boldsymbol{q}$ (i.e. with even characteristic)
occurs because the Riemann theta functions with odd values of $4 \boldsymbol{p} \cdot \boldsymbol{q}$ (i.e. with odd characteristic)
vanish identically.

In order to determine the R\'enyi entropies of two disjoint intervals 
for $\mathcal{T}_{\mathbb{Z}_2}$ and $\mathcal{T}_{\textrm{su}(2)_2}$,
and therefore extend (\ref{f22}) to a generic value of the R\'enyi index, 
we have to specify the characters of the form (\ref{charactersIsing}) 
that must be included in the corresponding partition functions.  
Our procedure is based on a generalization of (\ref{torusprojector})
to $\mathscr{R}_n$ with $n \geqslant 2$.

\begin{figure}[t!]
\begin{subfigure}{.5\textwidth}
	\centering
        \includegraphics[width=1.15\textwidth]{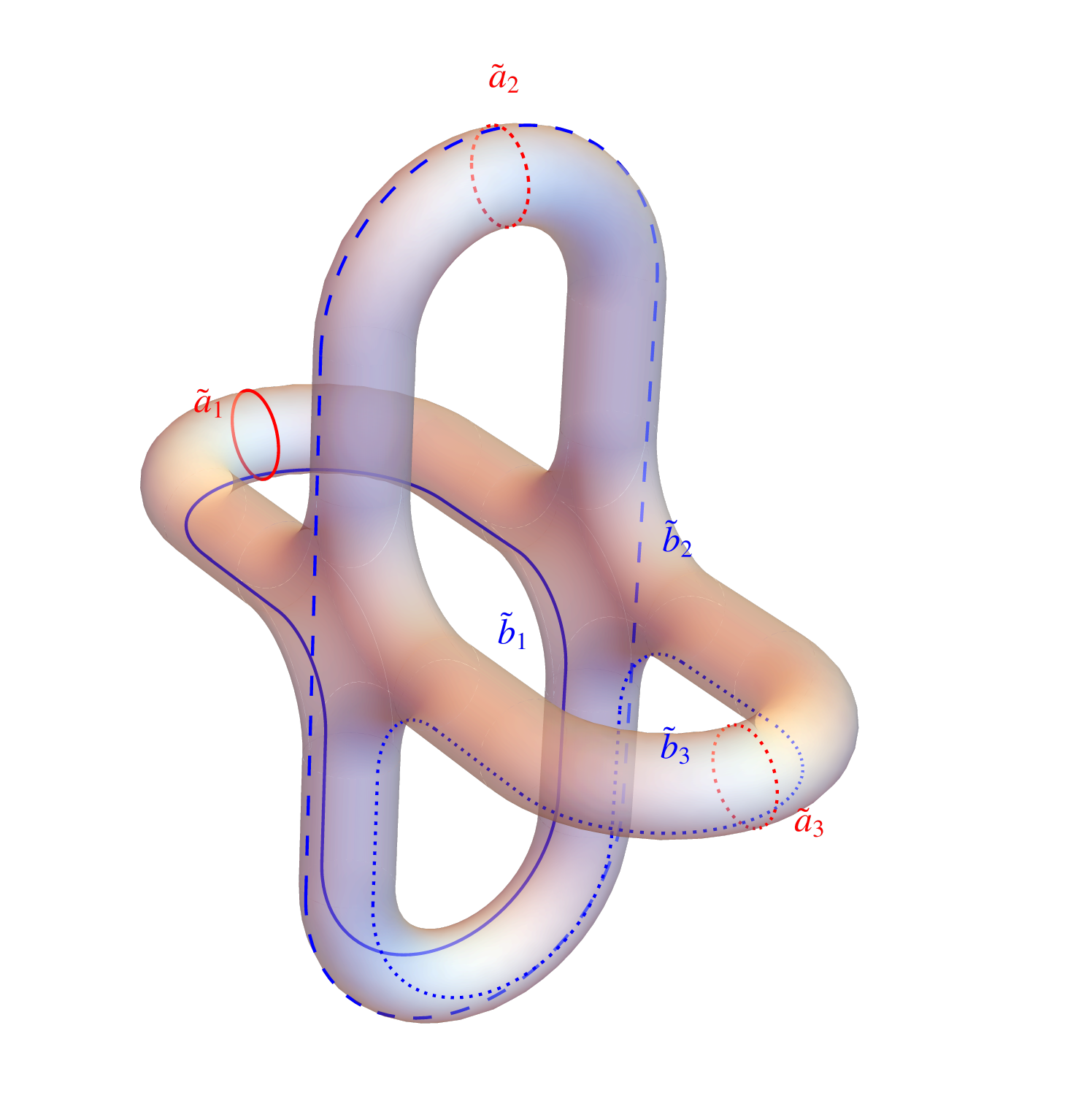}
  \end{subfigure}
  \begin{subfigure}{.5\textwidth}
\centering
\begin{tikzpicture}[scale=1.2,>=Stealth,thick]

\draw[fill=csheet, draw=black] (-3.3,0.7) -- (0.7,0.7) -- (1,2.2) -- (-3,2.2) -- cycle;
\draw[orange, thick] (-2.5,1.5) -- (-1.5,1.5);
\draw[teal, thick] (-2.5,1.4) -- (-1.5,1.4);
\draw[orange, thick] (-0.8,1.5) -- (0.2,1.5); 
\draw[teal, thick] (-0.8,1.4) -- (0.2,1.4); 
\node[red, thick] at (-2.68,1.45) {$\tilde{a}_1$};
\draw[red, thick] (-2,1.45) ellipse (0.9cm and 0.42cm);
\draw[blue, thick]
  (-1.75,1.5) arc[start angle=180, end angle=0, x radius=0.6cm, y radius=0.32cm];
\node[blue, thick] at (-1.1,2.01) {\small{$\tilde{b}_1$}};

\begin{scope}[shift={(0,0.3)}]
\draw[fill=csheet, draw=black] (-3.3,-1.3) -- (0.7,-1.3) -- (1,0.2) -- (-3,0.2) -- cycle;
\draw[cyan, thick] (-2.5,-0.5) -- (-1.5,-0.5);
\draw[orange, thick] (-2.5,-0.6) -- (-1.5,-0.6);
\draw[cyan, thick] (-0.8,-0.5) -- (0.2,-0.5); 
\draw[orange, thick] (-0.8,-0.6) -- (0.2,-0.6); 
\node[red, thick] at (-2.68,-0.55) {$\tilde{a}_2$};
\draw[red, dashed] (-2.0,-0.55) ellipse (0.9cm and 0.42cm);
\draw[blue, dashed]
  (-1.75,-0.5) arc[start angle=180, end angle=0, x radius=0.6cm, y radius=0.32cm];
\node[blue, thick] at (-1.1,0.01) {\small{$\tilde{b}_2$}};
\draw[blue, thick] (-0.7,-0.5) arc[start angle=20, end angle=340, x radius=0.15cm, y radius=0.15cm];
\draw[blue, thick] (-1.6,-0.6) arc[start angle=200, end angle=520, x radius=0.15cm, y radius=0.15cm];
\end{scope}

\begin{scope}[shift={(-5,-3.4)}]
\draw[fill=csheet, draw=black] (1.7,0.7) -- (5.7,0.7) -- (6,2.2) -- (2,2.2) -- cycle;
\draw[magenta, thick] (2.5,1.5) -- (3.5,1.5);
\draw[cyan, thick] (2.5,1.4) -- (3.5,1.4);
\draw[magenta, thick] (4.2,1.5) -- (5.2,1.5); 
\draw[cyan, thick] (4.2,1.4) -- (5.2,1.4); 
\node[red, dotted] at (2.32,1.45) {$\tilde{a}_3$};
\draw[red, dotted] (3,1.45) ellipse (0.9cm and 0.42cm);
\draw[blue, dotted]
  (3.25,1.5) arc[start angle=180, end angle=0, x radius=0.6cm, y radius=0.32cm];
\node[blue, thick] at (3.9,2.01) {\small{$\tilde{b}_3$}};
\draw[blue, thick] (4.3,1.5) arc[start angle=20, end angle=340, x radius=0.15cm, y radius=0.15cm];
\draw[blue, thick] (3.4,1.4) arc[start angle=200, end angle=520, x radius=0.15cm, y radius=0.15cm];
\draw[blue, dashed] (4.35,1.52) arc[start angle=20, end angle=340, x radius=0.2cm, y radius=0.2cm];
\draw[blue, dashed] (3.35,1.38) arc[start angle=200, end angle=520, x radius=0.2cm, y radius=0.2cm];
\end{scope}

\begin{scope}[shift={(-5,-3.1)}]
\draw[fill=csheet, draw=black] (1.7,-1.3) -- (5.7,-1.3) -- (6,0.2) -- (2,0.2) -- cycle;
\draw[teal, thick] (2.5,-0.5) -- (3.5,-0.5);
\draw[magenta, thick] (2.5,-0.6) -- (3.5,-0.6);
\draw[teal, thick] (4.2,-0.5) -- (5.2,-0.5); 
\draw[magenta, thick] (4.2,-0.6) -- (5.2,-0.6); 
\draw[blue, thick]
  (3.35,-0.6) arc[start angle=180, end angle=360, x radius=0.5cm, y radius=0.3cm];
  \draw[blue, dashed]
  (3.3,-0.6) arc[start angle=180, end angle=360, x radius=0.55cm, y radius=0.35cm];
  \draw[blue, dotted]
  (3.25,-0.6) arc[start angle=180, end angle=360, x radius=0.6cm, y radius=0.4cm];
\end{scope}

\end{tikzpicture}
  \end{subfigure}%
  \\
  \caption{
   The Riemann surface $\mathscr{R}_4$ 
   (see (\ref{eq:replicaSurface})) 
   with a specific canonical homology basis that provides the period matrix (\ref{periodma}). 
  }
\label{replica4}
\end{figure}
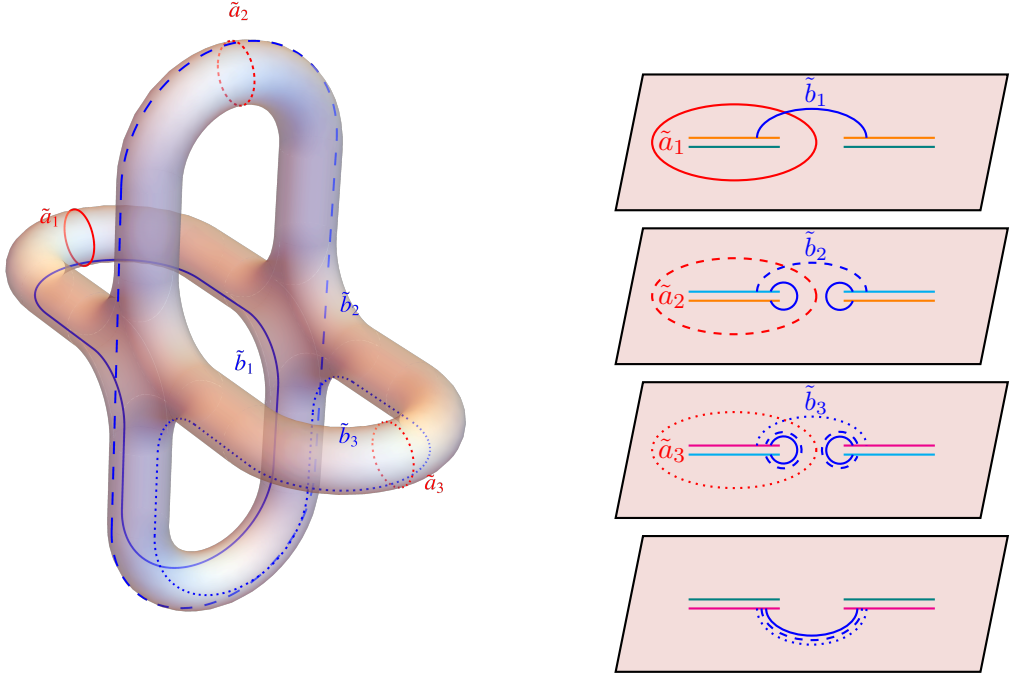

It is worth considering a canonical homology basis
$\big\{ (\tilde{a}_j, \tilde{b}_j) \, ; 1 \leqslant j \leqslant n-1 \big\}$
of $\mathscr{R}_n$ where the cycles $\tilde{a}_j$ 
enclose the cut corresponding to the interval $R_1$ 
on $n-1$ consecutive copies, in the multi-sheet representation of 
$\mathscr{R}_n$. 
Indeed, in this basis, we find it more natural to provide a general rule for the insertion of the proper projectors. 
In the special case of $n=4$, these cycles are shown in 
Fig.\,\ref{replica4}, 
for two equivalent representations of $\mathscr{R}_4$.
This canonical homology basis has been employed in 
\cite{GravaAbel,Headrick:2012fk,Coser:2013qda}.
The period matrix $\tilde{\tau}_n$ in this 
canonical homology basis is obtained from the period matrix 
$\tau_n$ in  (\ref{periodm}) as follows \cite{Coser:2013qda}
\be 
 \tilde{\tau}_n\,=\,\iup \,\tau_n\,\ilow 
 \label{periodma}
\ee
where $\iup$ and $\ilow$ are respectively 
the upper and lower triangular $(n-1)\times(n-1)$ matrices 
whose elements are 
\be
 \big[\iup \big]_{ij} \equiv
\left\{
\begin{array}{ll}
1 \hspace{.5cm} & i \leqslant j \\
\rule{0pt}{.5cm}
0 & i> j
\end{array}
\right.
\;\;\;\qquad\;\;\;\;
 \big[\ilow \big]_{ij} \equiv
\left\{
\begin{array}{ll}
1 \hspace{.5cm} & i \geqslant j \\
\rule{0pt}{.5cm}
0 & i < j
\end{array}
\right.
\ee
hence they satisfy $\iup = \ilow^{\textrm{t}}$.
Notice that $\iup$ and $\ilow$ simplify to $1$ when $n=2$,
implying that (\ref{periodma}) becomes 
$\tilde{\tau}_2 = \tau_2$ in this case.
In the appendix\,\ref{modular}, we also show that
\be
\tau_n(1-\x) \,=\, -\,\tilde{\tau}_n(\x)^{-1} \,.
\label{inverse-g-tilde}
\ee
In the special case of $n=2$, this relation 
and the observation made in the previous sentence 
lead to $\tau_2(1-\x) = -1/\tau_2(\x)$, 
as one easily finds also from (\ref{periodm2}).

We  construct the partition functions 
for the models $\mathcal{T}_{\mathbb{Z}_2}$ 
and $\mathcal{T}_{\textrm{su}(2)_2}$
by implementing the fusion rules characterising these models. 
The resulting expressions have a form similar to (\ref{partg}), 
where $\pmb{i}$ and $\pmb{k}$ take values 
in specific subsets of the allowed values established in (\ref{im-kr-values}).

As for $\mathcal{T}_{\mathbb{Z}_2}$, 
since (\ref{fusionrules}) and (\ref{theories}) 
tell us that the field $\sigma$ does not occur, 
we define the partition function obtained from (\ref{partg})
by imposing that $i_m$ and $k_r$ cannot take the value $1/16$.
\\
From (\ref{charactersIsing})-(\ref{im-kr-values}),
for a generic genus and a generic period matrix $\tau$,
the characters where $i_m$ and $k_m$ 
cannot take the value $\tfrac{1}{16}$ read
\be
\chi_{\pmb{i}}^{\pmb{k}} (\tau)=\frac{1}{2^{g} \sqrt{\parti_0}}  \sum_{\boldsymbol{q}} (-1)^{2\sum_{m=1}^g q_m \delta_{i_m,\e}} \Big(\ThetaF{\pmb{0}}{\boldsymbol{q}}(\tau) \Big)^{\frac{1}{2}} .
\label{charactersIsing2}
\ee
By applying this observation to our case, 
where $g=n-1$ and the period matrix is (\ref{periodma}), 
for the partition function of $\mathcal{T}_{\mathbb{Z}_2}$ we find 
\bea
& & \parti_{\mathbb{Z}_2,\,n-1}
\,=
\sum_{\pmb{i},\pmb{k} \,\in \, \mathsf{S}} 
\!
\big|
\chi_{\pmb{i}}^{\pmb{k}}(\tilde{\tau}_n)
\big|^2
=\,
\frac{1}{2^{n-1}\parti_0}\sum_{\boldsymbol{q}}
\left|
\ThetaF{\pmb{0}}{\boldsymbol{q}}(\tilde{\tau}_n) 
\right| 
\label{partgz2} 
\eea
where 
\be
\mathsf{S} \equiv 
\Big\{\,
i_m \, , k_r \in  \big\{ 0\, , 1/2 \, \big\}\,;
1\leqslant m \leqslant n-1\; ;
1 \leqslant r \leqslant n-2
\,\Big\} \label{S-set}
\ee
and the elements of the vector $\boldsymbol{q}$ 
take all the allowed values.
From the first expression in (\ref{characters-Ising-details}), 
it is evident that $\boldsymbol{p}=\boldsymbol{0}$
when $i_m \neq \tfrac{1}{16}$ is imposed. 
Combining (\ref{partgz2}) with the same normalization 
holding for  (\ref{FnIsing}), 
for the function of $\x$ in (\ref{Fn}) for $\mathcal{T}_{\mathbb{Z}_2}$ 
we find 
\be
\mathcal{F}_{\mathbb{Z}_2,\,n}(\x)
\,=\, 
\frac{1}{2^{n-1}\,
\Big|\ThetaF{\pmb{0}}{\pmb{0}}\big(\tilde{\tau}_n\big)\Big|} \;
\sum_{\boldsymbol{q}}
\left|\ThetaF{\pmb{0}}{\boldsymbol{q}}
\big(\tilde{\tau}_n\big) 
\right| \,.
\label{Fnz2}
\ee
Plugging this expression into (\ref{asymetrydefintro}),  
for the corresponding crossing asymmetry, we obtain 
\be
\label{asyft-z2}
A_{\mathbb{Z}_2,\,n}(\x)
= 
\frac{1}{n-1}\,\log\!
\left(\frac{\mathcal{F}_{\mathbb{Z}_2,\,n}(\x)}{\mathcal{F}_{\mathbb{Z}_2,\,n}(1-\x)}\right) .
\ee

As for $\mathcal{T}_{\textrm{su}(2)_2}$, we follow the same procedure.
In particular, 
since (\ref{fusionrules}) and (\ref{theories}) 
tell us that only the identity field is allowed,  
the corresponding partition function is obtained from (\ref{partg}),
by removing from the summation all the characters 
whose diagrammatic representative (\ref{multi}) contains 
at least either one $\tfrac{1}{16}$ or one $\tfrac{1}{2}$.
This procedure selects only the character (\ref{charactersIsing2})
corresponding to $\pmb{i}=\pmb{0}$ and $\pmb{k}=\pmb{0}$, namely
\be
\chi_{\pmb{0}}^{\pmb{0}} (\tau)=\frac{1}{2^{g} \sqrt{\parti_0}} \sum_{\boldsymbol{q}} \Big(\ThetaF{\pmb{0}}{\boldsymbol{q}}(\tau) \Big)^{\frac{1}{2}} .
\label{charactersIsing3}
\ee
Specializing this expression to 
$g=n-1$ and to the period matrix (\ref{periodma}),
we find that the partition function of $\mathcal{T}_{\textrm{su}(2)_2}$ is
\be
\label{partgsu22}
\parti_{\textrm{su}(2)_2,\,n-1}
\,=\, 
\big|
\chi_{\pmb{0}}^{\pmb{0}}(\tilde{\tau}_n)
\big|^2\,=\,
\frac{1}{2^{2n-2}\, \parti_0}\, 
\bigg|
\sum_{\boldsymbol{q}}
\Big(\ThetaF{\pmb{0}}{\boldsymbol{q}}(\tilde{\tau}_n)\Big)^\frac{1}{2} 
\,\bigg|^2 .
\ee 
From this result and the normalization condition employed  
(\ref{FnIsing}), for the function of $\x$ in (\ref{Fn}) 
of the $\mathcal{T}_{\textrm{su}(2)_2}$ model we obtain 
\be
\mathcal{F}_{\textrm{su}(2)_2,n}(\x)
\,=\,
\frac{1}{2^{2n-2}\,
\Big|\ThetaF{\pmb{0}}{\pmb{0}}\big(\tilde{\tau}_n\big) \Big|}\;
\Bigg|
\sum_{\boldsymbol{q}}\Big(\ThetaF{\pmb{0}}{\boldsymbol{q}}
\big(\tilde{\tau}_n\big)\Big)^\frac{1}{2} \,
\Bigg|^2\label{Fnsu22}
\ee
which provides the crossing asymmetry (\ref{asymetrydefintro}) as follows 
\be
\label{asyft-su2}
A_{\textrm{su}(2)_2,\,n}(\x)
= 
\frac{1}{n-1}\,\log\!
\left(\frac{\mathcal{F}_{\textrm{su}(2)_2,\,n}(\x)}{\mathcal{F}_{\textrm{su}(2)_2,\,n}(1-\x)}\right) .
\ee

The expressions (\ref{asyft-su2}) and (\ref{asyft-z2}) 
are the main results of this manuscript
and their properties are discussed in Sec.\,\ref{properties}.

It is worth discussing the invariance 
of the partition functions (\ref{partgz2}) and (\ref{partgsu22}).
A crucial feature of the partition function of a complete CFT$_2$ on a Riemann surface of genus $g$ is its invariance under the 
symplectic modular group $\textrm{Sp}(2g,\mathbb{Z})$, 
whose action on the period matrix is given by a symplectic $(2g) \times (2g)$ matrix made by integer numbers,
as discussed in more details in the appendix\;\ref{modular}.
This symmetry encodes the freedom to choose 
different canonical homology bases and, in general, 
it mixes the $a$-type and $b$-type cycles (see (\ref{cycletansform})).
Considering the transformation rule of the Riemann theta function with characteristic (\ref{Thetafunc}) under these transformations
\cite{Faygenus,Alvarez-Gaume:1986rcs,Verlinde:1986kw},
in the appendix\;\ref{modular}
we observe that (\ref{partgz2}) and (\ref{partgsu22}) 
are invariant under the transformations corresponding to 
the semidirect product 
$\textrm{GL}(g,\mathbb{Z})\ltimes \textrm{Sym}(g,\mathbb{Z})$,
that is a Siegel parabolic subgroup of $\textrm{Sp}(2g,\mathbb{Z})$ \cite{Siegelsym,Hulek1998,Geer2006,Maloney:2020nni},
whose elements act on the period matrix as follows 
\be 
\tau_n \,\mapsto\, A\,\tau_n \,A^{\textrm{t}}
+
B A^{\textrm{t}}  
\;\;\; \qquad \;\;\;
  (A^{-\textrm{1}} B)^{\textrm{t}}= A^{-\textrm{1}}B 
\label{submodular}
\ee
where $A$ and $B$ are $g\times g$ matrices made by integer numbers. 
Notice that the second condition in (\ref{submodular}) corresponds to imposing that $B A^{\textrm{t}}$ is symmetric. 
More details about this subgroup and its action are provided 
in appendix\;\ref{modular}.
The crucial feature of 
$\textrm{GL}(g,\mathbb{Z})\ltimes \textrm{Sym}(g,\mathbb{Z})$ 
in our analysis is that it preserves the condition 
 $\boldsymbol{p} = \boldsymbol{0}$ 
in the characteristic of the Riemann theta function.
Moreover, these transformations are such that the transformed $a$-type cycles are entirely  determined by the original $a$-type cycles
(see (\ref{cycletansform}) with vanishing $C$).
Following \cite{Coser:2013qda}, we observe that (\ref{periodma}) and its inverse transformation 
correspond 
to the special cases of (\ref{submodular}) 
given by 
$(A, B)=\big(\iup, \boldsymbol{0}_g\big)$ 
and  $(A, B)=\big((\iup)^{-1}, \boldsymbol{0}_g \big)$.
When $g=1$, 
the transformations (\ref{submodular}) reduce to 
the first map in (\ref{tsinv}) and its powers, 
whose action is $\tau \mapsto \tau + m$ with $m \in \mathbb{Z}$.

We find it instructive to write 
the expressions (\ref{asyft-z2}) and (\ref{asyft-su2}) 
for the crossing asymmetry, also in a different form. 
The transformation (\ref{Haag-exchange-UV-def}),
that exchanges $\x$ and $1-\x$,
is not a symmetry for the partition functions 
(\ref{partgz2}) and (\ref{partgsu22})
and therefore the corresponding symplectic matrix does not belong to 
$\textrm{GL}(g,\mathbb{Z})\ltimes \textrm{Sym}(g,\mathbb{Z})$.
Its action on the period matrix (\ref{periodm}) reads 
\be 
{\tau}_n(1-\x)
=
- \,(\ilow)^{-1} 
\,\tau_n(\x)^{-1} \,
(\iup)^{-1}
\label{x1x}
\ee
(see also Eq.\,(C.23) in \cite{Coser:2013qda} for $N=2$).

By using (\ref{x1x}) 
and the transformation property of the Riemann theta function with characteristic, 
in the appendix\;\ref{modular} we show that 
\be 
\Big(\ThetaF{\pmb{0}}{\pmb{0}}\big({\tau}_n(1-\x)\big)\Big)^{-1} 
\,
\ThetaF{\pmb{0}}{\boldsymbol{q}}\big({\tau}_n(1-\x)\big) 
=
\Big(\ThetaF{\pmb{0}}{\pmb{0}}\big({\tau}_n(\x)\big)\Big)^{-1} 
\,
\ThetaF{-\iup\boldsymbol{q}}{\pmb{0}}\big({\tau}_n(\x)\big) 
\label{transform00m}
\ee
which allows us to write (\ref{asyft-z2}) and (\ref{asyft-su2})
respectively as 
\bea
A_{\mathbb{Z}_2,n}(\x)
&=&
\frac{1}{n-1}\,
\log\!\left[
\left({\sum_{\boldsymbol{q_1}}\left|\ThetaF{\pmb{0}}{\boldsymbol{q_1}}\big({\tau}_n\big)\right|}\right)
\Bigg/
\left({\sum_{\boldsymbol{q_2}}\left|\ThetaF{-\iup\boldsymbol{q_2}}{\pmb{0}}\big({\tau}_n\big) \right|}\right)
\right]
\nonumber
\\
\rule{0pt}{.9cm}
\label{asygz2}
&=&
\frac{1}{n-1}\,
\log\!\left[
\left({\sum_{\boldsymbol{q}}\left|\ThetaF{\pmb{0}}{\boldsymbol{q}}\big({\tau}_n\big)\right|}\right)
\Bigg/
\left({\sum_{\boldsymbol{p}}
\left|\ThetaF{\boldsymbol{p}}{\pmb{0}}\big({\tau}_n\big) \right|}\right)
\right]
\eea
and
\bea
A_{\textrm{su}(2)_2,n}(\x)
&=&
\frac{2}{n-1}\,
\log\!\left[\;
\left(\,\sum_{\boldsymbol{q_1}}
\sqrt{
\ThetaF{\pmb{0}}{\boldsymbol{q_1}}\big({\tau}_n\big) 
}\;
\right)\Bigg/\left(\,
\sum_{\boldsymbol{q_2}}
\sqrt{
\ThetaF{-\iup\boldsymbol{q_2}}{\pmb{0}}\big({\tau}_n\big) 
}
\,\right)
\;\right]
\nonumber
\\
\rule{0pt}{.9cm}
&=&
\frac{2}{n-1}\,
\log\!\left[\;
\left(\,\sum_{\boldsymbol{q}}
\sqrt{
\ThetaF{\pmb{0}}{\boldsymbol{q}}\big({\tau}_n\big) 
}\;
\right)\Bigg/\left(\,
\sum_{\boldsymbol{p}}
\sqrt{
\ThetaF{\boldsymbol{p}}{\pmb{0}}\big({\tau}_n\big) 
}
\,\right)
\;\right] .
\label{asygsu22}
\eea

The partitions functions (\ref{partgz2}) and (\ref{partgsu22}) 
can also be obtained by inserting  the proper projectors, 
extending to $\mathscr{R}_n$ the discussion reported in the final part 
of Sec.\,\ref{JonesIsing2} for the $n=2$ case. 
It is crucial to establish the position of these projectors on $\mathscr{R}_n$.
Inserting projectors along lines encircling each of the intervals in all the copies is redundant because the topological nature of these line operators allows us to move them on $\mathscr{R}_n$ and fuse them according to their fusion rule.
A natural choice is given by the $n$ lines encircling only the first interval on each sheet, but one of these line operators is still redundant 
(as it happens also in the $n=2$ case); 
hence we choose to ignore the closed line in the $n$-th sheet
without loss of generality. 
Thus, in the case of $n=4$,
we insert projectors only along all the cycles of $a$-type in Fig.\,\ref{replica4}.
These projectors generalise (\ref{projectors-ising}) and are defined 
through the generalization of the Verlinde lines on the $a$-type cycles,
\cite{MooreSeiberg,Verlinde:1986kw,Gaiotto:2014lma,Chang:2018iay}. 
Notice that the 
$\textrm{GL}(g,\mathbb{Z})\ltimes \textrm{Sym}(g,\mathbb{Z})$ 
invariance of the partition functions (\ref{partgz2}) and (\ref{partgsu22}) allows to consider equivalently the projectors along 
all the $a$-type cycles obtained by applying a generic transformation 
(\ref{submodular}) to the $a$-type cycles
associated to (\ref{periodma}) or (\ref{periodm}).
Indeed, these two period matrices are also related through a transformation (\ref{submodular}), as already remarked above. 
A similar discussion occurs also in Sec.\,\ref{Fermions}.

\subsection{Properties of the crossing asymmetry and Jones index}
\label{properties}

It is worth discussing 
some properties of the crossing asymmetries 
$A_{\mathbb{Z}_2,n}(\x)$ and $A_{\textrm{su}(2)_2,n}(\x)$,
whose explicit expressions 
are given by (\ref{asyft-z2}) and (\ref{asyft-su2}), 
or by (\ref{asygz2}) and (\ref{asygsu22}), respectively.

\begin{figure}[t!]
\centering
    \begin{minipage}{1\textwidth}
            \centering
		\includegraphics[width=1\textwidth]{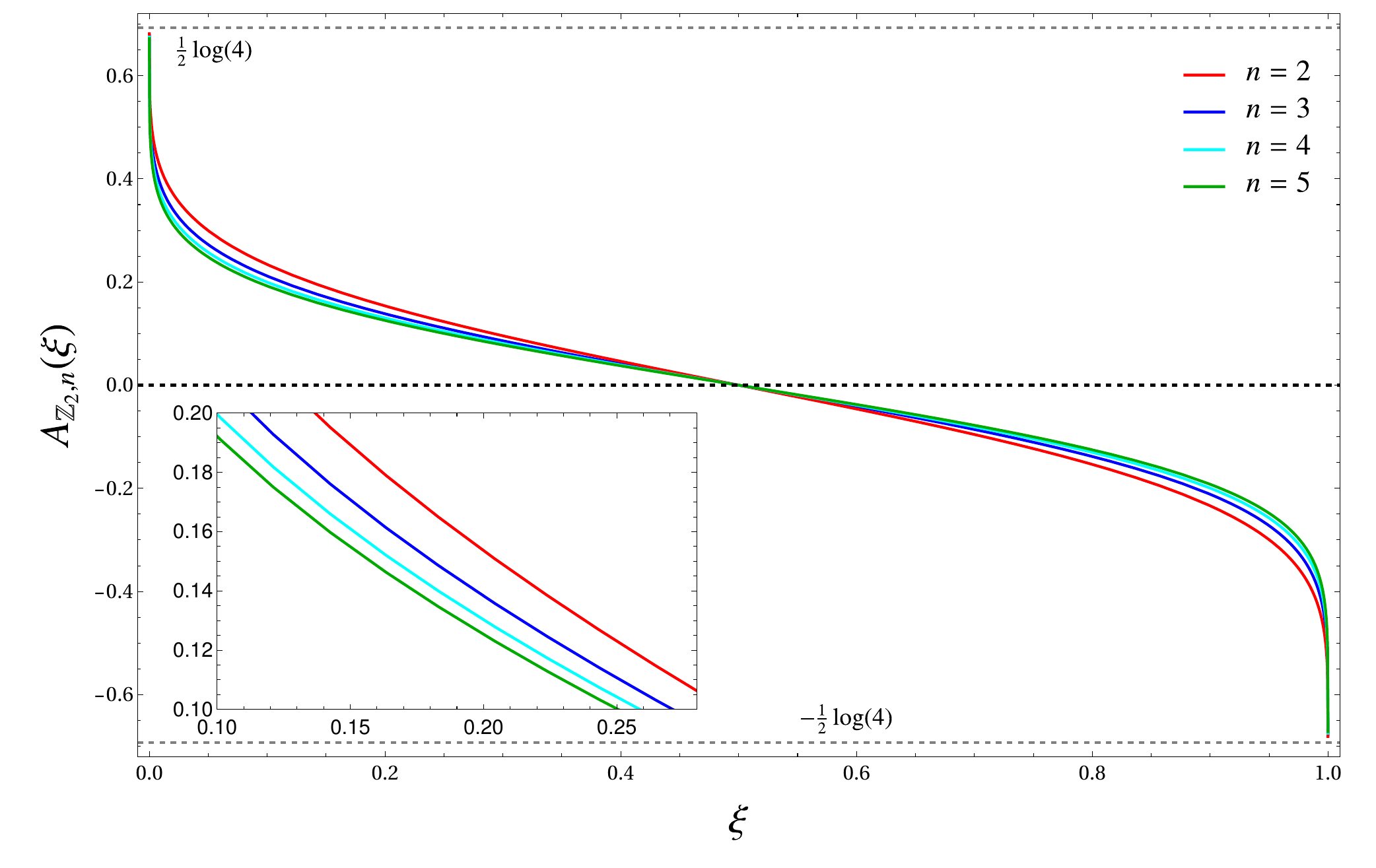}
	\end{minipage}
	\\
	\begin{minipage}{1\textwidth}
            \centering
		\includegraphics[width=1\linewidth]{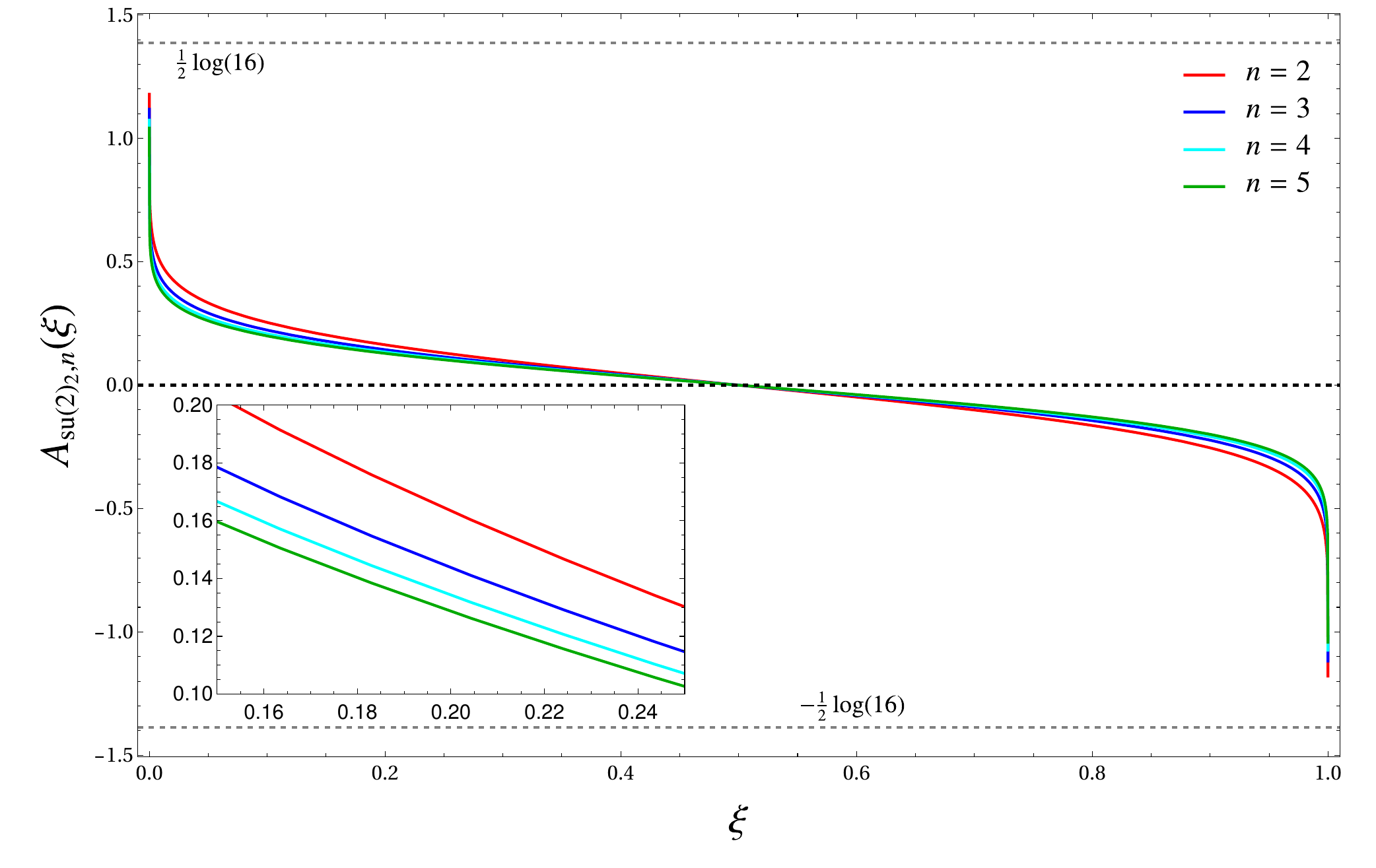}
	\end{minipage}
    \caption{
    The crossing asymmetries 
    ${A}_{\mathbb{Z}_2\!,\,n} (\x)$ and ${A}_{\textrm{su}(2)_2\!,\,n} (\x)$
    as functions of the cross ratio $\x$ 
    (in the top and bottom panel respectively),
    for different values of $n$,
    evaluated through their analytic expressions 
    given in (\ref{asyft-z2}) and (\ref{asyft-su2}) respectively.
    The horizontal dashed gray lines highlight the value of the corresponding global Jones index (see (\ref{limtsmu})).
    }
    \label{fig:Asymmetry}
\end{figure}

The numerical evaluation of these crossing asymmetries is
very demanding as $n$ increases. 
This is because the occurrence of the 
Riemann theta function with characteristic (\ref{Thetafunc}) gives them a non algebraic form. 
In our numerical analysis, 
we have employed the method discussed in \cite{Grava:2021yjp}.
This method allows to write the analytic expressions for
$A_{\mathbb{Z}_2,n}(\x)$ and $A_{\textrm{su}(2)_2,n}(\x)$
in an algebraic form,
that has been used to plot the solid curves 
displayed in Fig.\,\ref{fig:Asymmetry}, representing these quantities.

The main property of the crossing asymmetries 
$A_{\mathbb{Z}_2,n}(\x)$ and $A_{\textrm{su}(2)_2,n}(\x)$
is that the global Jones index for the corresponding model
is obtained through the limit (\ref{jones-index-limit-asymmetry-intro}).
In particular, 
for $\mathcal{T}_{\mathbb{Z}_2}$ and $\mathcal{T}_{\textrm{su}(2)_2}$,
we find respectively that  
\be 
\lim_{\x \to 1} 
A_{\mathbb{Z}_2,n}(\x)
=
-\frac{1}{2}\log(4)
\;\;\;\;\qquad\;\;\;
\lim_{\x \to 1} 
A_{\textrm{su}(2)_2,n}(\x)
=
-\frac{1}{2}\log(16)
\label{limtsmu}
\ee
whose values are highlighted by 
the horizontal dashed gray lines in Fig.\,\ref{fig:Asymmetry}.
The results (\ref{limtsmu}) have been observed numerically 
for all the values of $n$ that we have been able to explore,
and this provides an important consistency check 
for the analytic expressions (\ref{asyft-z2}) and (\ref{asyft-su2}).

The limits in (\ref{limtsmu}) can be derived also analytically, 
by studying the expansions of the analytic expressions 
(\ref{asygz2}) and (\ref{asygsu22})  as $\x \to 0^+$.
This analysis can be carried out by adapting to our case
in a straightforward way the results found 
in Sec.\,5 of \cite{Calabrese:2010he}
about the expansion of (\ref{Thetafunc}) 
with the period matrix (\ref{periodm})
as $\x \to 0^+$.
In particular, 
from (\ref{asygz2}) and (\ref{asygsu22}),
by introducing \cite{Calabrese:2010he}
\be
\label{S-alha-def}
\mathcal{S}_\alpha(n) 
\equiv 
\sum_{j=1}^{n-1} 
\frac{n}{\big[ \sin(\pi j/n)\big]^{2\alpha}}
\ee
we find that 
for $\mathcal{T}_{\mathbb{Z}_2}$ 
we need the following expansions
\be
\label{expansion-xi-0-z2}
\sum_{\pmb q}
\Big| \,
\ThetaF{\pmb 0}{\pmb q}\big({\tau}_n\big)
\,\Big|
\,=\,
2^{n-1}
\big( 1 + O(\x) \big)
\;\qquad\;
\sum_{\pmb p}
\Big| \,
\ThetaF{\pmb p}{\pmb 0}\big({\tau}_n\big)
\,\Big|
\,=\,
1 
+ 
\frac{\mathcal{S}_{1/4}(n) }{\sqrt{2n}}
\, \x^{1/4}  
+\dots 
\ee
while for $\mathcal{T}_{\textrm{su}(2)_2}$ we need
\be
\label{expansion-xi-0-su2}
\sum_{\pmb q}
\sqrt{
\ThetaF{\pmb 0}{\pmb q}\big({\tau}_n\big)
}
\,=\,
2^{(n-1)}\big( 1 + O(\x) \big)
\qquad
\sum_{\pmb p}
\sqrt{
\ThetaF{\pmb p}{\pmb 0}\big({\tau}_n\big)
} 
\,=\,
1 
+ 
\frac{\mathcal{S}_{1/8}(n) }{\sqrt[4]{8n}}
\, \x^{1/8}
+\dots 
\ee
where the dots denote subleading terms. 
Plugging the expansions in 
(\ref{expansion-xi-0-z2}) and (\ref{expansion-xi-0-su2})
into (\ref{asygz2}) and (\ref{asygsu22}) respectively, 
the expansions of the crossing asymmetries as $\x \to 0$ read
\be
\label{Asym-xi-0-expansion-z2}
A_{\mathbb{Z}_2,n}(\x)
=
\frac{1}{2} \log(4) 
-
\frac{\mathcal{S}_{1/4}(n) }{\sqrt{2}\, (n-1)\,\sqrt{n}}
\; \x^{1/4}
+\dots
\ee
and
\be
\label{Asym-xi-0-expansion-su2}
A_{\textrm{su}(2)_2,n}(\x)
=
\frac{1}{2} \log(16) 
-
\frac{\sqrt[4]{2}\;\mathcal{S}_{1/8}(n) }{(n-1)\,\sqrt[4]{n}}
\; \x^{1/8}
+\dots \;.
\ee
The leading term of these two expansions gives 
the global Jones index of the corresponding model
for any finite value of the R\'enyi index $n$,
as previously stated in 
(\ref{jones-index-limit-asymmetry-intro}) and (\ref{limtsmu}).

We remark that the next subleading term depends on the R\'enyi index 
in a non trivial way. 
However, the exponent of $\x$ in these terms is independent of $n$.
This kind of term appears in the expansion of (\ref{Fn}) and its power is determined by the smallest conformal dimension in the spectrum of the primaries via $2(h+\bar{h})$, as discussed in 
\cite{Calabrese:2010he}.
In our analysis, the subleading term in the expansions 
(\ref{Asym-xi-0-expansion-z2})-(\ref{Asym-xi-0-expansion-su2})
comes from the expansion of 
$\parti_{\mathcal{T} \!, \, n-1}  (\tau_n)\big|_{R \leftrightarrow R'}$.
By expressing this partition functions in terms of the characters (\ref{charactersIsing}), one observes that it does not correspond to 
 a local model because it includes primary fields that are non local 
 with respect to each other or themselves \cite{Benedetti:2024dku}.
 For instance, considering $\mathcal{T}_{\mathbb{Z}_2}$ 
for $n=2$, we have that 
\bea
\label{part-ill-1}
\parti_{\mathbb{Z}_2,1}(\tau_2) 
\big|_{R \leftrightarrow R'}
&=&
\parti_{\mathbb{Z}_2,1}(-1/\tau_2) 
=\,\,  
\frac{1}{2} \bigg(\left|\frac{\theta_3(\tau_2)}{\eta(\tau_2)}\right|+ \left|\frac{\theta_2(\tau_2)}{\eta(\tau_2)}\right|\bigg)
\\
\rule{0pt}{.5cm}
& & 
\hspace{-2cm}
\propto\,
\chi_0(\tau_2) \, \bar{\chi}_0(\bar{\tau}_2)+
2\,\chi_{\frac{1}{16}}(\tau_2) \,\bar{\chi}_{\frac{1}{16}}(\bar{\tau}_2)
+
\chi_{0}(\tau_2) \,\bar{\chi}_{\frac{1}{2}}(\bar{\tau}_2)+
\chi_{\frac{1}{2}}(\tau_2) \,\bar{\chi}_{0}(\bar{\tau}_2)+ \chi_{\frac{1}{2}}(\tau_2) \,\bar{\chi}_{\frac{1}{2}}(\bar{\tau}_2)
\nonumber
\eea
where $\sigma$  provides the lowest scaling dimension,  
and therefore we have $2(1/16+1/16)=1/4$. 
\\
Instead, in the case of $\mathcal{T}_{\textrm{su}(2)_2}$ and $n=2$, 
one finds 
\bea
\label{part-ill-2}
\parti_{\textrm{su}(2)_2,1}(\tau_2) \big|_{R \leftrightarrow R'}
&=&
\parti_{\textrm{su}(2)_2,1}(-1/{\tau}_2) 
\,=\,  
\frac{\big| \big( \sqrt{\theta_3(\tau_2)}+\sqrt{\theta_2(\tau_2)}\,\big)\big|^2
}{4\, |{\eta(\tau_2)}|}
\\ 
\rule{0pt}{.6cm}
& &
\hspace{-2.3cm}
\propto\,
\chi_0(\tau_2) \, \bar{\chi}_0(\bar{\tau}_2)
+ \chi_{0}(\tau_2) \,\bar{\chi}_{\frac{1}{2}}(\bar{\tau}_2)+ \chi_{\frac{1}{2}}(\tau_2) \,\bar{\chi}_{0}(\bar{\tau}_2)\, + \chi_{\frac{1}{2}}(\tau_2) \,\bar{\chi}_{\frac{1}{2}}(\bar{\tau}_2) + 2\,\chi_{\frac{1}{16}}(\tau_2) \bar{\chi}_{\frac{1}{16}}(\bar{\tau}_2)
\nonumber
\\
\rule{0pt}{.5cm}
& &
\hspace{-1.8cm}
+\,\sqrt{2}\Big(\chi_{0}(\tau_2) \,\bar{\chi}_{\frac{1}{16}}(\bar{\tau}_2)+\,
\chi_{\frac{1}{16}}(\tau_2) \,\bar{\chi}_{0}(\bar{\tau}_2)+ \chi_{\frac{1}{2}}(\tau_2) \,\bar{\chi}_{\frac{1}{16}}(\bar{\tau}_2)+\chi_{\frac{1}{16}}(\tau_2) \,\bar{\chi}_{\frac{1}{2}}(\bar{\tau}_2) \Big)
\nonumber
\eea
where the subleading term in (\ref{Asym-xi-0-expansion-su2}) 
is determined by 
$(h, \bar{h}) = (1/16,0)$ and $(h, \bar{h}) =(0,1/16)$, 
that correspond to fields in the twisted Hilbert space of the $\mathcal{L}_\sigma$ line \cite{Petkova:2000ip,Chang:2018iay},
and therefore yield $2(1/16+0)=1/8$. 
In order to further check this non locality, 
we observe that, by applying (\ref{global2}) 
to (\ref{part-ill-1}) and (\ref{part-ill-2}),
for the values of the Jones index one finds $1/4$ and $1/16$ 
respectively, 
which are smaller than one, and therefore they are not allowed.
It would be interesting to apply these considerations also to other non complete CFT$_2$ models.

In the replica limit $n \to 1$, 
by using that $\mathcal{S}_\alpha(1)=0$ 
and that \cite{Calabrese:2010he}
\be
\mathcal{S}'_\alpha(1)
=
\frac{\sqrt{\pi}\; 
\Gamma\big( \alpha +1\big)}{
2\,\Gamma\big( \alpha +\tfrac{3}{2}\big)}
\ee
it is straightforward to find that 
(\ref{Asym-xi-0-expansion-z2}) and 
(\ref{Asym-xi-0-expansion-su2}) 
become respectively
\be
\label{Asym-xi-0-expansion-n=1-z2}
A_{\mathbb{Z}_2,1}(\x)
=
\frac{1}{2} \log(4) 
-
\frac{\sqrt{\pi}\; 
\Gamma\big( \tfrac{5}{4}\big)}{
2\sqrt{2}\;\Gamma\big ( \tfrac{7}{4}\big)}
\; \x^{1/4}
+\dots
\ee
and 
\be
\label{Asym-xi-0-expansion-n=1-su2}
A_{\textrm{su}(2)_2,1}(\x)
=
\frac{1}{2} \log(16) 
-
\frac{\sqrt{\pi}\; \sqrt[4]{2}\;
\Gamma\big( \tfrac{9}{8}\big)}{
2\,\Gamma\big( \tfrac{13}{8}\big)}
\; \x^{1/8}
+\dots \; .
\ee

Beside the limiting regime (\ref{jones-index-limit-asymmetry-intro}), 
that provides the global Jones index, 
it is interesting to establish whether the asymmetry 
is a monotonic function of the cross ratio. 
From the definition (\ref{asymetrydefintro}), 
we have that ${A}_{\mathcal{T}\!,\,n} (\x=1/2)=0$.
However, 
since the Haag duality violations only depend on the topology of the region and not on its specific geometry \cite{Casini:2020rgj}
(as one can formalize through the transportability of DHR sectors \cite{Casini:2025lfn}),
we do not expect that ${A}_{\mathcal{T}\!,\,n}(\x)$ 
vanishes at any other value of $\x$. 
In the limiting case given by $n \to 1$, 
the monotonicity of the crossing asymmetry has been proven in 
\cite{Benedetti:2024dku}, by relating the difference of mutual informations in (\ref{asymetrydef-mutual}) to a relative entropy and exploiting its 
monotonicity property \cite{Lindblad:1975kmh,Hollands:2022dem}.
However, proving the monotonicity condition 
${A}_{\mathcal{T}\!,\,n}(\x)' \leqslant 0$ 
for a generic $n \geqslant 2$ is still an open problem.
This is because
the monotonicity of Rényi relative entropies with $n \geqslant 2$
extends to the property defined in 
\cite{Muller-Lennert:2013liu,Frank:2013rov,Casini:2018cxg}, 
which cannot be related to (\ref{asymetrydef-mutual}).
In the specific models given by 
$\mathcal{T}_{\mathbb{Z}_2}$ and $\mathcal{T}_{\textrm{su}(2)_2}$,
the curves in Fig.\,\ref{fig:Asymmetry} show that
${A'}_{\mathcal{T}\!,\,n}(\x)\leqslant 0$ for these models
when $2 \leqslant n \leqslant 5$.

\begin{figure}[t!]
\centering
 \includegraphics[width=\linewidth]{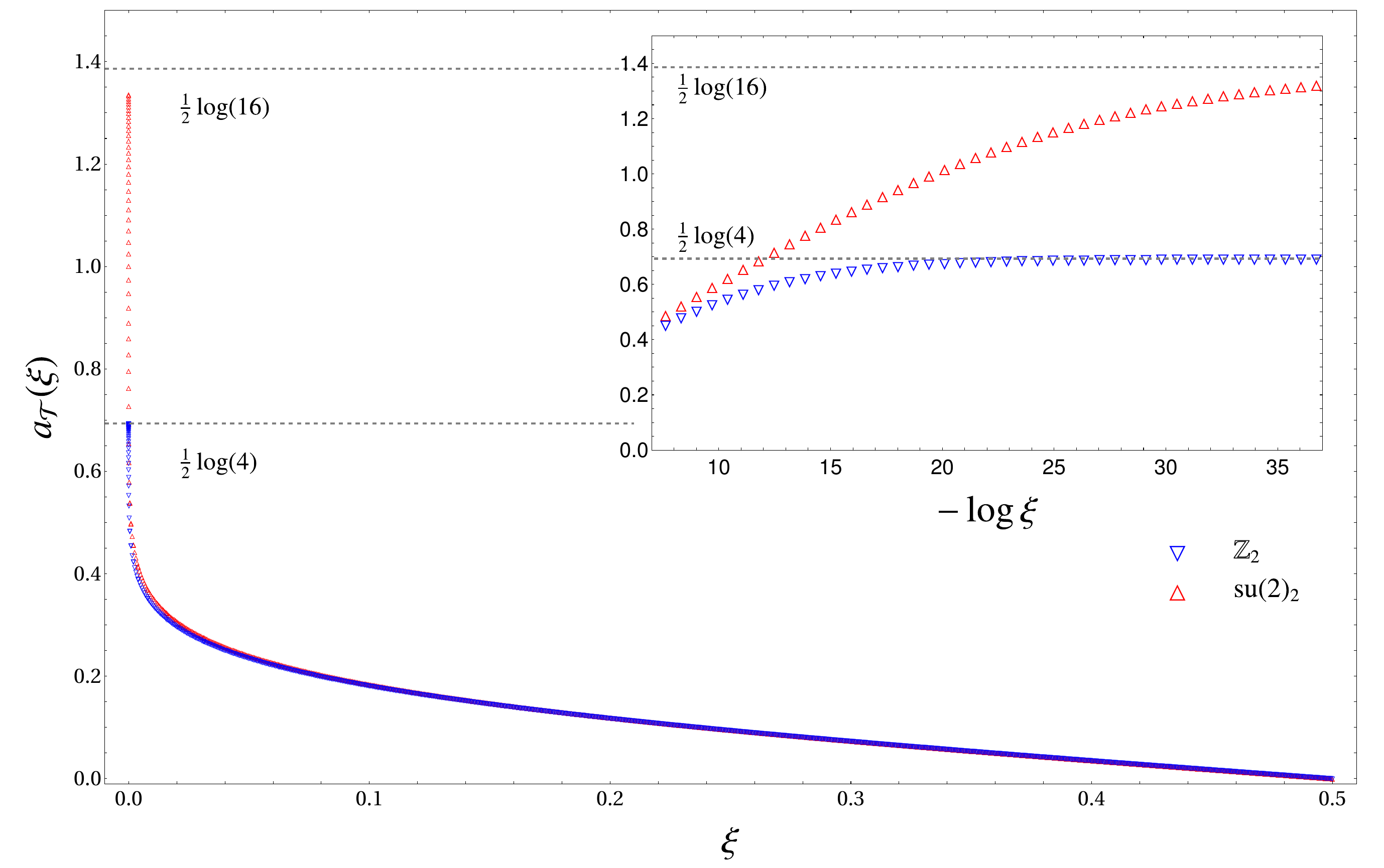}
  \caption{
  The term $a_{\mathcal{T}}(\x)$ in (\ref{nfit}) for 
  $\mathcal{T}_{\mathbb{Z}_2}$ (blue markers)
  and $\mathcal{T}_{\textrm{su}(2)_2}$ (red markers).
} 
\label{fig:ninfinity}
\end{figure}

It is also worth exploring the limit 
of the crossing asymmetries $n \to +\infty$
(\ref{asygz2}) and (\ref{asygsu22}).
Taking this limit for the R\'enyi entropies gives the largest eigenvalue of the entanglement spectrum, and it is often called single-copy entanglement \cite{Peschel:2005zml}.
We perform a numerical analysis 
by first evaluating 
(\ref{asygz2}) and (\ref{asygsu22}) 
at a given value of $\x$,
for $2\leqslant n\leqslant 8$,
and then fitting these seven numerical values labeled by $n$ through the linear combination of a constant parameter and $1/n$ multiplied by a multiplicative parameter. 
This leads to introduce the following approximation 
\be 
\tilde{A}_{\mathcal{T}\!,\,n}(\x)
\equiv
a_{\mathcal{T}}(\x)
+
\frac{b_{\mathcal{T}}(\x)}{n} 
\label{nfit}
\ee
where the functions $a_{\mathcal{T}}(\x)$ and $b_{\mathcal{T}}(\x)$ 
are obtained through this fitting procedure. 
At a given value of $\x$, we observe that 
the approximation (\ref{nfit}) nicely captures the few values corresponding to $2\leqslant n\leqslant 8$, 
although we know that the exact dependence on $n$ of 
$A_{\mathcal{T}\!,\,n}(\x)$ is more complicated. 
This approximation has also been employed in
\cite{Agon:2013iva, DeNobili:2015dla} to study the replica limit. 
By using (\ref{nfit}), we have that $a_{\mathcal{T}}(\x)$ provides an
approximation of $\lim_{n \to \infty} A_{\mathcal{T}\!,\,n}(\x)$.

In Fig.\,\ref{fig:ninfinity} we report the numerical results of $a_{\mathbb{Z}_2}(\x)$ and $a_{\textrm{su}(2)_2}(\x)$ for  $\x\in \mathsf{D}_1\cup \mathsf{D}_2$,
where
$\mathsf{D}_1 \equiv 
\big\{\,\x=0.0005\, n_1\,;n_1\in \mathbb{Z}\; ;
1\leqslant n_1\leqslant 1000\, \big\}$
and 
$\mathsf{D}_2 \equiv \big\{\,\x=2^{-n_2}\,; n_2\in \mathbb{Z}\; ;
2\leqslant n_2\leqslant 52\, \big\}$.
 In the inset, 
 we show the set of points in $\mathsf{D}_2 $ 
as a function of $-\log(\x)$, in order to highlight that,  
 when $\x\to 0^+$, 
 the function $a_{\mathbb{Z}_2}(\x)$ converges to the expected value, 
 while for $a_{\textrm{su}(2)_2}(\x)$ the expected asymptotic value has not been reached yet.

It is worth considering the asymptotic behaviors of the coefficients 
in (\ref{Asym-xi-0-expansion-z2}) and (\ref{Asym-xi-0-expansion-su2}) 
as $n \to +\infty$. 
This is easily obtained from the fact that $\mathcal{S}_\alpha(n) \sim n^2$ in this limit. Indeed, taking $n \to +\infty$ in the definition 
(\ref{S-alha-def}), we find
\be
\mathcal{S}_\alpha(n) \,\sim\,
\frac{n^2}{\pi} \int_0^{\pi} \big[\sin(x)\big]^{-2\alpha} \,\textrm{d}x
\,=\,
\frac{\Gamma(1/2-\alpha)}{\sqrt{\pi}\;\Gamma(1-\alpha)}\; n^2
\;\;\;\qquad\;\;\;
\alpha < \frac{1}{2} \;.
\ee
Specialising this result to the cases 
given by $\alpha =1/4$ and $\alpha =1/8$, 
we obtain that the coefficient of the subleading term 
in (\ref{Asym-xi-0-expansion-z2}) and (\ref{Asym-xi-0-expansion-su2})
diverges like $n^{1/2}$ and $n^{3/4}$ respectively.
This is in contrast with our numerical results and the expectation that 
the crossing asymmetry provides a smooth function vanishing only at $\x =1/2$.
Hence, we conclude that the two limits $\x \to 0$ and $n \to +\infty$ 
do not commute.

\section{Free  fermion: 
Jones index from the $n\geqslant 2$ R\'enyi entropies}
\label{Fermions}

In this section, we explore the subalgebras of 
the free Majorana fermion whose entanglement entropies 
have been studied in \cite{Casini:2005rm,Casini:2009vk,Headrick:2012fk}.
This is a complete model that is not invariant 
under all the modular transformations.
In particular, its partition function on the torus is invariant 
under $\tau \mapsto -1/\tau$ but not invariant under 
$\tau \mapsto \tau+1$.

The action of the CFT$_2$ model with $c=1/2$
that we are going to consider reads
\be
S 
\propto
\int
\Big(
\maj_L (\partial_t-\partial_x) \maj_L + \maj_R (\partial_t+\partial_x)  \maj_R 
\Big)\,
\textrm{d} t \, \textrm{d} x
\label{majoact}
\ee
where $\maj_L $ and $\maj_R$ 
are the real Majorana spinors 
providing the left and right moving components of the field.
Let us consider the transformations given by 
\be
\label{tm-full}
(-1)^F :
\left\{\begin{array}{l}
\maj_L (t,x) \to -\maj_L (t,x) \\
\rule{0pt}{.5cm}
\maj_R (t,x) \to -\maj_R (t,x) 
\end{array}\right.
\ee
and
\be
\label{tm-LR}
(-1)^{F_L} :
\left\{\begin{array}{l}
\maj_L (t,x) \to -\maj_L (t,x) \\
\rule{0pt}{.5cm}
\maj_R (t,x) \to \maj_R (t,x) 
\end{array}\right.
\;\;\;\qquad\;\;\;
(-1)^{F_R} :
\left\{\begin{array}{l}
\maj_L (t,x) \to \maj_L (t,x) \\
\rule{0pt}{.5cm}
\maj_R (t,x) \to -\maj_R (t,x) 
\end{array}\right.
\ee
which leave the action (\ref{majoact}) invariant. 
These transformations are not independent; indeed
\bea
&(-1)^{F}\times(-1)^{F_L}=(-1)^{F_L}\times(-1)^{F}=(-1)^{F_R} \label{fusionFF1}
\\
&(-1)^{F}\times(-1)^{F_R}=(-1)^{F_R}\times(-1)^{F}=(-1)^{F_L} \label{fusionFF2}
\\
&(-1)^{F_L}\times(-1)^{F_R}=(-1)^{F_R}\times(-1)^{F_L}=(-1)^{F} \label{fusionFF}
\eea
where $\times$ denotes  the composition of different transformations. 
Hence, $(-1)^F\,,\, (-1)^{F_L}$ and $(-1)^{F_R}$ form a $\mathbb{Z}_2\times \mathbb{Z}_2$ symmetry (see \cite{Seiberg:2023cdc,Shao:2023gho} for a complete characterization of symmetries for different boundary conditions). 
Let us consider the following projectors
\be 
\mathcal{P}_{\mathbb{Z}_2}^F=
\frac{1}{2}
\big(1+(-1)^{F}\big)
\;\;\qquad\;\;
\mathcal{P}_{\mathbb{Z}_2^L}^{F}=
\frac{1}{2}\big(1+(-1)^{F_L}\big) 
\;\;\qquad\;\;
\mathcal{P}_{\mathbb{Z}_2^R}^{F}=
\frac{1}{2}\big(1+(-1)^{F_R}\big) \,.
\label{projectors-fermion}
\ee
By using (\ref{fusionFF1})-(\ref{fusionFF}), 
we find that any product of two projectors in (\ref{projectors-fermion}) becomes
\be 
\mathcal{P}_{\mathbb{Z}_2\times \mathbb{Z}_2}^F=\mathcal{P}_{\mathbb{Z}_2}^{F}\mathcal{P}_{\mathbb{Z}_2^L}^{F}=\mathcal{P}_{\mathbb{Z}_2}^{F}\mathcal{P}_{\mathbb{Z}_2^R}^{F}=\mathcal{P}_{\mathbb{Z}_2^L}^{F}\mathcal{P}_{\mathbb{Z}_2^R}^{F}=\frac{1}{4}\big(1+(-1)^F+(-1)^{F_L}+(-1)^{F_R}\big) \,.
\label{projectorproduct}
\ee

The field content of the complete model 
that we are going to explore in this section 
is made by the primary fields given by 
$1$, $\varepsilon$ and the two spin $1/2$ primaries 
constructed from the chiral representations as follows \cite{Ginsparg:1988ui}
\be
\psi\equiv 
\bigg(\frac{1}{2}\, ,0\bigg)
\;\;\;\qquad\;\;\;
\bar{\psi}\equiv 
\bigg(0\, ,\frac{1}{2}\bigg)\label{Majoranaprimaries}
\ee
that are represented in (\ref{majoact}) by $\maj_L $ and  $\maj_R$ respectively, while $\varepsilon$ is given by their product. 
The fusion rules (\ref{fusionrules}) 
and the conservation of fermion numbers in (\ref{tm-full}) and (\ref{tm-LR})
lead to \cite{Ginsparg:1988ui}
\be
\label{fusionmajorana}
\psi \times \psi = \bar{\psi} \times \bar{\psi}= 1\, 
\;\qquad\;\;
{\psi} \times \bar{\psi}= \bar{\psi}  \times {\psi} = \varepsilon 
\;\qquad\;
\psi \times \varepsilon=  \varepsilon \times {\psi} = \bar{\psi}
\;\qquad\;
\bar{\psi}\times \varepsilon=  \varepsilon \times\bar{\psi}= {\psi}  \,.
\hspace{.5cm}
\ee
By adding to these fusion rules also
$\varepsilon \times \varepsilon =1$,
that occurs also in (\ref{fusionrules}), one finds that  
$\big\{ 1,\psi,\bar{\psi} , \varepsilon\big\}$ 
forms a closed model under fusion.

All the possible subalgebras of this complete theory 
that are closed under fusion are 
\be 
\label{algebras-list}
\mathcal{T}_{\mathbb{Z}_2}
\equiv
\big\{1,\varepsilon \big\}
\;\;\qquad \;\;
\mathcal{T}_{\mathbb{Z}^L_2}
\equiv
\big\{1,\psi \big\}
\;\;\qquad \;\;
\mathcal{T}_{\mathbb{Z}^R_2}\equiv
\big\{1,\bar{\psi} \,\big\}
\;\;\qquad \;\;
\mathcal{T}_{\textrm{su}(2)_2}\equiv
\big\{1 \big\} \,.
\ee
The models $\mathcal{T}_{\mathbb{Z}_2}$ 
and $\mathcal{T}_{\textrm{su}(2)_2}$,
whose global Jones index are given by 
(\ref{mu4}) and (\ref{mu16}) respectively, 
have been widely discussed (see (\ref{theories})) 
in Sec.\,\ref{Ising}, Sec.\,\ref{JonesIsing2}
and Sec.\,\ref{JonesIsingn}.
As for the submodels $\mathcal{T}_{\mathbb{Z}^L_2}$ and $\mathcal{T}_{\mathbb{Z}^R_2}$, that have a fermionic nature, 
their global Jones index reads 
\be 
\mu^{\phantom{x}}_{\mathbb{Z}^L_2}
=
\mu^{\phantom{x}}_{\mathbb{Z}^R_2}
= 
\big(1^2 + 1^2\big)^2 
= \left(\frac{1^2 + 1^2+ 1^2 +1^2}{1^2 + 1^2} \right)^2=4 \,.
\label{mufermion}
\ee

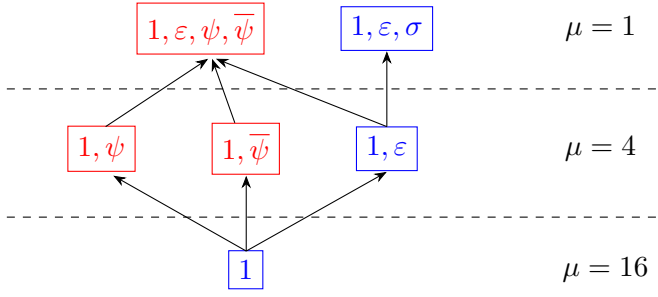
\begin{figure}[t!]
\centering
\begin{tikzpicture}[>=Stealth, node distance=1.0cm, every node/.style={font=\normalsize}]
\node[draw, rectangle,blue] (top) {$1,\varepsilon,\sigma$};
\node[draw, rectangle, below=of top,blue] (mid) {$1,\varepsilon$};
\node[draw, rectangle, left=of mid,red] (midl) {$1,\overline{\psi}$};
\node[draw, rectangle, left=of midl,red] (midr) {$1,\psi$};
\node[draw, rectangle, left=of top,red] (mayo) { $1,\varepsilon,\psi,\overline{\psi}$};
\node[draw, rectangle, below=of midl,blue] (bot) {$1$};
\draw[->] (0,-1.3)-- (0,-0.3);
\draw[->] (0,-1.3)-- (-2.25,-0.4);
\draw[->] (-2,-1.25)-- (-2.3,-0.4);
\draw[->] (-3.7,-1.3)-- (-2.35,-0.4);
\draw[->] (-1.85,-2.95)-- (0,-1.9);
\draw[->] (-1.85,-2.95)-- (-1.85,-1.95);
\draw[->] (-1.85,-2.95)-- (-3.6,-1.95);
\draw[dashed] (-5,-0.8)-- (5,-0.8);
\draw[dashed] (-5,-2.5)-- (5,-2.5);
\node[right=1.6cm of top] {$\mu = 1$};
\node[right=1.8cm of mid] {$\mu = 4$};
\node[right=3.8cm of bot] {$\mu = 16$};
\end{tikzpicture}
\caption{
Subalgebras of the Ising CFT$_2$ and of the free Majorana fermion  
explored in this manuscript, with the corresponding values of the 
global Jones index. 
The models highlighted in blue include only spin zero primaries, 
while the ones highlighted in red include primaries with 
conformal spin $\pm 1/2$.}
\label{sub2}
\end{figure}

In Fig.\,\ref{sub2}, we summarize the models 
considered in this manuscript, 
highlighting in red the ones that include also primary fields with 
conformal spin $\pm 1/2$, discussed in this section.

Consider the model (\ref{majoact}) on the torus where 
NS boundary conditions are imposed on both $\Psi_L$ and $\Psi_R$
along both the cycles of a canonical homology basis.
The partition function reads  
\cite{Alvarez-Gaume:1986rcs,Verlinde:1986kw,Ginsparg:1988ui,DiFrancesco:1997nk,MooreSeiberg}
\be 
\parti_{\text{\tiny Majorana},\,1}
\,=\,
\mathrm{Tr}_{\mathcal{H}_{\text{NS}}}
\!\left(q^{L_0-1/48}\,\bar{q}^{\bar{L}_0-1/48}\right)= \left|\frac{\theta_3(\tau_2)}{\eta(\tau_2)}\right| 
\;\;\;\qquad\;\;\; 
q=\textrm{e}^{2\pi \textrm{i} \tau_2}
\label{majornafreepart}
\ee
where $\mathrm{Tr}_{\mathcal{H}_{\text{NS}}}$ 
denotes the trace over the Hilbert space associated to 
NS boundary conditions. 
From (\ref{transf-jacobi-T})-(\ref{transf-eta-dedekind}), 
we remark that,
while (\ref{majornafreepart}) is invariant under $\tau \mapsto -1/\tau$, 
it is not invariant under $\tau \mapsto 1+\tau$.
However,
the invariance under $\tau \mapsto -1/\tau$ is enough to guarantee the completeness of the model (see (\ref{global2}))
\cite{LongoKawaMuger,Benedetti:2024dku},
as discussed in Sec.\,\ref{completenessqft}.
The partition function (\ref{majornafreepart})
can be written equivalently in terms of the Virasoro characters (\ref{idchar}) and (\ref{epchar}), 
by taking into account that
$h-\bar{h} \in \mathbb{Z}$ does not hold in this case
because the weaker condition 
$2(h-\bar{h}) \in \mathbb{Z}$ is obeyed. 
Indeed, the primary fields (\ref{Majoranaprimaries})
with $h-\bar{h}=\pm1/2$ occur.
This form reads 
\be 
\parti_{\text{\tiny Majorana},1}=\chi_0(\tau_2)\,\bar{\chi}_0(\bar{\tau_2})+\chi_{\frac{1}{2}}(\tau_2)\,\bar{\chi}_{\frac{1}{2}}(\bar{\tau_2})+\chi_{0}(\tau_2)\,\bar{\chi}_{\frac{1}{2}}(\bar{\tau_2})+\chi_{\frac{1}{2}}(\tau_2)\,\bar{\chi}_{0}(\bar{\tau_2})=  \left|\frac{\theta_3(\tau_2)}{\eta(\tau_2)}\right| \,.
\phantom{xxx}
\label{partmajorana}
\ee
By imposing the proper normalization,
the partition function reported 
either in (\ref{majornafreepart}) or in (\ref{partmajorana})
provides the $n=2$ R\'enyi entropy  
of the union of two disjoint intervals on the line 
when the whole system is in the ground state, 
which is given by (\ref{entropydisjoint}) with $c=1/2$ and 
$\mathcal{F}_{\text{\tiny Majorana},2}(\x)=1$ identically. 
This result has been first obtained in \cite{Casini:2005rm} 
and differs in a substantial way from the one obtained in
\cite{Calabrese:2010he}, as discussed in 
\cite{Headrick:2012fk,Lokhande:2015zma, Coser:2015dvp}.

It is worth observing that adding the field $\sigma$, 
i.e. considering the model whose field content is
$\big\{ 1,\sigma, \psi,\bar{\psi} , \varepsilon\big\}$ 
provides a model that is closed under the fusion rules 
(\ref{fusionrules}) and (\ref{fusionmajorana}).
However, this is not a well defined CFT$_2$ because 
$\sigma$  and $\psi$  are not mutually local fields. 
Indeed, for this model the global index is 
\be
\mu^{\phantom{x}}_{\{ 1,\sigma, \psi,\bar{\psi} , \varepsilon\}}
=
\left(\frac{ 1^2 + \sqrt{2}^2 +  1^2}{ 1^2  + \sqrt{2}^2+  1^2+ 1^2+ 1^2} \right)^2
\!=\,
\frac{4}{9}
\ee
which is smaller than one and therefore it is not allowed \cite{Jones83}. 

From (\ref{partmajorana}) and (\ref{algebras-list}),
one realises that the partition functions for the submodels $\mathcal{T}_{\mathbb{Z}^L_2}$ and $\mathcal{T}_{\mathbb{Z}^R_2}$ 
on the torus 
can be written in terms of the characters 
(\ref{idchar}) and (\ref{epchar})
respectively as 
\be
\parti_{\mathbb{Z}^L_2,1}
=
\chi_0(\tau_2)\,\bar{\chi}_0(\bar{\tau_2})+\chi_{\frac{1}{2}}(\tau_2)\,\bar{\chi}_{0}(\bar{\tau}_2)
=
\frac{1}{2}\Bigg(
\left|\frac{\theta_3(\tau_2)}{\eta(\tau_2)}\right| 
+  
\frac{\sqrt{\theta_3(\tau_2)\,\theta_4(\bar{\tau}_2)}}
{\left|\eta(\tau_2)\right|} \,\Bigg)
\label{partmaj1}
\ee
and
\be
\parti_{\mathbb{Z}^R_2,1}
=
\chi_0(\tau_2)\,\bar{\chi}_0(\bar{\tau}_2)+\chi_{0}(\tau_2)\,\bar{\chi}_{\frac{1}{2}}(\bar{\tau}_2)=\frac{1}{2}\Bigg(\left|\frac{\theta_3(\tau_2)}{\eta(\tau_2)}\right| 
+  
\frac{\sqrt{\theta_4(\tau_2) \, \theta_3(\bar{\tau}_2)}}
{\left|\eta(\tau_2)\right|} \, \Bigg) 
\label{partmaj2}
\ee
which coincide because $\tau_2$ in (\ref{periodm2})
is purely imaginary. 
Combining (\ref{partmaj1}) and (\ref{partmaj2}) 
with the proper normalization and the fact that 
$\tau_2$ in (\ref{periodm2}) is purely imaginary, 
the corresponding functions of $\xi$ in the R\'enyi entropies
(\ref{entropydisjoint}) read 
\be 
\mathcal{F}_{\mathbb{Z}^L_2,2}(\x)
=
\mathcal{F}_{\mathbb{Z}^R_2,2}(\x)=\frac{1}{2}\Bigg(1 +  \frac{\sqrt{\theta_3(\tau_2) \,\theta_4({\tau}_2)}}
{\left|\theta_3 (\tau_2)\right|} \, \Bigg) \,.
\ee
Hence, the crossing asymmetry (\ref{asymmetrypart}) for these models is 
\be
\label{asymmetry-z2-LR}
A_{\mathbb{Z}^L_2,2}(\x)
= 
A_{\mathbb{Z}^R_2,2}(\x)
= \log\!
\left(\frac{\sqrt{\theta_3(\tau_2)}+\sqrt{\theta_4(\tau_2)}}{\sqrt{\theta_3(\tau_2)}+\sqrt{\theta_2(\tau_2)}} \,\right)
\ee
which is equal to $A_{\textrm{su}(2)_2,2}(\x)/2$, from (\ref{asyz22}).
We remark that, 
by specialising (\ref{jones-index-limit-asymmetry-intro}) to (\ref{asymmetry-z2-LR}),
the global Jones index (\ref{mufermion}) is recovered.

It is instructive to recover the partition functions 
(\ref{z21}), (\ref{z22}), (\ref{partmaj1}) and (\ref{partmaj2}) 
through arguments analogue to the ones discussed in Sec.\,\ref{JonesIsing2}, that are based on the topological defects.
By introducing topological defects along the $a$-type cycle of the torus,
which can be understood also as topological operators implementing the symmetries of the action (\ref{majoact}), 
from (\ref{tm-full}) and (\ref{tm-LR}) one finds that 
the insertion of these operators produces a change in the boundary conditions along the $b$-type cycle 
and therefore induces a change of the corresponding spin structure.
In particular, we have that 
\bea
\label{proj-F-z2-full}
 \Tr_{\mathcal{H}_{\text{NS}}}
 \Big((-1)^F q^{L_0-\frac{1}{48}}\,\bar{q}^{\bar{L}_0-\frac{1}{48}} \Big)
 \,&=&\, 
\begin{tikzpicture}[baseline={(current bounding box.center)}]
  \draw[thick] (0,0) rectangle (1.5,1.5);
  \draw[red, thick] (0,0.75) -- (1.5,0.75);
  \node at (0.75,1.05) {\textcolor{red}{\small{$(-1)^{F}$}}};
\end{tikzpicture} 
\;=\,
\left|\frac{\theta_4(\tau_2)}{\eta(\tau_2)}\right|
\\
\rule{0pt}{1cm}
\Tr_{\mathcal{H}_{\text{NS}}}\Big((-1)^{F_L} q^{L_0-\frac{1}{48}}\,\bar{q}^{\bar{L}_0-\frac{1}{48}} \Big) &=& \,
\begin{tikzpicture}[baseline={(current bounding box.center)}]
  \draw[thick] (0,0) rectangle (1.5,1.5);
  \draw[blue, thick] (0,0.75) -- (1.5,0.75);
  \node at (0.75,1.05) {\textcolor{blue}{\small{$(-1)^{F_L}$}}};
\end{tikzpicture} 
\;=\,
\displaystyle{\frac{\sqrt{\theta_4(\tau_2)\,\theta_3(\bar{\tau}_2)}}{\left|\eta(\tau_2)\right|}}
\\
\label{proj-F-z2-R}
\rule{0pt}{1cm}
\Tr_{\mathcal{H}_{\text{NS}}}\Big((-1)^{F_R} q^{L_0-\frac{1}{48}}\,\bar{q}^{\bar{L}_0-\frac{1}{48}} \Big) &=& \,
\begin{tikzpicture}[baseline={(current bounding box.center)}]
  \draw[thick] (0,0) rectangle (1.5,1.5);
  \draw[blue, thick] (0,0.75) -- (1.5,0.75);
  \node at (0.75,1.05) {\textcolor{blue}{$(-1)^{F_R}$}};
\end{tikzpicture} 
\;=\,
\displaystyle{\frac{\sqrt{\theta_3(\tau_2)\,\theta_4(\bar{\tau}_2)}}
{\left|\eta(\tau_2)\right|}} \;.
\eea
By inserting the projectors (\ref{projectors-fermion})  
into the partition function (\ref{majornafreepart})
(see also (\ref{torusprojector}))
and using (\ref{proj-F-z2-full})-(\ref{proj-F-z2-R}),
one recovers 
(\ref{z21}), (\ref{partmaj1}) and (\ref{partmaj2}) respectively;
indeed
\bea
\parti_{\mathbb{Z}_2,1} 
&=& \frac{1}{2}\,
\begin{tikzpicture}[baseline={(current bounding box.center)}]
  \draw[thick] (0,0) rectangle (1.5,1.5);
\end{tikzpicture}
\;+\; \frac{1}{2}\,
\begin{tikzpicture}[baseline={(current bounding box.center)}]
  \draw[thick] (0,0) rectangle (1.5,1.5);
  \draw[red, thick] (0,0.75) -- (1.5,0.75);
  \node at (0.75,1.05) {\textcolor{red}{\small{$(-1)^{F}$}}};
\end{tikzpicture} 
\;=\,
\frac{1}{2}\Bigg(\left|\frac{\theta_3(\tau_2)}{\eta(\tau_2)}\right| +  \left|\frac{\theta_4(\tau_2)}{\eta(\tau_2)}\right|\Bigg)
\\
\rule{0pt}{1cm}
\parti_{\mathbb{Z}^R_2,1} 
&=& \frac{1}{2}\,
\begin{tikzpicture}[baseline={(current bounding box.center)}]
  \draw[thick] (0,0) rectangle (1.5,1.5);
\end{tikzpicture}
\;+\; \frac{1}{2}\,
\begin{tikzpicture}[baseline={(current bounding box.center)}]
  \draw[thick] (0,0) rectangle (1.5,1.5);
  \draw[blue, thick] (0,0.75) -- (1.5,0.75);
  \node at (0.75,1.05) {\textcolor{blue}{\small{$(-1)^{F_L}$}}};
\end{tikzpicture}  
\;=\,
\frac{1}{2}\Bigg(\left|\frac{\theta_3(\tau_2)}{\eta(\tau_2)}\right| +  \frac{\sqrt{\theta_4(\tau_2)\,\theta_3(\bar{\tau}_2)}}
{\left|\eta(\tau_2)\right|}\,\Bigg)
\\
\rule{0pt}{1cm}
\parti_{\mathbb{Z}^L_2,1} 
& = &
\frac{1}{2}\,
\begin{tikzpicture}[baseline={(current bounding box.center)}]
  \draw[thick] (0,0) rectangle (1.5,1.5);
\end{tikzpicture}
\;+\; \frac{1}{2}\,
\begin{tikzpicture}[baseline={(current bounding box.center)}]
  \draw[thick] (0,0) rectangle (1.5,1.5);
  \draw[blue, thick] (0,0.75) -- (1.5,0.75);
  \node at (0.75,1.05) {\textcolor{blue}{\small{$(-1)^{F_R}$}}};
\end{tikzpicture}
\;=\,
\frac{1}{2}\Bigg( \left|\frac{\theta_3(\tau_2)}{\eta(\tau_2)}\right| +  \frac{\sqrt{\theta_3(\bar{\tau}_2)\,\theta_4(\tau_2)}}
{\left|\eta(\tau_2)\right|}\,\Bigg) \,.
\eea
As for $\mathcal{T}_{\textrm{su}(2)_2}$, we can recover  the partition function (\ref{z22}) by projecting onto the invariant part of the $\mathbb{Z}_2\times \mathbb{Z}_2$ symmetry in (\ref{majornafreepart}) 
through the projector (\ref{projectorproduct}). This gives
\be
\parti_{\textrm{su}(2)_2,1} 
=
\frac{1}{4}\,
\begin{tikzpicture}[baseline={(current bounding box.center)}]
  \draw[thick] (0,0) rectangle (1.5,1.5);
\end{tikzpicture}
+ \frac{1}{4}\,
\begin{tikzpicture}[baseline={(current bounding box.center)}]
  \draw[thick] (0,0) rectangle (1.5,1.5);
  \draw[red, thick] (0,0.75) -- (1.5,0.75);
  \node at (0.75,1.05) {\textcolor{red}{\small{$(-1)^{F}$}}};
\end{tikzpicture} 
+ 
\frac{1}{4}\,
\begin{tikzpicture}[baseline={(current bounding box.center)}]
  \draw[thick] (0,0) rectangle (1.5,1.5);
  \draw[blue, thick] (0,0.75) -- (1.5,0.75);
  \node at (0.75,1.05) {\textcolor{blue}{\small{\small{$(-1)^{F_L}$}}}};
\end{tikzpicture} 
+ 
\frac{1}{4}\,
\begin{tikzpicture}[baseline={(current bounding box.center)}]
  \draw[thick] (0,0) rectangle (1.5,1.5);
  \draw[blue, thick] (0,0.75) -- (1.5,0.75);
  \node at (0.75,1.05) {\textcolor{blue}{\small{$(-1)^{F_R}$}}};
\end{tikzpicture}
\;=
\frac{\big(\sqrt{\theta_4(\tau_2)}+\sqrt{\theta_3(\tau_2)}\, \big)^2}
{4\left|\eta(\tau_2)\right|} \;.
\ee

Considering the partition function of the model  
(\ref{majoact}) on the Riemann surface $\mathscr{R}_n$ 
in (\ref{eq:replicaSurface})
when NS boundary conditions are imposed 
along all the cycles of the canonical homology basis, 
one obtains 
\cite{Alvarez-Gaume:1986rcs,Verlinde:1986kw}
\be
\parti_{\text{\tiny Majorana},\,n-1}\,
= \!\!
\sum_{\pmb{i},\pmb{k},\pmb{j},\pmb{r} \,\in \, \mathsf{S}} 
\!\!
\chi_{\pmb{i}}^{\pmb{k}}({\tau}_n)\,
\chi_{\pmb{j}}^{\pmb{r}}(\bar{\tau}_n)
\,=\,
\frac{1}{\parti_0}\left|\ThetaF{\pmb{0}}{\pmb{0}}(\tau_n) \right|\label{partgmajo}
\ee
where $\mathsf{S}$ has been defined in (\ref{S-set}).
This partition function is not invariant 
under the generic transformation of $\textrm{Sp}(2g,\mathbb{Z})$
but only under the stabilizer of the characteristic defined by 
$\pmb{p}=\pmb{q}=\pmb{0}$, which is known as the theta group 
\cite{Freitag1991}.
From (\ref{p-q-prime-def}), one finds that the properties characterising this subgroup of $\textrm{Sp}(2g, \mathbb{Z})$ are given by 
$\big(CD^{\textrm{t}}\big)_{\textrm{d}}=0 \,\textrm{mod}\, 2$ 
and 
$\big(AB^{\textrm{t}}\big)_{\textrm{d}}=0 \,\textrm{mod}\, 2$.
This subgroup includes the invariance under $\x \mapsto 1 -\x$,
that induces the change (\ref{x1x}) on the period matrix,
which implies (\ref{transform00m})
(see also the appendix\;\ref{modular}).
The Rényi entropies (\ref{entropydisjoint}) associated with (\ref{partgmajo}) have  $\mathcal{F}_{\text{\tiny Majorana},\,n}(\x)=1$ 
identically, 
as computed in \cite{Casini:2005rm,Casini:2009vk}
also by employing the entanglement Hamiltonian
(see \cite{Coser:2015dvp} for the contribution of 
other boundary conditions).
Hence, from (\ref{asymetrydefintro}) 
and (\ref{jones-index-limit-asymmetry-intro}),
one can verify
the completeness of the model whose partition function on $\mathscr{R}_n$  
is (\ref{partgmajo}).

The partition functions of the submodels 
$\mathcal{T}_{\mathbb{Z}^L_2}$ and $\mathcal{T}_{\mathbb{Z}^R_2}$
on $\mathscr{R}_n$ can be found by inserting 
the $\mathbb{Z}_2$ projectors (\ref{projectors-fermion})
built from $(-1)^F$, $(-1)^{F_L}$ and $(-1)^{F_R}$ 
in the proper way. 
The action of the fermion parity operators (\ref{tm-full})-(\ref{tm-LR})
along the cycle $a_i$ 
induces a change in the boundary conditions 
along the corresponding cycle $b_i$ in the canonical homology basis. 
In particular,
for the Riemann theta functions involved in our analysis 
we have that \cite{Alvarez-Gaume:1986rcs,Verlinde:1986kw}
\bea
\label{minus-one-F-theta}
&&
\big[(-1)^F \,\big]\big|_{a_i} 
\;\,:
\ThetaF{\boldsymbol{0}}{\boldsymbol{q}}(\tau_n) 
\,\longrightarrow\;
\ThetaF{\boldsymbol{0}}{f_i(\boldsymbol{q})}(\tau_n) 
\\
&&
\big[(-1)^{F_L}\,\big]\big|_{a_i}
:
\ThetaF{\boldsymbol{0}}{\boldsymbol{q}}(\tau_n) 
\,\longrightarrow
\Big(\ThetaF{\boldsymbol{0}}{f_i(\boldsymbol{q})}(\tau_n)\Big)^{\frac{1}{2}} \Big(\ThetaF{\boldsymbol{0}}{\boldsymbol{q}}(\bar{\tau}_n)\Big)^{\frac{1}{2}} 
\\
&&
\big[(-1)^{F_R}\,\big]\big|_{a_i}
:
\ThetaF{\boldsymbol{0}}{\boldsymbol{q}}(\tau_n) 
\,\longrightarrow
\Big(\ThetaF{\boldsymbol{0}}{\boldsymbol{q}}(\tau_n)\Big)^{\frac{1}{2}} \Big(\ThetaF{\boldsymbol{0}}{f_i(\boldsymbol{q})}(\bar{\tau}_n)\Big)^{\frac{1}{2}}
\eea
where $\big[(\,\dots)^\sharp\big]\big|_{a_i}$ 
indicates the insertion of the 
projector in (\ref{projectors-fermion}) 
corresponding to $(\,\dots)^\sharp$ along the 
$i$-th cycle of $a$-type
and  $f_i(\boldsymbol{q})^{\textrm{t}} \equiv (q_1,q_2,\dots, q_i+1/2,\dots, q_{n-1})$,
that affects only the $i$-th element of $\boldsymbol{q}$.

The modification of (\ref{partgmajo}) provided by the introduction of 
either $\mathcal{P}_{\mathbb{Z}_2}^{F}$ in (\ref{projectors-fermion}) 
or $\mathcal{P}_{\textrm{su}(2)_2}^F$ in (\ref{projectorproduct}) 
along all the $a$-type cycles 
gives either (\ref{partgz2}) or (\ref{partgsu22})
respectively, as discussed in Sec.\,\ref{IsingGenus}.
Instead, 
introducing either $\mathcal{P}_{\mathbb{Z}_2}^{F_L}$ 
or $\mathcal{P}_{\mathbb{Z}_2}^{F_R}$ in (\ref{projectors-fermion})
leads to the partition functions of 
$\mathcal{T}_{\mathbb{Z}^L_2}$ and $\mathcal{T}_{\mathbb{Z}^R_2}$ 
on $\mathscr{R}_n$ respectively, that read 
\be
\label{part-Z2-L}
\parti_{\mathbb{Z}^L_2,n-1}
\,=\!
\sum_{\pmb{i},\pmb{k} \,\in \, \mathsf{S}}
\!
\chi_{\pmb{i}}^{\pmb{k}}({\tau}_n)\,
\chi_{\pmb{0}}^{\pmb{0}}(\bar{\tau}_n)
\,=\,
\frac{1}{2^{n-1}\parti_0}
\Big(
\ThetaF{\pmb{0}}{\pmb{0}}(\tau_n) 
\Big)^\frac{1}{2}
\bigg[
\sum_{\boldsymbol{q}}
\Big(\ThetaF{\pmb{0}}{\boldsymbol{q}}(\bar{\tau}_n)\Big)^\frac{1}{2} 
\bigg]
\ee
and
\be
\label{part-Z2-R}
\parti_{\mathbb{Z}^R_2,n-1}
\,=\!
\sum_{\pmb{i},\pmb{k} \,\in \, \mathsf{S}} 
\! \chi_{\pmb{0}}^{\pmb{0}}({\tau}_n)\,
\chi_{\pmb{i}}^{\pmb{k}}(\bar{\tau}_n)
\,=\,
\frac{1}{2^{n-1}\parti_0}
\bigg[
\sum_{\boldsymbol{q}}
\Big(\ThetaF{\pmb{0}}{\boldsymbol{q}}({\tau_n}) \Big)^\frac{1}{2} 
\bigg] 
\Big(\ThetaF{\pmb{0}}{\pmb{0}}(\bar{\tau}_n) \Big)^\frac{1}{2}
\ee
which coincide because $\tau_n$ in (\ref{periodm}) is purely imaginary. 
As for the R\'enyi entropies of two disjoint intervals 
(\ref{entropydisjoint}) for these two models, 
by considering the proper normalization 
and  taking into account that the period matrix (\ref{periodm}) satisfies $\bar{\tau}_n(\x)=-{\tau}_n(\x)$, we have that
\be
\mathcal{F}_{\mathbb{Z}^L_2,n}(\x)
\,=\,
\mathcal{F}_{\mathbb{Z}^R_2,n}(\x)
\,=\,
\frac{\sum_{\boldsymbol{q}}
\Big(\ThetaF{\pmb{0}}{\boldsymbol{q}}({\tau_n})\Big)^\frac{1}{2}}{2^{n-1}\,
\Big(\ThetaF{\pmb{0}}{\pmb{0}}(\tau_n) \Big)^\frac{1}{2}} \;.
\ee

It is worth discussing the invariance of 
(\ref{part-Z2-L}) and (\ref{part-Z2-R}). 
This symmetry is given by a subgroup obtained by intersecting 
the theta group 
(introduced in the text below (\ref{partgmajo}))
and the subgroup 
$\textrm{GL}(g,\mathbb{Z})\ltimes \textrm{Sym}(g,\mathbb{Z})$ 
defined in (\ref{submodular}).
This intersection is given by the transformations (\ref{submodular})
with $\big(AB^{\textrm{t}}\big)_{\textrm{d}}=0 \,\textrm{mod}\, 2$.
Notice that (\ref{periodma}) is included in this set.
However, we remark that this set does not contain (\ref{x1x})
and this leads to the non completeness of the submodels 
$\mathcal{T}_{\mathbb{Z}^L_2}$ and $\mathcal{T}_{\mathbb{Z}^R_2}$.

Finally, from (\ref{asymmetrypart}), 
the corresponding crossing asymmetry reads
\be
A_{\mathbb{Z}^L_2,n}(\x)
\,=\,
A_{\mathbb{Z}^R_2,n}(\x)
\,=\,
\frac{1}{n-1}\,
\log\!\left[
\left(\sum_{\boldsymbol{p}}
\sqrt{
\ThetaF{\,\boldsymbol{p}}{\pmb{0}}\big({\tau}_n\big)
}\;\right)
\Bigg/
\left(\sum_{\boldsymbol{q}}
\sqrt{
\ThetaF{\pmb{0}}{\boldsymbol{q}}\big({\tau}_n\big) 
}\;\right)\right]
\label{asyl}
\ee
which is equal to (\ref{asygsu22}) multiplied by $1/2$.
The latter observation straightforwardly implies 
that the expansion of the crossing asymmetry (\ref{asyl})
as $\x \to 0$ is the expansion (\ref{Asym-xi-0-expansion-su2})
multiplied by a global factor $1/2$.
Its leading term combined with (\ref{jones-index-limit-asymmetry-intro}) give 
$\mu^{\phantom{x}}_{\mathbb{Z}^L_2}=
\mu^{\phantom{x}}_{\mathbb{Z}^R_2}= \tfrac{1}{4}\mu^{\phantom{x}}_{\textrm{su}(2)_2}$
for any finite value of the R\'enyi index $n$, 
in agreement with (\ref{mufermion}).

\section{Conclusions}
\label{sec-conclusions}

We have explored the submodels 
of the complete CFT$_2$ given by the Ising model 
and the free Majorana fermion discussed in \cite{Casini:2005rm}
that are closed under the fusion rules and non complete. 
They are shown pictorially in Fig.\,\ref{sub1} and Fig.\,\ref{sub2},
where the corresponding values of the global Jones index are also indicated. 
This Jones index can be found through the limit
(\ref{asymetrydefintro}) of the crossing asymmetry 
(\ref{jones-index-limit-asymmetry-intro}),
 constructed through the R\'enyi entropies of two disjoint intervals when the model is on the line and in its ground state \cite{Benedetti:2024dku}.
 
The main results of this paper are the analytic expressions 
of the crossing asymmetry for the non complete submodels 
occurring in Fig.\,\ref{sub2},
for a generic value of the R\'enyi index. 
As for the submodels $\mathcal{T}_{\mathbb{Z}_2}$ 
and $\mathcal{T}_{\textrm{su}(2)_2}$ of the Ising CFT$_2$,
their crossing asymmetries are given by (\ref{Fnz2})-(\ref{asyft-z2})
and (\ref{Fnsu22})-(\ref{asyft-su2}) respectively, 
or, equivalently, by (\ref{asygz2}) and (\ref{asygsu22}) respectively.
These analytic expressions provide the curves displayed in 
Fig.\,\ref{fig:Asymmetry}, that highlight the monotonic behavior of the crossing asymmetry as a function of the cross ratio, for the finite 
values of the R\'enyi index that we have explored. 
The expansions of these crossing asymmetries,
reported in (\ref{Asym-xi-0-expansion-z2}) and (\ref{Asym-xi-0-expansion-su2}), 
show that the leading term provides the global Jones index 
for any value of the R\'enyi index,
as also stated in (\ref{asymetrydefintro}) and argued in \cite{Benedetti:2024dku}.
As for the submodels of the free Majorana fermion 
in (\ref{algebras-list}),
which include $\mathcal{T}_{\mathbb{Z}^L_2}$ and $\mathcal{T}_{\mathbb{Z}^R_2}$
beside the previous ones, the crossing asymmetry (\ref{asyl})
are simply related to (\ref{asygsu22}); 
hence also the corresponding Jones indexes are obtained accordingly. 
The invariance of the partition functions of all these submodels 
under the proper subgroups of $\textrm{Sp}(2g, \mathbb{Z})$
is also discussed.

Various directions can be explored in future studies 
that involve the crossing asymmetry. 
For instance, the relation (\ref{jones-index-limit-asymmetry-intro}) 
with the global Jones index for a generic value of $n$
could be studied for states different from the ground state
(like e.g. the thermal state or excited states \cite{Berganza:2011mh})
or in other models (like e.g. the compact boson and the minimal models), also in the presence of boundaries \cite{Estienne:2021xzo,Estienne:2023ekf}.
It would also be very insightful  to find some connection 
between the crossing asymmetry and the renormalization group flow
\cite{BenedettiMinimal,Benedetti:2026drn}.

\vskip 20pt 
\centerline{\bf Acknowledgments } 
\vskip 5pt

We are grateful to Horacio Casini, 
Atish Dabholkar, Marina Huerta, Roberto Longo and
Javier Magan
for insightful discussions.
We thank Andrea Coser for the Mathematica code
employed to draw the Riemann surfaces.
V.B. is  supported by 
a RFA Fellowship from the Abdus Salam International Centre for Theoretical Physics (ICTP), Trieste, 
and by INFN Iniziativa Specifica ST$\&$FI.
E.T. thanks the Isaac Newton Institute (Cambridge),
within the program {\it Quantum field theory with boundaries, impurities, and defects}
(supported by EPSRC grant EP/Z000580/1)
for hospitality and financial support.
I.D. and E.T. thank
the Yukawa Institute for Theoretical Physics (Kyoto),
within the workshops {\it Extreme Universe 2025} (YITP-T-25-01)
and  {\it Progress of Theoretical Bootstrap},
for hospitality and financial support 
during the last part of this work.
This work was funded by the European Union -- NextGenerationEU, Mission 4, Component 2, Inv.1.3, in the framework of the PNRR Project National Quantum Science and Technology Institute (NQSTI) PE00023; CUP: G93C22001090006.

\appendix

\section{Details on the higher genus characters for the Ising CFT$_2$}
\label{app-details0higher-genus}

In this appendix, we give some details about 
the genus $g$ characters  (\ref{charactersIsing}). 
We describe the internal loops in Sec.\,\ref{Internal-loops}
and write the explicit expressions for $g=2$ in Sec.\,\ref{g2characters}.

\subsection{Internal loops}
\label{Internal-loops}

In the following, we provide all the internal loops allowed by the fusion rules (\ref{fusionrules}) of the Ising CFT$_2$. 
The characters $\chi_{\pmb{i},\pmb{j}}^{\pmb{k}}$ in (\ref{multi}) are 
labeled by the three vectors $\pmb{i}=(i_1,i_2\dots,i_{g})$, 
$\pmb{j}=(j_1=i_1,j_2\dots,j_{g}=i_{g})$ 
and  $\pmb{k}=(k_1,k_2\dots,k_{g-1})$. 
In the Ising CFT$_2$, only two of them are independent because of the constraints imposed by the fusion rules (\ref{fusionrules}). 
Indeed, considering an internal loop of the form 
\be
\begin{tikzpicture}[baseline={(current bounding box.center)}, scale=1.3]
  \draw[thick] (-0.45,0.5) -- (0.25,0.5);
  \draw[thick] (0.75,0.5) circle (0.5);
  \draw[thick] (1.25,0.5) -- (1.95,0.5); 
  \node at (-0.1,0.7) {$k_{m-1}$};
  \node at (0.75,1.2) {$i_m$};
  \node at (1.55,0.7) {$k_m$};
   \node at (0.75,-0.2) {$j_m$};
\end{tikzpicture}%
\ee
we have that 
e.g. the label $j_{m}$ can be obtained from the fusion rules (\ref{fusionrules}),
from $k_{m-1}$, $k_{m}$ and $i_{m}$. 
Hence, $\pmb{j}$ is determined from $\pmb{i}$ and $\pmb{k}$, 
that are the independent vectors in  (\ref{charactersIsing}).
The allowed internal loops are
\begin{equation*}
\renewcommand{\arraystretch}{2.5}
\setlength{\arraycolsep}{2.5em}
\begin{array}{cccc}
\loopi{$0$}{$0$}{$0$}{$0$} & 
\loopi{$0$}{$\frac{1}{2}$}{$\frac{1}{2}$}{$0$} & 
\loopi{$\frac{1}{2}$}{$\frac{1}{2}$}{$0$}{$\frac{1}{2}$} & 
\loopi{$\frac{1}{2}$}{$0$}{$\frac{1}{2}$}{$\frac{1}{2}$} \\

\loopi{$0$}{$\frac{1}{16}$}{$\frac{1}{16}$}{$0$} & 
\loopi{$\frac{1}{2}$}{$\frac{1}{16}$}{$\frac{1}{16}$}{$\frac{1}{2}$} & 
\loopi{$0$}{$\frac{1}{16}$}{$\frac{1}{16}$}{$\frac{1}{2}$} & 
\loopi{$\frac{1}{2}$}{$\frac{1}{16}$}{$\frac{1}{16}$}{$0$}
\end{array}
\label{loop-diagram-app-A}
\end{equation*}
Notice that in the first and last loop,
at $m=1$ and $m=g$ respectively,
where the diagram becomes a tadpole, 
we that $k_{0}=k_g=\id$ in (\ref{loop-diagram-app-A}).

\subsection{Genus two: Characters and partition functions}
\label{g2characters}

We find it instructive to discuss in detail the genus two case
by considering all the Ising characters and the partition functions
that are constructed through them in the main text. 
In this case, the characters (\ref{charactersIsing}) have only three labels 
and do not contain internal loops of the form  (\ref{multi}) 
in its most generic version. 
Indeed, they read
\be 
\chi_{i_1,i_2}^{k_1}\,=\genusii{$i_1$}{$k_1$}{$i_2$}
\ee
Let us first consider the genus two characters that do not include the chiral representation of conformal weight $1/16$.
They can be computed explicitly from (\ref{charactersIsing2}), finding 
(the dependence on the period matrix is removed hereafter, 
to enlighten the notation)
\bea
\label{ca1} 
& &
\chi_{\id,\id}^{\id}
\,=\,
\genusii{$\id$}{$\id$}{$\id$}
\,=\,
\frac{1}{4\parti_0^{1/2}}
\Bigg( 
\ThetaN{0\,0}{0\,0}^{\frac{1}{2}} \! + 
\ThetaN{0\,0}{\frac{1}{2}\,0}^{\frac{1}{2}} \!+ 
\ThetaN{0\,0}{0\,\frac{1}{2}}^{\frac{1}{2}} \!+
\ThetaN{0\,0}{\frac{1}{2}\,\frac{1}{2}}^{\frac{1}{2}}\, 
\Bigg) 
\\
\label{ca2} 
& &
\chi_{\e,\id}^{\id}
\,=\,
\genusii{$\e$}{$\id$}{$\id$}
\,=\,
\frac{1}{4\parti_0^{1/2}}
\Bigg( 
\ThetaN{0\,0}{0\,0}^{\frac{1}{2}} \!- \,
\ThetaN{0\,0}{\frac{1}{2}\,0}^{\frac{1}{2}} \!+ \,
\ThetaN{0\,0}{0\,\frac{1}{2}}^{\frac{1}{2}} \!-\,
\ThetaN{0\,0}{\frac{1}{2}\,\frac{1}{2}}^{\frac{1}{2}}
\, \Bigg)  
\\
& &
\label{ca3} \chi_{\id,\e}^{\id}
\,=\,
\genusii{$\id$}{$\id$}{$\e$}
\,=\,
\frac{1}{4\parti_0^{1/2}}
\Bigg( 
\ThetaN{0\,0}{0\,0}^{\frac{1}{2}} \!+ \,
\ThetaN{0\,0}{\frac{1}{2}\,0}^{\frac{1}{2}} \!- \,
\ThetaN{0\,0}{0\,\frac{1}{2}}^{\frac{1}{2}} \!- \,
\ThetaN{0\,0}{\frac{1}{2}\,\frac{1}{2}}^{\frac{1}{2}}
\, \Bigg)  
\\
& & \label{ca4} \chi_{\e,\e}^{\id}
\,=\,
\genusii{$\e$}{$\id$}{$\e$}
\,=\,
\frac{1}{4\parti_0^{1/2}}
\Bigg( 
\ThetaN{0\,0}{0\,0}^{\frac{1}{2}} \!- \,
\ThetaN{0\,0}{\frac{1}{2}\,0}^{\frac{1}{2}} \!- \,
\ThetaN{0\,0}{0\,\frac{1}{2}}^{\frac{1}{2}} \!+ \,
\ThetaN{0\,0}{\frac{1}{2}\,\frac{1}{2}}^{\frac{1}{2}}
\, \Bigg) 
\eea
where we remark that the upper components of the characteristics of 
the Riemann theta functions are equal to zero. 
As for the characteristics that include also the $1/16$ representation, 
from (\ref{charactersIsing}) we have
\bea
& & \hspace{-2cm}
\label{ca5} \chi_{\s,\id}^{\id}
\,=\,
\genusii{$\s$}{$\id$}{$\id$}
\,=\,
\frac{\sqrt{2}}{4\parti_0^{1/2}}
\Bigg(
\ThetaN{\frac{1}{2}\,0}{0\,0}^{\frac{1}{2}} \!+\,
\ThetaN{\frac{1}{2}\,0}{0\,\frac{1}{2}}^{\frac{1}{2}} 
\,\Bigg)  
\\
\label{ca6} 
& & \hspace{-2cm}
\chi_{\s,\e}^{\id}
\,=\,
\genusii{$\s$}{$\id$}{$\e$}
\,=\,
\frac{\sqrt{2}}{4\parti_0^{1/2}}
\Bigg(
\ThetaN{\frac{1}{2}\,0}{0\,0}^{\frac{1}{2}} \!-\,
\ThetaN{\frac{1}{2}\,0}{0\,\frac{1}{2}}^{\frac{1}{2}} \,\Bigg) 
\\
\label{ca7} 
& & \hspace{-2cm}
\chi_{\id,\s}^{\id}
\,=\,
\genusii{$\id$}{$\id$}{$\s$}
\,=\,
\frac{\sqrt{2}}{4\parti_0^{1/2}} 
\Bigg(
\ThetaN{0\,\frac{1}{2}}{0\,0}^{\frac{1}{2}} \!+\,
\ThetaN{0\,\frac{1}{2}}{\frac{1}{2}\,0}^{\frac{1}{2}} \,\Bigg)   
\eea
\bea
\label{ca8} 
& & \hspace{-2cm}
\chi_{\e,\s}^{\id}
\,=\,
\genusii{$\e$}{$\id$}{$\s$}
\,=\,
\frac{\sqrt{2}}{4\parti_0^{1/2}} 
\Bigg(\ThetaN{0\,\frac{1}{2}}{0\,0}^{\frac{1}{2}} \!-\,
\ThetaN{0\,\frac{1}{2}}{\frac{1}{2}\,0}^{\frac{1}{2}} 
\,\Bigg)   
\\
\label{ca9} 
& & \hspace{-2cm}
\chi_{\s,\s}^{\id}\,
\,=\,
\genusii{$\s$}{$\id$}{$\s$}
\,=\,
\frac{1}{2\parti_0^{1/2}} \; 
\ThetaN{\frac{1}{2}\,\frac{1}{2}}{0\,0}^{\frac{1}{2}}  
\\
\label{ca10}
& & \hspace{-2cm}
\chi_{\s,\s}^{\e}\,
\,=\,
\genusii{$\s$}{$\e$}{$\s$}
\,=\,
\frac{1}{2\parti_0^{1/2}}\;
\ThetaN{\frac{1}{2}\,\frac{1}{2}\vspace{.05cm}}{\frac{1}{2}\,\frac{1}{2}}^{\frac{1}{2}}  .
\eea

The explicit expressions reported above allow us to write the partition functions 
of the complete and non complete models at genus two that we have considered in Sec.\,\ref{JonesIsingn} and Sec.\,\ref{Fermions}.

As for the incomplete models, 
the partition function of $\mathcal{T}_{\textrm{su}(2)_2}$
in (\ref{partgsu22}) for $n=3$ reads
\be 
\label{p21}  \parti_{\textrm{su}(2)_2,2}
\,=\,
\big|\chi_{\id,\id}^{\id} \big|^2
\,=\,
\frac{1}{{4\parti_0}}\left|\; 
\ThetaN{0\,0}{0\,0}^{\frac{1}{2}} \!+\, 
\ThetaN{0\,0}{\frac{1}{2}\,0}^{\frac{1}{2}} \!+\, 
\ThetaN{0\,0}{0\,\frac{1}{2}}^{\frac{1}{2}} \!+\, 
\ThetaN{0\,0}{\frac{1}{2}\,\frac{1}{2}}^{\frac{1}{2}} \; \right|^2 
\ee
while the partition function of $\mathcal{T}_{\mathbb{Z}_2}$
in (\ref{partgz2}) for $n=3$ is
\bea
\label{p22}
\parti_{\mathbb{Z}_2,2} &=& \big|\chi_{\id,\id}^{\id} \big|^2+\big|\chi_{\e,\id}^{\id} \big|^2+\big|\chi_{\id,\e}^{\id} \big|^2+\big|\chi_{\e,\e}^{\id} \big|^2
\\
\rule{0pt}{.9cm}
&=& 
\frac{1}{{4\parti_0}}\Bigg( \left|\ThetaN{0\,0}{0\,0} \right|+\left|\ThetaN{0\,0}{\frac{1}{2}\,0} \right|+ \left|\ThetaN{0\,0}{0\,\frac{1}{2}} \right|+ \left|\ThetaN{0\,0}{\frac{1}{2}\,\frac{1}{2}}  \right|\Bigg) \,.
\nonumber   
\eea

As for the complete models, 
the $g=2$ partition function of the Ising CFT$_2$ in (\ref{partg}) 
reads
\bea 
\label{p23} 
\parti_{\text{\tiny Ising},2} 
&=& 
\big|\chi_{\id,\id}^{\id} \big|^2
+\big|\chi_{\e,\id}^{\id} \big|^2
+\big|\chi_{\id,\e}^{\id} \big|^2
+\big|\chi_{\e,\e}^{\id} \big|^2
\\
\rule{0pt}{.5cm}
& &
+\;\big|\chi_{\s,\id}^{\id} \big|^2
+\big|\chi_{\s,\e}^{\id} \big|^2 
+\big|\chi_{\id,\s}^{\id} \big|^2
+\big|\chi_{\s,\e}^{\id} \big|^2
+\big|\chi_{\s,\s}^{\id} \big|^2
+\big|\chi_{\s,\s}^{\e} \big|^2 
\nonumber
\\
\rule{0pt}{.9cm}
&=& 
\frac{1}{{4\parti_0}}\,
\Bigg(\, \left|\ThetaN{0\,0}{0\,0} \right|+\left|\ThetaN{0\,0}{\frac{1}{2}\,0} \right|+ \left|\ThetaN{0\,0}{0\,\frac{1}{2}} \right| + \left|\ThetaN{0\,0}{\frac{1}{2}\,\frac{1}{2}}  \right|+ \left|\ThetaN{\frac{1}{2}\,0}{0\,0} \right| 
\nonumber
\\ 
& & \hspace{1.3cm}
+\,\left|\ThetaN{\frac{1}{2}\,0}{0\,\frac{1}{2}} \right|+ \left|\ThetaN{0\,\frac{1}{2}}{0\,0} \right|+ \left|\ThetaN{0\,\frac{1}{2}}{\frac{1}{2}\,0}  \right|+ \left|\ThetaN{\frac{1}{2}\,\frac{1}{2}}{0\,0} \right|+\left|\ThetaN{\frac{1}{2}\,\frac{1}{2}\vspace{2pt}}{\frac{1}{2}\,\frac{1}{2}} \right|
\,\Bigg)
\nonumber
\eea
where we recall that the Riemann theta functions with odd characteristics vanish identically. 

From (\ref{partgmajo}),
the partition function of the free Majorana fermion 
(see Sec.\,\ref{Fermions}) on $\mathscr{R}_3$ reads
\bea
\label{p24} 
\parti_{\text{\tiny Majorana},2} 
&=& 
\big|\chi_{\id,\id}^{\id} \big|^2
+\big|\chi_{\e,\id}^{\id} \big|^2
+\big|\chi_{\id,\e}^{\id} \big|^2
+\big|\chi_{\e,\e}^{\id} \big|^2 
+\chi_{\id,\id}^{\id}  \, \bar{\chi}_{\id,\e}^{\id}
+\chi_{\id,\e}^{\id} \bar{\chi}_{\id,\id}^{\id}
\\ 
\rule{0pt}{.5cm}
& &  
+\; \chi_{\id,\id}^{\id} \, \bar{\chi}_{\e,\id}^{\id}
+\chi_{\e,\id}^{\id} \,\bar{\chi}_{\id,\id}^{\id} 
+ \chi_{\e,\e} ^{\id} \, \bar{\chi}_{\id,\e}^{\id}
+\chi_{\id,\e}^{\id} \,\bar{\chi}_{\e,\e}^{\id} 
+\chi_{\e,\e}^{\id} \, \bar{\chi}_{\e,\id}^{\id}
\nonumber
\\
\rule{0pt}{.5cm}
& &
+\;\chi_{\e,\id}^{\id} \, \bar{\chi}_{\e,\e}^{\id} 
+   \chi_{\e,\id}^{\id}   \,\bar{\chi}_{\id,\e}^{\id}
+\chi_{\id,\e}^{\id} \, \bar{\chi}_{\e,\id}^{\id} 
+ \chi_{\id,\id}^{\id}  \, \bar{\chi}_{\e,\e}^{\id}
+\chi_{\e,\e}^{\id} \, \bar{\chi}_{\id,\id}^{\id} 
\nonumber
\\
\rule{0pt}{.9cm}
&=& \frac{1}{\parti_0}\left|\ThetaN{0\,0}{0\,0} \right|
\nonumber
\eea 
where we adopted the shorthand notation given by  ${\chi}_{i_1,i_2}^{k_1}\equiv{\chi}_{i_1,i_2}^{k_1}({\tau_3})$ 
and  $\bar{\chi}_{i_1,i_2}^{k_1}\equiv\bar{\chi}_{i_1,i_2}^{k_1}(\bar{\tau}_3)$.

\section{
A Siegel parabolic subgroup of $\textrm{Sp}(2g, \mathbb{Z})$
}
\label{modular}

In this appendix, we explore some features of the integral symplectic group
$\textrm{Sp}(2g, \mathbb{Z})$ that determine the properties of the partition functions (\ref{partgz2}) and (\ref{partgsu22}).

The integral symplectic group $\textrm{Sp}(2g, \mathbb{Z})$ is made by the 
$(2g)\times (2g)$ matrices made by integers satisfying the following properties
\be
\label{M-matrix-symplectiv-g}
M=\bigg(
\begin{array}{cc} D\; & C \\ B\; & A \end{array}
\bigg)
\;\;\;\qquad\;\;\;
D^{\textrm{t}} A -B^{\textrm{t}} C=\boldsymbol{1}_{g}
\qquad
D^{\textrm{t}}B=B^{\textrm{t}} D
\qquad
C^{\textrm{t}}A=A^{\textrm{t}} C
\ee
where $A,B,C$ and $D$ are integer $g\times g$ matrices and $\boldsymbol{1}_{g}$ is the $g\times g$ identity matrix. 
Notice that the last two conditions in (\ref{M-matrix-symplectiv-g}) 
mean that $B^{\textrm{t}} D$ and $A^{\textrm{t}} C$ 
are symmetric matrices.
Under the transformation $M\in \textrm{Sp}(2g, \mathbb{Z})$ 
in (\ref{M-matrix-symplectiv-g}), 
the period matrix and the canonical homology basis 
of a Riemann surface of genus $g$ transform respectively as \cite{Siegelsym,Faygenus,Mumford91}
\be 
\tau'=(A\tau +B)(C\tau +D)^{-1}
\ee
(which becomes the generic modular transformation of the modular parameter 
of the torus in the special case of $g=1$) and 
\be 
\bigg(
\begin{array}{c} \boldsymbol{a}' \\ \boldsymbol{b}' \end{array}
\bigg)
=
M\bigg(
\begin{array}{c} \boldsymbol{a} \\ \boldsymbol{b} \end{array}
\bigg)
=
\bigg(
\begin{array}{c} D\, \boldsymbol{a}+ C\, \boldsymbol{b} 
\\ 
B\, \boldsymbol{a} + A\, \boldsymbol{b}
\end{array}
\bigg) 
\label{cycletansform}
\ee
where the $g$-dimensional vectors $\boldsymbol{a}$ and $\boldsymbol{b}$
collect the cycles of $a$-type and $b$-type respectively. 
A generic transformation (\ref{M-matrix-symplectiv-g}) induces 
on the Riemann theta function (\ref{Thetafunc}) the following change 
\cite{Faygenus,Alvarez-Gaume:1986rcs,Verlinde:1986kw,Mumford91} 
\be 
\ThetaF{\boldsymbol{p}'}{\boldsymbol{q}'}(\tau') 
\,=\,
\varphi(A,B,C,D)\;
\textrm{e}^{\textrm{i} (\boldsymbol{p}'\cdot \boldsymbol{q}'-\boldsymbol{p}\cdot \boldsymbol{q}+\boldsymbol{p}\cdot (AB)_{\textrm{d}})}\;
\sqrt{\det\left(C\tau +D\right)}\;
\ThetaF{\boldsymbol{p}}{\boldsymbol{q}}(\tau) 
\label{thetatransform}
\ee
where $\varphi(A,B,C,D) = \textrm{e}^{\textrm{i} \omega(A,B,C,D)}$ is a phase 
obeying $\varphi(A,B,C,D)^8=1$ and 
\be
\label{p-q-prime-def}
\boldsymbol{p}' = 
D\,\boldsymbol{p}-C\,\boldsymbol{q}
+ 
\frac{1}{2} \big(CD^{\textrm{t}}\big)_{\textrm{d}}
\;\;\;\qquad\;\;\;
\boldsymbol{q}' = 
-B\,\boldsymbol{p}+A\,\boldsymbol{q}
+ 
\frac{1}{2}  \big( AB^{\textrm{t}}\big)_{\textrm{d}}
\ee
where $X_{\textrm{d}}$ denotes the vector obtained by considering 
the diagonal elements of the matrix $X$. 
From (\ref{charactersIsing}) and (\ref{thetatransform}),
the characters of the Ising CFT$_2$
are shuffled among themselves by a generic transformation (\ref{M-matrix-symplectiv-g}).

In the partition functions of 
$\mathcal{T}_{\mathbb{Z}_2}$ and $\mathcal{T}_{\textrm{su}(2)_2}$
(see (\ref{partgz2}) and (\ref{partgsu22}) respectively),
only the characteristics with $\boldsymbol{p}=\pmb{0}$ occur.
From the first expression in (\ref{p-q-prime-def}), 
it is straightforward to realise that 
the condition $\boldsymbol{p}=\pmb{0}$ is preserved by the 
transformations (\ref{M-matrix-symplectiv-g}) where $C = \boldsymbol{0}_g$,
being $\boldsymbol{0}_g$ defined as the $g \times g$ vanishing matrix. 
Indeed, (\ref{thetatransform}) for this case becomes 
\be 
\ThetaF{\pmb{0}}{A\,\boldsymbol{q} + \tfrac{1}{2} (AB^{\textrm{t}})_{\textrm{d}}}(\tau') 
\,=\,
\varphi(A,B,0,D)\,\sqrt{\det (D)}\;
\ThetaF{\pmb{0}}{\boldsymbol{q}}(\tau)  \,.
\ee

The transformations (\ref{M-matrix-symplectiv-g}) with
$C = \boldsymbol{0}_g$ provide the 
lower Siegel parabolic subgroup of $\textrm{Sp}(2g, \mathbb{Z})$,
whose generic element reads
\cite{Siegelsym,Hulek1998,Geer2006,Maloney:2020nni}
\be 
M=\bigg(\begin{array}{cc} A^{-\textrm{t}} & \boldsymbol{0}_g\, \\ B & A\,\end{array}\bigg)
\;\;\;\;\qquad\;\;\;\; 
B = A\, W
\qquad
W \in \textrm{Sym}(g,\mathbb{Z})
\label{submodular2}
\ee
where $A^{-\textrm{t}} \equiv (A^{\textrm{t}})^{-1}$
and $W$ is a $g \times g$ symmetric matrix made by integers.
The matrix (\ref{submodular2}) 
can be written as follows
\be
    M = \bigg(\begin{array}{cc} A^{-\textrm{t}} & \boldsymbol{0}_g\, \\ \boldsymbol{0}_g & A\,\end{array}\bigg)\bigg(\begin{array}{cc} \boldsymbol{1}_{g} & \boldsymbol{0}_g\, \\ 
    A^{-\textrm{1}}B \; & \boldsymbol{1}_{g}\,\end{array}\bigg)
    \label{Levi-dec}
\ee
where the second matrix in the product occurring in the r.h.s. is 
unipotent\footnote{A matrix $N$ is unipotent when 
$(N-\boldsymbol{1})^k = 0$ for some $k>1$.}.
The matrix in (\ref{submodular2}) is the generic element of  
$\textrm{GL}(g,\mathbb{Z})\ltimes \textrm{Sym}(g,\mathbb{Z})$
and its factorization (\ref{Levi-dec}) is called Levi decomposition.
By using (\ref{cycletansform}), one finds that 
the corresponding transformation on the homology cycles is 
\be 
\bigg(
\begin{matrix} \boldsymbol{a}' \,
\\ \boldsymbol{b}' \,\end{matrix}
\bigg)
=
\bigg(
\begin{array}{c} 
A^{-\textrm{t}}\, \boldsymbol{a}
\\ 
B\, \boldsymbol{a} + A\, \boldsymbol{b}
\end{array}\bigg) \,.
\ee
In the special case of $g=1$, a transformation of $\textrm{SL}(2,\mathbb{Z})$ acts on the torus parameter $\tau$ as the modular transformation 
$\tau'=  \tfrac{a \tau + b }{c\tau+d } $, where all the parameters are integers and $ad-cb=1$. 
When $c=0$, since $ad=1$ for $a,d\in\mathbb{Z}$ implies $a=d=\pm 1$,
this transformation reduces to $\tau'=  \tau + m$ where $m\equiv b/d \in \mathbb{Z}$,
which is equivalent to the transformation obtained 
by applying $m$ times the $T$ transformation (see the text above (\ref{tsinv})).

The parabolic subgroup 
$\textrm{GL}(g,\mathbb{Z})\ltimes \textrm{Sym}(g,\mathbb{Z})$
that we are considering includes also the transformation between the canonical homology bases, introduced in Sec.\,\ref{JonesIsing} and Sec.\,\ref{IsingGenus},
that have been shown in Fig.\,\ref{replica4tilde} and Fig.\,\ref{replica4}
for a special case with $n=4$.
Indeed, these homology bases are related as follows \cite{Coser:2013qda}
\be 
\bigg(
\begin{array}{c} \tilde{\boldsymbol{a}} \\ \tilde{\boldsymbol{b}}\end{array}
\bigg)
=
\bigg(
\begin{array}{cc}  \ilow^{\!\!-1} & 0 \\ 0 & \iup \end{array}
\bigg)
\bigg(
\begin{array}{c} \boldsymbol{a} \\ \boldsymbol{b} \end{array}
\bigg)
\label{tautautilde}
\ee
and the relation between the corresponding 
period matrices is reported in (\ref{periodma}). 
A crucial observation in our analysis is that the parabolic subgroup 
$\textrm{GL}(g,\mathbb{Z})\ltimes \textrm{Sym}(g,\mathbb{Z})$
that we are discussing does not contain the transformation of $\textrm{Sp}(2g, \mathbb{Z})$ induced by the mapping
$\x \mapsto 1-\x$.
Indeed, the following transformation 
\be
M=\bigg(
\begin{array}{cc}  0 &  \iup\\  -(\ilow)^{-1}& 0\end{array}
\bigg)
\;\;\;\;\qquad\;\;\;\; 
{\tau}_n(1-\x)
\,=\,
-\ilow^{\!\!-1} \,{\tau}_n(\x)^{-1} (\iup)^{-1}
\label{x1x2}
\ee
is a special case of (\ref{M-matrix-symplectiv-g}), 
but it does not belong to the subgroup defined by (\ref{submodular2}).
Combining (\ref{periodma}) and (\ref{x1x2}), 
one easily finds (\ref{inverse-g-tilde}) and 
\be
\tilde{\tau}_n(1-\x) \,=\,
\iup\,\,\tilde{\tau}_n(\x)^{-1}\,\ilow \,.
\ee
Specifying (\ref{thetatransform})
to the transformation (\ref{x1x2}), we get
\be 
\ThetaF{\pmb{0}}{\boldsymbol{q}}\big({\tau}_n(1-\x)\big) 
\,=\,
\varphi\big(0,-(\ilow)^{-1},\iup,0\big)
\,\sqrt{\det\left(\tau_n \right)} \;
\ThetaF{-\iup\boldsymbol{q}\vspace{.02cm}}{\pmb{0}}\big({\tau}_n(\x)\big)
\label{theta-relation-app-B}
\ee
which can be employed 
to simplify the crossing asymmetries in (\ref{asyft-z2}) and (\ref{asyft-su2}).
Indeed, (\ref{theta-relation-app-B}) leads to (\ref{transform00m}),
which allows to write (\ref{asygz2}) and (\ref{asygsu22}).

We find it instructive to conclude this discussion by reporting 
the explicit transformations of the genus two characters 
described in the appendix\;\ref{g2characters}
under the transformation (\ref{tautautilde}).
First, notice that the period matrix for (\ref{periodm}) 
for $n=3$ simplifies to
\be 
{\tau_3}=\frac{w_3}{\sqrt{3}}
\bigg(
\begin{array}{cc}  -2 & \,1  \\ 1 & -2  \end{array}
\bigg) 
\;\;\;\qquad\;\;\;
w_3 \equiv -\,\textrm{i} \,\frac{{}_2F_1(1/3,2/3,1,1-\x)}{{}_2F_1(1/3,2/3,1,\x)} \;.
\ee
The transformation (\ref{tautautilde}), that provides (\ref{periodma}), 
in this case becomes
\be 
\tilde{\tau}_3
=
\bigg(\begin{array}{cc}  1\, & 1  \\ 1\, & 0  \end{array}\bigg)
\,\tau_3\,
\bigg(\begin{array}{cc}  0\, & 1  \\ 1\, & 1  \end{array}\bigg)
= 
-\frac{w_3}{\sqrt{3}} 
\bigg(\begin{array}{cc}  2\, & 1  \\ 1\, & 2  \end{array}\bigg)  \,.
\label{t22}
\ee
By employing (\ref{cycletansform}), for the cycles of the homology bases we find 
\be 
\left(
\begin{array}{c} 
\tilde{a}_1 \\ \tilde{a}_2 \\ \tilde{b}_1 \\ \tilde{b}_2
\end{array}
\right)
=
\left(
\begin{array}{cccc} 
1 \,& \,0 \,& \,0 \,& 0 \,\\ -1 & 1 & 0 & 0 \\ 0 & 0 & 1 & 1 \\ 0 & 0 & 1 & 0 
\end{array}
\right)
\left(
\begin{array}{c} {a}_1\\ {a}_2 \\ {b}_1\\ {b}_2 \end{array}
\right)
\ee
while (\ref{thetatransform}) specified to this case becomes 
\be 
\ThetaF{p_1 \;\;\; p_2-p_1}{q_1+q_2 \;\;\; q_2}(\tilde{\tau}_3) 
\,=\, 
\varphi\big(\iup,0,0,(\ilow)^{-1}\big)
\,\ThetaF{p_1 \; p_2}{q_1 \; q_2}({\tau_3})  \,.
\ee
This allows us to find that the transformations of the characters (\ref{ca1})-(\ref{ca4}) that do not involve the field $\sigma$ transform among themselves.
Indeed, we get
\be
\chi_{\id,\id}^{\id} (\tilde{\tau}_3)=\chi_{\id,\id}^{\id} (\tau_3)
\qquad
\chi_{\e,\id}^{\id} (\tilde{\tau}_3)=\chi_{\e,\e}^{\id} (\tau_3) 
\qquad
\chi_{\id,\e}^{\id} (\tilde{\tau}_3)=\chi_{\id,\e}^{\id} (\tau_3)
\qquad
\chi_{\e,\e}^{\id} (\tilde{\tau}_3)=\chi_{\e,\id}^{\id} (\tau_3)
\ee
while for the characters (\ref{ca5})-(\ref{ca10}) we get
\bea
& \hspace{-1.1cm}
\chi_{\s,\id}^{\id} (\tilde{\tau}_3)
\,=\,
\frac{1}{\sqrt{2}}
\big(\chi_{\s,\s}^{\id} (\tau_3)+\chi_{\s,\s}^{\e} (\tau_3)\big)
\qquad 
&
\chi_{\s,\e}^{\id} (\tilde{\tau}_3)
\,=\,
\frac{1}{\sqrt{2}}
\big(\chi_{\s,\s}^{\id} (\tau_3)-\chi_{\s,\s}^{\e} (\tau_3)\big) 
\nonumber 
\\ 
\rule{0pt}{.6cm}
& \hspace{-1.1cm}
\chi_{\id,\s}^{\id} (\tilde{\tau}_3)
\,=\,
\chi_{\id,\s}^{\id} (\tau_3) 
\qquad\qquad\qquad\qquad\qquad
&
\chi_{\s,\e}^{\id} (\tilde{\tau}_3)
\,=\,
\chi_{\s,\e}^{\id} (\tau_3)
\\ 
\rule{0pt}{.6cm}
& \hspace{-1.1cm}
\chi_{\s,\s} ^{\id}(\tilde{\tau}_3)
\,=\,
\sqrt{2}
\big(\chi_{\s,\id}^{\id} (\tau_3)+\chi_{\s,\e}^{\id} (\tau_3) \big)
\qquad 
&
\chi_{\s,\s}^{\e} (\tilde{\tau}_3)
\,=\,
\sqrt{2}\big(\chi_{\s,\id}^{\id} (\tau_3)-\chi_{\s,\e}^{\id} (\tau_3) 
\big) \,.
\nonumber
\eea
This shows that the genus two partition functions  (\ref{p21})-(\ref{p24}) are invariant under the transformation (\ref{t22}).


\bibliographystyle{nb}
\bibliography{refsJones}

\end{document}